\documentclass[11pt,epsf]{article}
\usepackage{hyperref}
\usepackage{subfigure}
\usepackage{graphics}
\usepackage{amssymb}
\usepackage{epsf}
\usepackage{rotating}
\usepackage{axodraw}
\usepackage{pstricks}

\def\g{\gamma}

\def\ds#1{#1\kern-1ex\hbox{/}}
\def\dsh{h\kern-1.2ex /}

\newcommand{\bea}{\begin{eqnarray}}
\newcommand{\eea}{\end{eqnarray}}

\def\beq{\begin{equation}}
\def\eeq{\end{equation}}

\def\beqn{\begin{eqnarray}}
\def\eeqn{\end{eqnarray}}
\def\ba{\begin{eqnarray}}
\def\ea{\end{eqnarray}}

\newcommand{\beqa}{\begin{eqnarray}}
\newcommand{\eeqa}{\end{eqnarray}}

\begin{document}
\begin{center}
\vspace{1.cm}

{\bf
An Anomalous Extra Z Prime from Intersecting Branes\\
with Drell-Yan and Direct Photons at the LHC}

\vspace{1.5cm}
{\bf\large Roberta Armillis$^{a}$, Claudio Corian\`{o} $^{a\,b}$,  Marco Guzzi $^{a}$ and Simone Morelli$^{a}$}

\vspace{1cm}

{\it  $^a$Dipartimento di Fisica, Universit\`{a} del Salento \\
and  INFN Sezione di Lecce,  Via Arnesano 73100 Lecce, Italy}\\
\vspace{.2cm}
{\it and} \\

{\it
$^b$Department of Physics and Institute of Plasma Physics \\
 University of Crete, 71003 Heraklion, Greece\\}
\vspace{.12in}
~\\

\begin{abstract}
We quantify the impact of gauge anomalies at the Large Hadron Collider by studying the
invariant mass distributions in Drell-Yan and in double prompt photon, using an extension
of the Standard Model characterized by an additional
anomalous $U(1)$ derived from intersecting branes.
The approach is rather general and applies to any anomalous abelian gauge current.
Anomalies are cancelled using either  the Wess-Zumino
mechanism with suitable Peccei-Quinn-like interactions
and a St\"uckelberg axion, or by the Green-Schwarz mechanism.
We compare predictions for the corresponding extra Z prime to anomaly-free
realizations such as those involving $U(1)_{B-L}$.
We identify the leading anomalous corrections to both channels,
which contribute at higher orders, and compare them against the
next-to-next-to-leading order (NNLO) QCD background.
Anomalous effects in these inclusive observables are found to be very
small, far below the percent level and below the size of the typical
QCD corrections quantified by NNLO $K$-factors.
\end{abstract}
\end{center}
\newpage

%\tableofcontents

\section{Introduction}
The study of anomalous gauge interactions at the
LHC and at future linear colliders is for sure a difficult topic,
but also an open possibility that deserves close theoretical and experimental attention.
Hopefully, these studies will be able to establish if an
additional anomalous extra $Z^{\prime}$ is present in the spectrum, introduced by an
abelian extension of the gauge structure of the Standard Model (SM), assuming that
extra neutral currents will be found in the next several years of
running of the LHC \cite{Langacker:2008yv}. The interactions that we discuss are characterized by
 {\em genuine anomalous vertices}  in which gauge anomalies
cancel in some non trivial way, not by a suitable (anomaly-free)
charge assignment of the chiral fermion spectrum for each generation.

The phenomenological investigation of this topic is rather new, while
various mechanisms of cancellation of the gauge anomalies involving
axions have been around for quite some time. Global anomalies,
for instance, introduced for the solution of the strong CP problem,
such as the Peccei-Quinn (PQ) solution \cite{Peccei:1977hh, Peccei:1977ur, Weinberg:1977ma, Wilczek:1977pj, Kim:1979if, Shifman:1979if, Dine:1981rt, Zhitnitsky:1980tq}
(reviewed in \cite{Peccei:2006as}) require an axion, while local anomalies, cancelled by a Wess-Zumino counterterm,
allow an axion-like particle in the spectrum, whose mass and gauge coupling
- differently from PQ axions - are independent. Similar constructions
hold also in the supersymmetric case and a generalization
of the WZ mechanism is the Green-Schwarz mechanism (GS) of string theory.
The two mechanisms  are related but not identical \cite{Coriano:2008pg},
the first of them being characterized by a unitarity bound.
Details on the relation between the two at the level of effective field theory can be found in
\cite{Coriano:2008pg, Armillis:2008bg}.

Intersecting brane models, in which several anomalous $U(1)$'s
and St\"uckelberg mass terms are present, may offer a realization of these constructions
\cite{Antoniadis:2001np, Ibanez:2001nd, Kiritsis:2003mc, Blumenhagen:2006ci},
which can also be investigated in a bottom-up approach by using effective
lagrangeans built out of the requirements of gauge invariance of the
1-loop effective action \cite{Coriano:2005js}.
In our analysis we will consider the simplest extension of these anomalous abelian gauge factors, which
involves a single anomalous $U(1)$, denoted as $U(1)_B$. The corresponding anomalous gauge boson
($B$) gets its mass via a combination of the Higgs and of St\"uckelberg mechanisms. Axions play a key role
in the cancellation of the anomalies in these theories although they may appear in other constructions
as well, due to the decoupling of a chiral fermion  in anomaly-free theories \cite{Coriano:2006xh}.

The presence of an anomalous $U(1)$ in effective models derived from string theory is quite common, although in all the previous literature before \cite{Coriano:2005js} and \cite{Coriano:2006xh,Coriano:2007fw,Coriano:2007xg} the phenomenological relevance of the anomalous $U(1)$ had not been worked out in any detail.
In particular, the dynamics of the anomalous extra gauge interaction had been neglected, by invoking a decoupling of the anomalous sector on the assumption of a large mass of the extra gauge boson.
In \cite{Coriano:2005js} it was shown that only one physical axion appears in the spectrum of these models, independently of the number of abelian factors, which is the most important feature of these realizations.
In our case, the axion can be massless or massive, depending on the structure of the scalar potential.
Recent developements in the study of these models include their supersymmetric extensions \cite{Anastasopoulos:2008jt} and their derivations as symplectic forms of supergravity \cite{DeRydt:2007vg,
DeRydt:2008hw}. Other interesting variants include the St\"uckelberg extensions considered in
\cite{Feldman:2006wb,Feldman:2006ce,Feldman:2007wj} which depart significantly from the Minimal Low Scale Orientifold Model (MLSOM) introduced in \cite{Coriano:2005js} and discussed below. Specifically these models are also characterized by the presence of two mechanisms of symmetry breaking (Higgs and
St\"uckelberg) but do not share the anomalous structure.  As such they do not describe the anomalous
$U(1)$'s of these special vacua of string theory.

Axion-like particles, beside being a natural candidate for dark matter, may play a role in explaining some puzzling results regarding the propagation of high energy gamma rays  \cite{DeAngelis:2007dy,DeAngelis:2008sk} due to the oscillations of photons into axions in the presence of intergalactic magnetic field. In general, the presence of independent mass/coupling relations for these particles allows to evade most of the experimental bounds coming from \textsc{CAST} and other experiments on the detection of PQ axions, characterized by a suppression of both mass and gauge couplings of this particle by the same large scale ($10^{10}-10^{12}$ GeV), (see \cite{Jaeckel:2006xm,Ahlers:2007qf}).
While a phenomenological study of St\"uckelberg axions is underway in a related work, here
we focus our attention on the gauge sector, quantifying the rates for the detection of anomalous neutral currents at the LHC in some specific and very important channels.

\begin{itemize}
\item{\bf Drell-Yan}
\end{itemize}
Being leptoproduction the best way to search for extra neutral
interactions, it is then obvious that the study of the anomalous vertices and
of possible anomalous extra $Z^\prime$ should seriously consider the investigation of this process.
We describe the modifications induced on Drell-Yan computed in the Standard Model (SM)
starting from the description of some of the properties of the new anomalous vertices and of the corresponding 1-loop counterterms, before moving to the analysis of the corrections. These appear - both in the WZ and GS cases in the relevant partonic channels at NNLO in the strong coupling constant
($O(\alpha_s^2)$). We perform several comparisons between anomalous
and non-anomalous extra $Z^{\prime}$ models
and quantify the differences with high accuracy.
\begin{itemize}
\item{\bf Direct Photons (Di-photon, DP)}
\end{itemize}
Double prompt (direct) photons offer an interesting signal which
is deprived of the fragmentation contributions especially at large
values of their invariant mass $Q$, due to the steep falling of the
photon fragmentation functions. In addition, photon isolation may
provide an additional help in selecting those events coming from channels in which
the contribution of the anomaly is more sizeable, such as gluon fusion.
Also in this case we perform a detailed investigation of this sector.
For direct photons, the anomaly appears in gluon fusion -at parton level
- in a class of amplitudes which are characterized by two-triangles graphs
- or BIM amplitudes - using the definitions of \cite{Coriano:2008pg}.

In both cases the quantification of the background needs extreme care,
due to the small signal, and the investigation of the renormalization/factorization
scale dependence of the predictions is of outmost importance. In particular,
we consider all the sources of scale-dependence in the analysis, including
those coming from the evolution of the parton densities (\textsc{Pdf}'s)
which are just by themeselves enough to overshadow the anomalous corrections.
For this reason we have used the program \textsc{Candia} in the evolution
of the  \textsc{Pdf}'s,  which has been documented in \cite{Cafarella:2008du}.
The implementations of DY and DP are part of two programs $\textsc{Candia}_{DY}$
and $\textsc{Candia}_{Axion}$ for the study of the QCD background with the
modifications induced by the anomalous signal. The QCD background in DP
is computed using \textsc{Diphox} \cite{Binoth:1999qq} and \textsc{Gamma2MC}
based on Ref. \cite{Bern:2002jx}. The NLO corrections to DP before the
implementation of \textsc{Diphox} have been computed
by Gordon and one of the authors back in 1995 \cite{Coriano:1996us}
and implemented in a Monte Carlo based on the phase space slicing method.

In the numerical analysis that we present we have included all the
contributions coming from the two mechanisms as separate cases,
corrections that are implemented in DY and DP. We will start
analyzing the contributions to these processes in more detail in the next
sections, discussing the specific features of the anomalous contributions and of the corresponding
counterterms at a phenomenological level. To make our treatment self-contained we have summarized
some of the properties of effective vertices in these models, relevant for our analysis.

Our work is organized as follows.
After a brief description of the anomalous interactions
and of the counterterms that appear either at lagrangean level
(WZ case) or at the level of the trilinear gauge vertex (GS case),
we discuss the main properties of these vertices and we address the
structure of the corrections in DY and in DP. Our study is mainly focused
on the invariant mass distributions in the two cases. The need for performing
these types of analysis in parallel will be explained below, and there is the
hope that it may be extended to other processes and observables  in the future,
such as rapidity distributions and rapidity correlations \cite{Chang:1997sn}.
We present high precision estimates of the QCD background at NNLO, which is
the order where, in these processes, the anomalous corrections start
to appear. Other analysis, of course, are also possible, such as those
involving 4-fermion decays in trilinear gauge interactions which could,
in principle, be sensitive to Chern-Simons terms
\cite{Coriano:2005js, Anastasopoulos:2006cz} if at least two anomalous $U(1)$'s
are present in the spectrum. These additional interactions are allowed
\cite{Armillis:2007tb} whenever the distribution of the partial anomalies
on a diagram is not fixed by symmetry requirements.
A complete description of these vertices has been carried out in
\cite{Armillis:2007tb}, useful for direct phenomenological studies.
This search, however, is expected to be experimentally also very difficult.

As we are going to show, the search for effects due to anomalous $U(1)$ at the LHC in
$pp$ collisions cannot avoid an analysis of the QCD background at NNLO.
DY and DP are the only two cases where this level of precision has been
obtained in perturbation theory. As we are going to show, the anomalous
effects at the LHC in these two key processes are tiny, since the invariant
mass distributions are down by a factor of $10^{3}-10^{4}$ compared to the
NNLO (QCD) background. The accuracy required at the LHC to identify these
effects on these observables should be of a fraction of a percent
($0.1 \%$ and below), which is beyond reach at a hadron collider due
to the larger indetermination intrinsic in QCD factorization and the parton model.

\section{Anomaly-free versus an anomalous extra $Z^{\prime}$ in Drell-Yan}

As we have already mentioned, the best mode to search for extra $Z^{\prime}$
at the LHC is in the production of a lepton pair via the Drell-Yan mechanism
($q\bar{q}$ annihilation) mediated by neutral currents. The final state is
easily tagged and resonant due to the $s$-channel exchange of the extra gauge boson.
In particular, a new heavier gauge boson modifies the invariant mass distribution
also on the $Z$ peak due to the small modifications induced on the couplings and
to the $Z-Z^{\prime}$ interference. In the anomalous model
that we have investigated, though based on a specific charge assignment, we find
larger rates for these distributions both on the peak of the $Z$ and of the
$Z^{\prime}$ compared to the other models investigated, if the extra resonance
is around 1 TeV. This correlation is expected to drop as the mass of the extra
$Z^{\prime}$ increases. In our case, as we will specify below, the mass of the extra resonance is
given by the St\"uckelberg ($M_1$) mass, which appears also (as a suppression scale)
in the interaction of the physical axion to the gluons and is essentially a free parameter.

In DY, the investigation of the NNLO hard scatterings goes
back to \cite{Hamberg:1990np}, with a complete computation of the invariant
mass distributions, made before that the NNLO corrections to the DGLAP evolution
had been fully completed. In our analysis we will compare three anomaly-free
models against a model of intersecting brane with a single anomalous $U(1)$.
The anomaly-free charge assignments come from a gauged $B-L$ abelian symmetry,
a ``$q+u$'' model -both described in \cite{Carena:2004xs} - and the free fermionic model analyzed in
\cite{Coriano:2008wf}. We start by summarizing our definitions and conventions.

In the anomaly-free case we address abelian extensions of the gauge structure
of the form $SU(3)\times SU(2) \times U(1)_Y \times U(1)_z$,
with a covariant derivative in the $W_{\mu}^{3},B_{Y}^{\mu},B_{z}^{\mu}$
(interaction) basis defined as
\ba
\hat{D}_{\mu}=\left[\partial_{\mu} -i g_2 \left( W_{\mu}^{1}T^{1} +
W_{\mu}^{2}T^{2} + W_{\mu}^{3}T^{3} \right) -i\frac{g_{Y}}{2}\hat{Y}
B_{Y}^{\mu}-i\frac{g_{z}}{2}\hat{z} B_{z}^{\mu} \right]
\ea
where we denote with $g_2,g_Y, g_z$ the couplings of $SU(2)$, $U(1)_Y$ and
$U(1)_z$, with  $\tan\theta_W=g_Y/g_2$. After the diagonalization of the
mass matrix we have
\ba
\left( \begin{array}{c}
A_{\mu} \\
Z_{\mu}  \\
Z^{\prime}_{\mu}
\end{array} \right)
=
\left( \begin{array}{ccc}
\sin\theta_W & \cos\theta_W & 0\\
\cos\theta_W & -\sin\theta_W & \varepsilon \\
-\varepsilon\sin\theta_W& \varepsilon\sin\theta_W & 1
\end{array} \right)
\left( \begin{array}{c}
W^{3}_{\mu} \\
B^{Y}_{\mu}  \\
B^{z}_{\mu}
\end{array} \right)
\label{massmatrixnoanomaly}
\ea
where $\varepsilon$ is a perturbative parameter which is
around $10^{-3}$ for the models analyzed, introduced in
\cite{Carena:2004xs} and \cite{Coriano:2008wf}. It  is defined as
\ba
\varepsilon=\frac{\delta M^2_{Z
Z^{\prime}}}{M^2_{Z^{\prime}}-M^2_{Z}}
\ea
while the mass of the $Z$ boson and of the extra $Z^{\prime}$ are
\ba
&&M_Z^2=\frac{g_2^2}{4
\cos^2\theta_W}(v_{H_1}^2+v_{H_2}^2)\left[1+O(\varepsilon^2)\right]
\nonumber\\
&&M_{Z^{\prime}}^2=\frac{g_z^2}{4}(z_{H_1}^2
v_{H_1}^2+z_{H_2}^2v_{H_2}^2+z_{\phi}^2
v_{\phi}^2)\left[1+O(\varepsilon^2)\right]
\nonumber\\
&&\delta M^2_{Z Z^{\prime}}=-\frac{g_2 g_z}{4\cos\theta_W}(z_{H_1}^2
v_{H_1}^2+z_{H_2}^2 v_{H_2}^2).
\ea
In this class of models we have two Higgs doublet $H_1$ and $H_2$,
whose vevs are $v_{H_1}$ and $v_{H_2}$ and an extra $SU(2)_{W}$ singlet $\phi$
whose vev is $v_{\phi}$. The extra $U(1)_z$ charges of the Higgs doublet
are respectively $z_{H_1}$ and $z_{H_2}$, while for the singlet this is denoted as $z_{\phi}$.
Taking the value of $v_{H_2}$ of the order of the electroweak scale ($\approx 246$ GeV),
we fix $v_{H_1}$ with $\tan\beta = v_{H_2}/v_{H_1}$, and we still have one free parameter, $v_{\phi}$, which enters in the calculation of the mass of the extra $Z^{\prime}$.
Then it is obvious that we can take the mass $M_{Z^{\prime}}$ and the coupling constant
$g_z$ as free parameters. We choose $\tan\beta\approx 40$ in order to reproduce the
mass of the $Z$ boson at $91.187$ GeV, choice that is performed, for consistency, also in the anomalous model. In this last case the Higgs sector is characterized only by 2 Higgs doublets, with the vev of the extra singlet being replaced by the St\"uckelberg mass. We define $g_2 \sin\theta_W  =g_Y \cos\theta_W =e$
and construct the $W^{\pm}$ charge eigenstates and the corresponding
generators $T^{\pm}$ as usual
\ba
&&W^{\pm}=\frac{W_1\mp iW_2}{\sqrt{2}}\nonumber\\
&&T^{\pm}=\frac{T_1\pm iT_2}{\sqrt{2}},
\ea
while in the neutral sector we introduce the rotation matrix
\ba
\left( \begin{array}{c}
W^{3}_{\mu} \\
B^{Y}_{\mu}  \\
B^{z}_{\mu}
\end{array} \right)
=
\left( \begin{array}{ccc}
\frac{\sin\theta_W (1+\varepsilon^2)}{1+\varepsilon^2} &
\frac{\cos\theta_W}{1+\varepsilon^2} & \varepsilon
\frac{\cos\theta_W}{1+\varepsilon^2}\\
\frac{\cos\theta_W(1+\varepsilon^2)}{1+\varepsilon^2} &
-\frac{\sin\theta_W}{1+\varepsilon^2} & \varepsilon
\frac{\sin\theta_W}{1+\varepsilon^2}\\
0 & \frac{\varepsilon}{1+\varepsilon^2}& \frac{1}{1+\varepsilon^2}
\end{array} \right)
\left( \begin{array}{c}
A_{\mu} \\
Z_{\mu}  \\
Z^{\prime}_{\mu}
\end{array} \right)
\ea
which relates the interaction and the mass eigenstates.
Substituting these expression in the covariant derivative we obtain
\ba
&&\hat{D}_{\mu}=\left[\partial_{\mu} -i A_{\mu} \left(g_2 T_3\sin\theta_W+
g_Y\cos\theta_W \frac{\hat{Y}}{2}\right)
-ig_2\left(W^{-}_{\mu}T^{-}+ W^{+}_{\mu}T^{+}\right)\right.
\nonumber\\
&&\hspace{1cm}\left.-iZ_{\mu}\left( g_2\cos\theta_W T_{3} -g_Y \sin\theta_W
\frac{\hat{Y}}{2}
+g_z \varepsilon\frac{\hat{z}}{2}\right)\right.
\nonumber\\
&&\hspace{1cm}\left.-iZ^{\prime}_{\mu}\left(-g_2\cos\theta_W
T_{3}\varepsilon +g_Y\sin\theta_W \frac{\hat{Y}}{2}\varepsilon
+g_z\frac{\hat{z}}{2}\right)\right]
\ea
where we have neglected all the $O(\varepsilon^2)$ terms.
Sending $g_z\rightarrow 0$ and $\varepsilon\rightarrow 0$
we obtain the SM expression.
The vector and the
axial couplings of the $Z$ and $Z^{\prime}$ to the fermions
are expressed equivalently in terms of the left - ($z_L$) and right
- ($z_R$) $U(1)_z$ chiral charges  and hypercharges ($Y_R$, $Y_L$)
of the  models that we have implemented. These can be found in
\cite{Coriano:2008wf} for the free fermionic case and in
\cite{Carena:2004xs} for the remaining models with a V-A structure given by

\ba
&&\frac{-i g_2}{4 c_w}\gamma^{\mu} {g_V}^{Z,j}=\frac{-i g_2}{c_w}
\frac{1}{2}\left[c_w^2
T_3^{L,j}-s_w^2(\frac{\hat{Y}^{j}_L}{2}+\frac{\hat{Y}^{j}_R}{2})
+\varepsilon \frac{g_z}{g_2} c_w
(\frac{\hat{z}_{L,j}}{2}+\frac{\hat{z}_{R,j}}{2})\right]\gamma^{\mu}
\nonumber\\
&&\frac{-i g_2}{4 c_w}\gamma^{\mu}\gamma^{5} {g_A}^{Z,j}=\frac{-i g_2}
{c_w}\frac{1}{2}\left[-c_w^2 T_3^{L,j}
-s_w^2(\frac{\hat{Y}^{j}_R}{2}-\frac{\hat{Y}^{j}_L}{2})
+\varepsilon \frac{g_z}{g_2} c_w
(\frac{\hat{z}_{R,j}}{2}-\frac{\hat{z}_{L,j}}{2})\right]\gamma^{\mu}\gamma^{5}
\nonumber\\
&&\frac{-i g_2}{4 c_w}\gamma^{\mu} {g_V}^{Z^{\prime},j}=\frac{-i g_2}{c_w}
\frac{1}{2}\left[ -\varepsilon c_w^2 T_3^{L,j}
+\varepsilon s_w^2(\frac{\hat{Y}^{j}_L}{2}+\frac{\hat{Y}^{j}_R}{2})
+\frac{g_z}{g_2}c_w(\frac{\hat{z}_{L,j}}{2}+
\frac{\hat{z}_{R,j}}{2})\right]\gamma^{\mu}
\nonumber\\
&&\frac{-i g_2}{4 c_w}\gamma^{\mu}\gamma^{5} {g_A}^{Z^{\prime},j}=\frac{-i
g_2}{c_w} \frac{1}{2}\left[ \varepsilon c_w^2 T_3^{L,j}
+\varepsilon s_w^2(\frac{\hat{Y}^{j}_R}{2}-\frac{\hat{Y}^{j}_L}{2})
+\frac{g_z}{g_2}c_w(\frac{\hat{z}_{R,j}}{2}-
\frac{\hat{z}_{L,j}}{2})\right]\gamma^{\mu}\gamma^{5},
\nonumber\\
\ea
where $j$ is an index which represents the quark or the lepton and we have
set $\sin\theta_W=s_w,\cos\theta_W=c_w$ for brevity.

\subsection{An anomalous extra $Z^{\prime}$}

In presence of anomalous interactions we can use the same formalism developed so far
for anomaly-free models with some appropriate changes.
Since the effective lagrangean of the class of the anomalous
models that we are investigating includes both a St\"uckelberg and a two-Higgs doublet sector,
the masses of the neutral gauge bosons are provided by a combination
of these two mechanisms. In this case we take as free parameters
the St\"ueckelberg mass $M_1$ and the anomalous coupling
constant $g_B$, with $\tan\beta$ as in the remaining anomaly-free models. As we have
already stressed, the analysis does not depend significantly on the choice of this parameter.
The value of the St\"uckelberg mass $M_1$
is loosely constrained by the D-brane model in terms of suitable wrappings ($n$) of the 4-branes
which define the charge embedding \cite{Ghilencea:2002da, Ibanez:2001nd}
reported in Tabs. \ref{parameters},\ref{charge_higgs},\ref{tabpssm} and \ref{charges}.

\begin{figure}[t]
\begin{center}
\includegraphics[width=6.5cm,angle=-90]{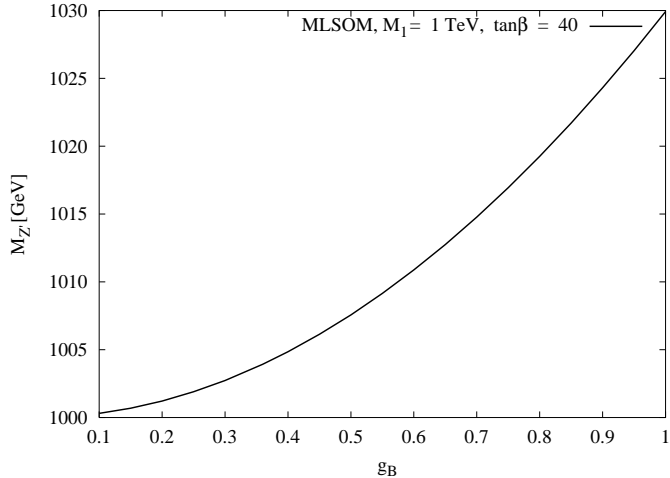}
\caption{\small Anomalous $M_{Z'}$ as a function of the coupling $g_B$.}
\label{MZp_mass_gB}
\end{center}
\end{figure}

The mass-matrix in the neutral gauge sector is given by
\begin{displaymath}
{\cal L}_{mass}=\left(W_3,~Y,~B\right){\bf M}^2\left(\begin{array}{c}
W_3\\
Y\\
B\end{array}\right),
\end{displaymath}
where $B$ is the St\"uckelberg field and the mass matrix is defined as
\bea
{\bf M}^2 = {1\over 4} \pmatrix{
{g^{}_2}^{2} v^2 & - {g^{}_2} \, {g^{}_Y} v^2 &  - {g^{}_2} \,  x^{}_B \cr
 - {g^{}_2} \,{g^{}_Y} v^2 &  {g^{}_Y}^{2} v^2 & {g^{}_Y}  x^{}_B \cr
 -{g^{}_2} \, x^{}_B &{g^{}_Y}  x^{}_B  & 2 M_1^2 + N^{}_{BB}}
\label{massmatrix}
\eea
with
\bea
N^{}_{BB}=  \left( q_u^{B\,2} \,{v^{\,2}_u} + q_d^{B\,2} \,{v^{\,2}_d} \right)\, g_B^{\,2},
&& x^{}_B=  \left(q_u^B {v^{\,2}_u} + q_d^B {v^{\,2}_d}  \right)\, g^{}_B.
\eea
Here $v_u$ and $v_d$ denote the vevs of the two Higgs fields $H_u,H_d$ while
$q_u^{B}$ and $q_d^{B}$ are the Higgs charges under the extra anomalous $U(1)_B$.
We have also defined $v=\sqrt{v_u^2+v_d^2}$ and $g=\sqrt{g_2^2+g_Y^2}$.
The massless eigenvalue of the mass matrix is associated to the photon $A^{}_{\gamma}$, while the two non-zero mass
eigenvalues denote the masses of the $Z$ and of the ${Z}^{\prime}$ vector bosons. These are given by
\bea
M_{Z}^2 &=&  \frac{1}{4} \left( 2 M_1^2 + g^2 v^2 + N^{}_{BB}
- \sqrt{\left(2  M_{1}^2 - g^2 v^2 + N^{}_{BB} \right)^2 + 4
   g^2 x_{B}^2} \right)    \\
&\simeq&     \frac{g^2 v^2}{2} - \frac{1}{M_{1}^2} \frac{g^2 x_{B}^{2}}{4}
 + \frac{1}{M_{1}^4}\frac{g^2 x_{B}^2}{8 } (N^{}_{BB} - g^2 v^2) , \nonumber\\
\nonumber\\
M_{{Z}^\prime}^2 &=&   \frac{1}{4} \left( 2 M_1^2 + g^2 v^2 + N^{}_{BB}
+ \sqrt{\left(2  M_{1}^2 - g^2 v^2 + N^{}_{BB} \right)^2 + 4   g^2 x_{B}^2} \right)   \\
&\simeq&    M^{2}_{1} +  \frac{N^{}_{BB}}{2} . \nonumber
\label{corr}
\eea
The mass of the $Z$ gauge boson gets corrected by terms
of the order $v^{2}/M_1$, see Fig.~\ref{MZp_mass_gB}, converging to the SM value as $M_1\to \infty$, while the mass of the $Z'$ gauge boson can grow large with $M_1$.
The physical gauge fields can be obtained from the rotation matrix $O^A$
\ba
\pmatrix{A_\g \cr Z \cr {{Z^\prime}}} =
O^A\, \pmatrix{W_3 \cr A^Y \cr B}  \label{OA}
\ea
which can be approximated at the first order as

\bea
O^A  \simeq  \pmatrix{
\frac{g^{}_Y}{g}           &     \frac{g^{}_2}{g}         &      0   \cr
\frac{g^{}_2}{g} + O(\epsilon_1^2)          &     -\frac{g^{}_Y}{g} + O(\epsilon_1^2) &      \frac{g}{2} \epsilon_1 \cr
-\frac{g^{}_2}{2}\epsilon_1     &     \frac{g^{}_Y}{2}\epsilon_1  &   1 + O(\epsilon_1^2) }
\label{matrixO}
\eea
which is the analogue of the matrix in Eq.~(\ref{massmatrixnoanomaly}) for the anomaly-free models, but here the role of the mixing parameter $\epsilon_1$ is taken by  the expression
\ba
\epsilon_1=\frac{x_B}{M_1^2}.
\label{alogue}
\ea
A relation between the two expansion parameters can be easily obtained in an approximate way by a direct comparison. For simplicity we take all the charges to be O(1) in all the models obtaining

\ba
&&M_Z^2\sim {g_2^2} v^2
\nonumber\\
&&M_{Z^{\prime}}^2 - M_Z^2 \sim g_z^2 v_{\phi }^2 \nonumber \\
&& \delta M^2_{Z Z^{\prime}}\sim{g_2 g_z} v^2
\ea
giving
\beq
\epsilon_1\sim \frac{v^2}{M_1^2},
\eeq
which is the analogue of Eq. (\ref{alogue}), having identified the St\"uckelberg mass with the vev of the extra singlet Higgs, $M_1\sim g_z v_{\phi}$. This is natural since the St\"uckelberg mechanism can be thought of as the low energy remnant of an extra Higgs whose radial fluctuations have been frozen and with the imaginary phase surviving at low energy as a CP-odd scalar \cite{Coriano:2006xh}.

Concerning the charge assignments, the corresponding model is obtained form the
intersection of 4 branes $(a,b,c,d)$ with generators $(q_a,q_b,q_c,q_d)$ which are rotated to the hypercharge basis, with an anomaly free hypercharge.  The $U(1)_{a}$ and $U(1)_{d}$ symmetries are proportional to the baryon number and the lepton number respectively. The $U(1)_{c}$ symmetry can be considered as the third component of the right-handed weak isospin; the $U(1)_{b}$ is a PQ-like symmetry. A discussion of the construction can be found in \cite{Ibanez:2001nd} and
\cite{Ghilencea:2002da}. The identification of the generators involve the solution of some constraint
equations, solutions which for a $T^6$ compactification are parametrized by a phase $\epsilon =\pm1$; the Neveu-Schwarz background
on the first two tori $\beta_i=1-b_i=1,1/2$, four integers
$n_{a2},n_{b1},n_{c1},n_{d2}$, which are the wrapping numbers of the branes around the extra (toroidal) manifolds of the compactification, and a parameter $\rho=1,1/3$, with an additional constraint in order to obtain the correct massless hypercharge.  One of the choice for these parameters is reported in Table
\ref{parameters}.

\begin{table}[t]
\begin{center}
\begin{tabular}{|c|c|c|c|c|c|c|c|c|}
\hline
    $\nu$  & $\beta_1$ & $\beta_2 $ & $n_{a2}$  &  $n_{b1}$ & $n_{c1}$ & $n_{d2}$ \\
\hline  1/3 & 1/2  & $  1 $ &  $n_{a2}$ &  -1 & 1 & 1 - $n_{a2}$\\
\hline
\end{tabular}
\end{center}
\caption{\small Parameters for a Class A model with a D6-brane.}
\label{parameters}
\end{table}

\begin{table}[t]
\begin{center}
\begin{tabular}{|c|c|c|c|c|}
\hline
   &  Y &$ X^{}_A$  & $X^{}_{B} $    \\
\hline $H^{}_{u}$ & 1/2 & 0  & 2     \\
\hline  $H^{}_{d}$   & 1/2  &0  & -2     \\
\hline \end{tabular}
\end{center}
 \caption{\small Higgs charges in the Madrid model.}
\label{charge_higgs}
\end{table}

\begin{table}[t]
\begin{center}
\begin{tabular}{|c|c|c|c|c|c|}
\hline
   &   $q_a$  & $q_b $ & $q_c $ & $q_d$   \\
\hline $Q_L$ & 1  & -1 & 0 & 0   \\
\hline  $u_R$   &  -1  & 0  & 1  & 0   \\
\hline   $d_R$  &  -1  & 0  & -1  & 0   \\
\hline  $ L$    &  0   & -1   & 0  & -1    \\
\hline   $e_R$  &  0  & 0  & -1  & 1      \\
\hline   $N_R$  &  0  & 0  & 1  & 1    \\
\hline
\end{tabular}
\end{center}
 \caption{\small SM spectrum charges in the $D$-brane basis for the Madrid model.}
\label{tabpssm}
\end{table}
\begin{table}[t]
\begin{center}
\begin{tabular}{|c|c|c|c|c|c|c|}
\hline
   &   $Q_L$  & $u_R $ & $d_R $ & $L$  &  $e_R$ & $N_R$ \\
\hline  $q_{Y}$  &  1/6    & - 2/3  & 1/3   &  -1/2   & 1 &  0  \\
\hline   $q_{B}$  & -1    & 0  & 0   & -1   & 0  & 0 \\
\hline \end{tabular}
\end{center}
 \caption{\small Fermion spectrum charges in the $Y$-basis for the Madrid model \cite{Ghilencea:2002da}.}
\label{charges}
\end{table}

\begin{figure}[t]
\begin{center}
\includegraphics[scale=1.0]{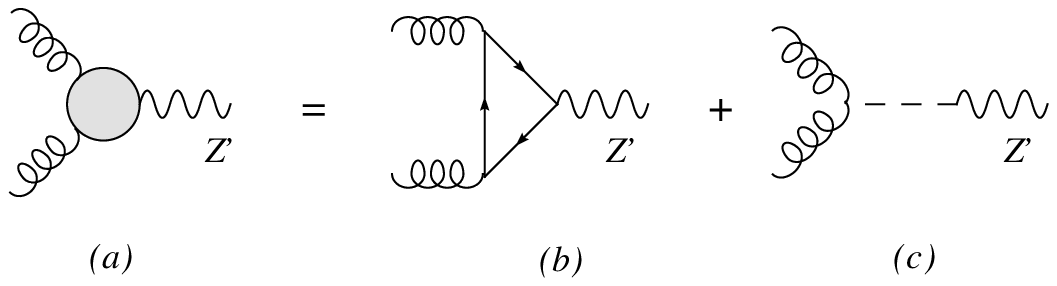}
\caption{\small A  gauge invariant GS vertex of the AVV type, composed of an AVV triangle  and a single counterterm of the Dolgov-Zakharov form.}
\label{GS_AVV}
\end{center}
\end{figure}

\section{ The GS and WZ vertices and gluon fusion}

As we have mentioned above in the previous sections, the two available
mechanisms that enforce at the level of the effective lagrangean the cancellation
of the anomalies involve either PQ-like (axion-like) interactions
- in the WZ case -  or the subtraction of the anomaly pole (for the GS case).
In a related analysis \cite{Armillis:2008bg} we have presented some of the main
features of the two mechanisms taking as an example an axial (anomalous) version of
QED  to illustrate the cancellation of the
anomaly in the two cases. In the GS case, the anomaly of a given diagram is removed
by subtracting the longitudinal pole of the triangle amplitude in the chiral limit.
We have stressed in \cite{Armillis:2007tb}
that the counterterm (the pole subtraction) amounts to the removal of one of the
invariant amplitudes of the anomaly vertex (the longitudinal component) and
corresponds to a vertex re-definition.

The procedure is exemplified in Fig.~\ref{GS_AVV} where we show the triangle
anomaly and the pole counterterm which is subtracted from the first amplitude.
The combination of the two contributions defines the GS vertex, which is made
of purely transverse components in the chiral limit \cite{Armillis:2008bg} and
satisfies an ordinary Ward identity. Notice that the vertex does not require
an axion as an asymptotic state in the related
S-matrix; for a non-zero fermion mass in the triangle diagram, the vertex
satisfies a broken Ward identity. We now proceed and summarize some of
these properties, working in the chiral limit.

Processes such as $g g\to \gamma\gamma$,  mediated by an
anomalous gauge boson $Z^{\prime}$, can be expressed in a simplified
form in which only the longitudinal component of the anomaly appears. We therefore set
$k_1^2=k_2^2=0$  and $m_f=0$,
which are the correct kinematical conditions to obtain the anomaly pole,
necessary for a parton model (factorized) description of the cross section
in a $pp$ collision at the LHC, where the initial state of the partonic hard-scatterings are on-shell.

We start from the Rosenberg form of the $AVV$ amplitude, which is given by
\ba
&&T^{\lambda\mu\nu}=A_1\varepsilon[k_1,\lambda,\mu,\nu]+A_2\varepsilon[k_2,\lambda,\mu,\nu]
+A_3 k_1^{\mu}\varepsilon[k_1,k_2,\nu,\lambda]
\nonumber\\
&&\hspace{1cm}+A_4 k_2^{\mu}\varepsilon[k_1,k_2,\nu,\lambda]
+A_5 k_1^{\nu}\varepsilon[k_1,k_2,\mu,\lambda]+A_6 k_2^{\nu}\varepsilon[k_1,k_2,\mu,\lambda]\,,
\ea
and imposing the Ward identities to bring all the anomaly on the axial-vector vertex, we obtain the usual conditions
\ba
&&A_1 = k_2^2 A_4 + k_1\cdot k_2 A_3
\nonumber\\
&&A_2 = k_1^2 A_5 + k_1\cdot k_2 A_6
\nonumber\\
&&A_3(k_1,k_2)=-A_6(k_1,k_2)
\nonumber\\
&&A_4(k_1,k_2)=-A_5(k_1,k_2),
\ea
where the invariant amplitudes $A_3,\dots,A_6$ are free from kinematical singularities for off-shell external lines. We set $k^2=(k_1+k_2)^2=s$.
As we have mentioned, in the parton model we take the initial gluons to be on-shell, while the hadronic cross section is obtained by convoluting the hard scattering given above (corrected by a color factor)  with the \textsc{Pdf}'s. The amplitude simplifies drastically in this case and takes the form
\ba
T^{\mu\nu\lambda}=A_6 k^{\lambda}\varepsilon[k_1,k_2,\nu,\mu]+
\left(A_4 + A_6\right)\left(k_2^{\nu}\varepsilon[k_1,k_2,\mu,\lambda]
-k_1^{\mu}\varepsilon[k_1,k_2,\nu,\lambda]\right),\,
\ea
in which the second piece drops off
for physical on-shell photon/gluon lines, leaving only a single invariant
amplitude to contribute to the final result
\ba
&&T^{\mu\nu\lambda}=A_6^{f}(s)(k_{1}+k_{2})^{\lambda}
\varepsilon\left[k_1,k_2,\nu,\mu\right]
\label{massiveT}
\ea
where
\ba
&&A_6^{f}(s)=\frac{1}{2\pi^2 s}\left(1 +\frac{m_f^2}{s}\log^{2}\frac{\rho^{}_f + 1 }{\rho^{}_f - 1 }\right),
\hspace{1cm} \rho^{}_f = \sqrt{1 - 4\frac{m_f^2}{s}}, \,\,\,\,\, s<0.
\label{a6}
\ea
The anomaly pole is given by the first term of  Eq. (\ref{a6})
\ba
&& T_c^{\mu\nu\lambda}\equiv
\frac{1}{2\pi^2 s} (k_{1}+k_{2})^{\lambda}
\varepsilon\left[k_1,k_2,\nu,\mu\right].
\label{tc}
\ea
The logarithmic functions in the expression above are continued
in the following way in the various region
\ba
&&0 < s < 4 m_f^2:
\nonumber\\
&&\rho_f\rightarrow i \sqrt{-\rho_f^2}; \hspace{0.5cm}
\frac{1}{2}\log\left(\frac{\rho_f +1}{\rho_f - 1}\right)\rightarrow
-i\arctan\frac{\sqrt{s}}{\sqrt{4 m_f^2 - s}},
\nonumber\\
&& s > 4 m_f^2 > 0:
\nonumber\\
&&\sqrt{-\rho_f^2} \rightarrow -i\rho_f; \hspace{0.5cm}
\arctan\frac{1}{\sqrt {- \rho_f^2}}\rightarrow \frac{\pi}{2}
+\frac{i}{2}\log\left(\frac{\sqrt{s - 4 m_f^2}+\sqrt{s}}{\sqrt{s}-\sqrt{s - 4 m_f^2}} \right).
\ea

\begin{figure}[t]
\begin{center}
\includegraphics[scale=1.0]{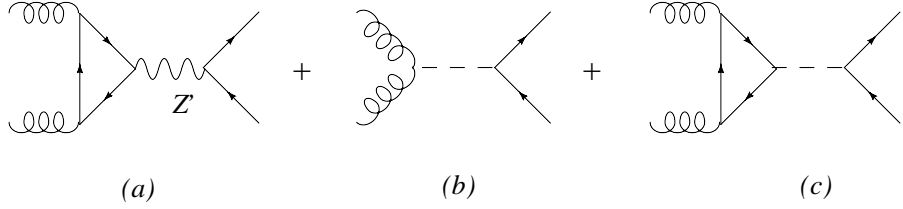}
\caption{\small One loop vertices and counterterms for the WZ mechanism.}
\label{onefig}
\end{center}
\end{figure}

\begin{figure}[t]
\begin{center}
\includegraphics[scale=1.0]{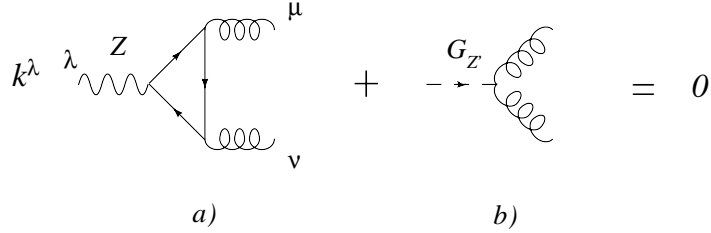}
\caption{\small  Ward identity in the WZ case in the chiral limit.}
\label{STIggch}
\end{center}
\end{figure}

Notice that the surviving amplitude $A_6$ multiplies a longitudinal
momentum exchange and, as discussed in the literature on
the chiral anomaly in QCD \cite{Achasov:1992bu,Dolgov:1971ri},
is characterized by a massless pole in $s$, which is the anomaly pole,
as one can clearly conclude from Eq. (\ref{a6}). This equation shows also how
chiral symmetry breaking effects appear in this amplitude at this special
kinematical point by the $m_f$ terms.

The subtraction of the anomaly pole is shown in
Fig.~\ref{GS_AVV} and is represented by diagram c).
The combination of diagrams b) and c) defines the GS vertex
of the theory \cite{Armillis:2008bg}, with diagram c) described by Eq. (\ref{tc}) ($- T_c$).
It is easily verified that in the massless fermion limit and for on-shell gluon lines,
the GS vertex is trivially vanishing by construction. In general, for any asymmetric
configuration of the external lines in the vertex, even in the massless limit,
the vertex has non-zero transverse components \cite{Knecht:2003xy,Jegerlehner:2005fs}.
The expression is well known \cite{Knecht:2002hr, Jegerlehner:2005fs} in the chiral
limit and has been shown to satisfy the Adler-Bardeen theorem \cite{Jegerlehner:2005fs}.

For a non-vanishing $m_f$ the GS vertex, for generic virtualities,
can be defined to be the general $AVV$ vertex, for instance extracted
from \cite{Kniehl:1989qu} or, in the longitudinal/transverse formulation,
by the amplitudes given in \cite{Jegerlehner:2005fs}, with the subtraction of the anomaly pole, as given in
\cite{Armillis:2008bg}. We will refer to the anomaly (subtraction)
counterterm of diagram b) as to the Dolgov-Zakharov \cite{Dolgov:1971ri}
(DZ) counterterm. The anomaly diagram reduces to its DZ form for two on-shell
gauge lines (photons/gluons) and in this case the transverse components
completely disappear. There are other cases in which, instead,
the longitudinal components cancel. This occurs if, for instance,
a conserved current is attached to the anomalous line, rendering the anomaly
"harmless", as explained in \cite{Armillis:2008bg}.

The analogous interaction in the WZ case is shown in Fig.~\ref{onefig},
where we have attached a fermion pair in the final state to better
identify the contributions. In this case, beside the anomalous contribution
of diagram a), the mechanism will require the  exchange of a physical axion,
shown in diagram b) and c). Diagram b) is the usual WZ counterterm
(or generalized PQ interaction)  while the third diagram is non-vanishing
only in the presence of fermions of non-zero mass. This third contribution
is numerically irrelevant and in DY is usually omitted. The WZ mechanism
re-establish gauge invariance of the effective lagrangean but is not based
on a vertex re-definition and, furthermore, involves an asymptotic axion state.
As shown in \cite{Coriano:2008pg} the presence of a unitarity bound in this
mechanism is a signal of its limitation as an effective theory
(see also the discussion in \cite{Armillis:2008bg}). We have summarized
in an appendix the discussion of this point
in a simple case.

\subsection{Ward identities}
\begin{figure}[t]
\begin{center}
\includegraphics[scale=1.0]{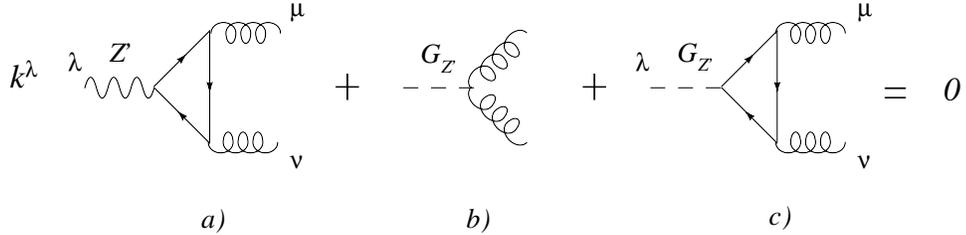}
\caption{\small Generalized Ward identity in the WZ case. }
\label{STInochiral}
\end{center}
\end{figure}

Both vertices satisfy ordinary Ward identities in the chiral limit and
generalized Ward identities away from it. In the chiral limit, for instance,
the WZ mechanism adds to the effective action of the anomalous theory an
interaction of the St\"uckelberg field ($b$) with the gluons $(b\, G\wedge G)$,
shown in diagram b) of Fig.~\ref{STIggch}. In this figure we have shown a diagrammatic realization
of the Ward identity for this case.

In WZ, being the cancellation based on a local field theory, the derivation of the generalized Ward identity can be formally obtained from the requirement of BRST invariance of the gauge-fixed effective
action, as we have shown in \cite{Armillis:2007tb}. This is illustrated in Fig.~\ref{STInochiral},  in the case of an anomalous $Z^{\prime}$, where we show the coupling of the goldstone - in the broken Higgs phase -  to the gluons (diagram b)) and to
the massive fermion (diagram c)) \cite{Armillis:2007tb}. The normalization of the counterterm in b) can be chosen to remove
the anomaly of diagram a) when a single fermion runs inside the anomaly loop. Alternatively, the same graphical representation holds if in the first and the last diagram we sum over the entire generation. In this case the counterterm is normalized to cancel the entire anomaly of the complete vertex.
\begin{figure}[t]
\begin{center}
\includegraphics[scale=0.8]{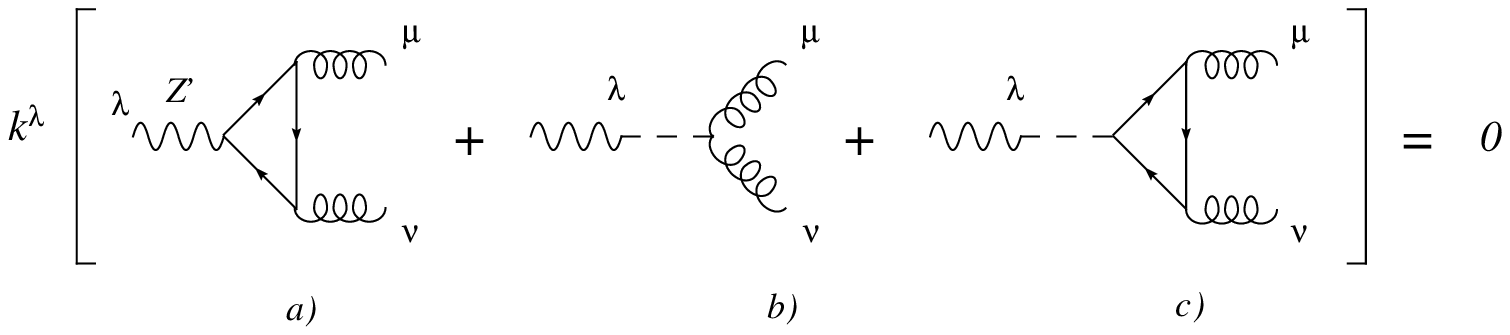}
\caption{\small Generalized Ward identity in the GS case.}
\label{STIgg}
\end{center}
\end{figure}

\begin{figure}[t]
\begin{center}
\includegraphics[scale=0.8]{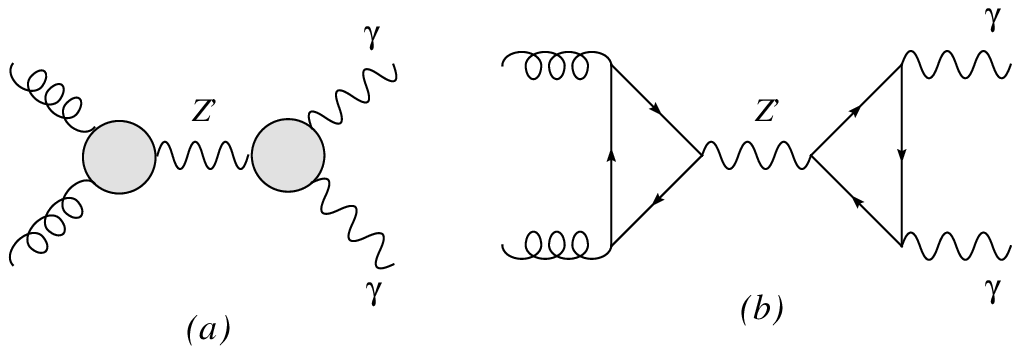}
\caption{\small  BIM amplitude with an anomalous $Z^\prime$ exchange.}
\label{ggBIM}
\end{center}
\end{figure}

The analysis in the GS case is slightly more subtle. We show in Fig.~\ref{STIgg}
the generalized Ward identity satisfied by the vertex in the massive case.
In the massless case only diagrams a) and b) survive, while  the contribution
of diagram c) comes from a direct computation. It is obtained by multiplying
typical pseudoscalar interaction - such as the one shown in Fig.~\ref{STInochiral},
diagram c) - by a massless pole. If we denote by
$T^{\mu\nu} $  the diagram describing the decay of a pseudoscalar
into two gluons, diagram c) takes the form
$k^{\lambda}/k^2 T_{\mu\nu}$, with a factorized pole on the anomalous external line.
We refer to \cite{Armillis:2008bg} for a detailed discussion of these points.

In our analysis we encounter a class of amplitudes
(\textsc{BIM} ) \cite{Coriano:2008pg} which are characterized
by two anomaly vertices connected by an $s$-channel exchange
of the anomalos gauge boson. These amplitudes grow quadratically with
the energy and are not eliminated by fine tuning Fig.~\ref{ggBIM}.
The true \textsc{BIM} amplitude is the one shown in diagram b) and
appears in the gluon fusion sector in the WZ case.  In the SM a similar
graph contributes only if heavy fermions run in the loop. They are comparable
in size to the anomalous BIM amplitude. Obviously, this contribution would be
identically vanishing if all the fermions of a given generation would be mass-degenerate.

Diagram a)
shown in the same figure, instead, is the GS version of the \textsc{BIM} amplitude and is identically zero in the chiral limit for on-shell gluon lines, as is the case in the parton model. For this reason the gluon fusion sector
disappears completely  for DP (in the GS mechanism), since the BIM amplitude in this case is obtained by replacing diagram b) of this figure with diagram a) which is indeed vanishing.

We can summarize the basic features of the anomalous sectors in anomaly free-models after QCD factorization, for generic virtualities of the external gauge lines, according to the following points:

1) In the SM the residual contributions coming from anomalous diagrams, such as in the $V V Z, V V Z^{\prime}$ vertices, where $V$ is a gauge field, are proportional to the mass of the heavy quarks in the anomaly loop. In the chiral limit, instead, both the anomaly pole contribution
and the transverse component of the anomaly cancel by charge assignment.

2) In the GS case, as we have just discussed, the anomaly pole is absent by definition, while the transverse
contributions are allowed. This separation between longitudinal and transverse components is less transparent for a heavy fermion mass, which induce a longitudinal component, proportional to
$m_f^2/s^2 $ times a small logarithmic correction of the ratio of the same variables,
away from the chiral limit. This longitudinal component, however, should not be confused
with the anomaly pole and {\em is not shifted or corrected} perturbatively in any way. It can couple, for instance, to a $t\bar t $ (top) quark current because of a broken Ward identity and can be interpreted as a manifestation of the GS mechanism at the LHC, but can be easily overshadowed by SM contributions.
This point will be re-addressed more formally below in Eq.~(\ref{translong}).

At some special kinematical points (two massless gauge lines, or three massless
gauge lines of the same virtualities) where the anomalous vertex takes its DZ form,
the GS vertex is identically vanishing in the massless case. In the presence of a
heavy fermion the logarithmic correction shown in
Eq.~(\ref{a6}) reappears.

3) In the WZ case the anomaly pole is not cancelled. A second sector (the exchange of the axion) is
needed to restore the gauge invariance of the effective action.
In a hadronic collision the BIM amplitudes induce very small deviations
from the SM behaviour after the convolution with the gluon density.
They are absent in DY at NNLO. In DP they affect the invariant mass distributions
- at large $Q$ - of the photon pair, for a given center of mass energy of the two
colliding protons. As such they are sensitive to large (Bjorken) $x$-values of the
gluon \textsc{Pdf}'s, region where the gluon density is rapidly decreasing.
In particular, in DP their contribution becomes more sizeable via intereference
with some box-like amplitudes ($gg \to \gamma \gamma$). In previous NLO study of this process
\cite{Coriano:1996us} they had been included even though they exceed the NLO accuracy, being truly NNLO
contributions. These amplitudes and vertices are the
basic building blocks of our numerical analysis and are responsible
for all the anomalous signal both in DY and in DP. We will try to quantify
their impact in the invariant mass distributions in both cases.

\section{ Invariant mass distributions in Drell-Yan }

Our NNLO analysis of the invariant mass distributions for lepton pair production,
for the computation of the QCD sectors, is based on the hard scatterings of
\cite{Hamberg:1990np}, and the NNLO evolution of the parton distributions
(\textsc{Pdf}'s) has been obtained with \textsc{Candia} \cite{Cafarella:2008du}.
The anomalous corrections to the invariant mass distributions have been evaluated
separately, since at NNLO they appear in DY in the
interference with the lowest order graph, and added to the standard QCD background.
It is important to recall that lepton pair production at low $Q$ via Drell-Yan is
sensitive to the \textsc{Pdf}'s at small-$x$ values, while in the high mass region
this process is essential in the search of
additional neutral currents. In our analysis we have selected a mass of 1 TeV for
the extra gauge boson and analyzed the signal and the background both on the peaks
of the $Z$ and of the of the new resonance.

At hadron level the colour-averaged inclusive differential cross section
for the reaction $H_1 +H_2 \rightarrow l_1 +l_2 +X $, is given by the expression \cite{Hamberg:1990np}
\ba
\frac{d\sigma}{dQ^2}=\tau \sigma_{\cal Z}(Q^2,M_{\cal Z}^2) W_{\cal Z}(\tau,Q^2)\hspace{1cm} \tau=\frac{Q^2}{S},
\label{factor}
\ea
where ${\cal Z}\equiv Z, Z^{\prime}$ is the point-like cross section and
all the information from the hadronic initial state is contained in the \textsc{Pdf}'s.
The hadronic structure function $W_{\cal Z}(\tau,Q^2)$ is given by a convolution product
between the parton luminosities $\Phi_{i j}(x,\mu_R^2,\mu_F^2)$ and the Wilson coefficients
$\Delta_{i j}(x,Q^2,\mu_R^2,\mu_F^2)$
\ba
W_{\cal Z}(\tau,Q^2, \mu_R^2,\mu_F^2)&=&\sum_{i, j}\int_{\tau}^{1}\frac{d x}{x}\Phi_{i j}(x,\mu_R^2,\mu_F^2)
\Delta_{i j}(\frac{\tau}{x},Q^2,\mu_F^2),
\ea
where the luminosities are given by
\ba
\Phi_{i j}(x,\mu_R^2,\mu_F^2)=\int_{x}^{1}\frac{d y}{y}f_{i}(y,\mu_R^2,\mu_F^2) f_{j}\left(\frac{x}{y},\mu_R^2,\mu_F^2\right)
\equiv  \left[f_{i}\otimes f_{j}\right](x,\mu_R^2,\mu_F^2)
\ea
and the Wilson coefficients (hard scatterings) depend on both the factorization ($\mu_F$) and renormalization scales $(\mu_R)$, formally expanded in the strong coupling $\alpha_s$ as
\ba
\Delta_{i j}(x,Q^2,\mu_F^2)=
\sum_{n=0}^{\infty}\alpha_s^n(\mu_R^2)\Delta^{(n)}_{i j}(x,Q^2,\mu_F^2,\mu_R^2).
\ea
We will vary $\mu_F$ and $\mu_R$ independently in order to determine the sensitivity of the prediciton on their variations and their optimal choice.

The anomalous corrections to the hard scatterings computed in the SM will be
discussed below. We just recall that the relevant point-like cross sections
appearing in the factorization formula (\ref{factor}) and which are part
of our analysis include, beside the Z and the $Z^{\prime}$ resonance, also
the contributions due to the photon and the $\gamma-Z,\, \gamma -Z^{\prime}$
interferences. For instance in the $Z^{\prime}$ case we have

\ba
&&\sigma_{\gamma}(Q^2)=\frac{4\pi\alpha_{em}^2}{3 Q^4}\frac{1}{N_c}
\nonumber\\
&&\sigma_{{Z^{\prime}}}(Q^2)=\frac{\pi\alpha_{em}}{4
M_{{Z^{\prime}}}\sin^2\theta_W \cos^2\theta_W N_c}
\frac{\Gamma_{{Z^{\prime}}\rightarrow \bar{l}
l}}{(Q^2-M_{Z^{\prime}}^2)^2 + M_{Z^{\prime}}^2 \Gamma_{Z^{\prime}}^2}
\nonumber\\
&&\sigma_{{Z^{\prime}},\gamma}(Q^2)=\frac{\pi\alpha_{em}^2}{6 N_c}
\frac{g_V^{Z^{\prime},l}g_V^{\gamma,l}}{\sin^2{\theta_W}\cos^2{\theta_W}}
\frac{(Q^2-M_{Z^{\prime}}^2)}{Q^2(Q^2-M_{Z^{\prime}}^2)^2+
M_{Z^{\prime}}^2\Gamma_{Z^{\prime}}^2},\nonumber\\
&&\sigma_{{Z^{\prime}},Z}(Q^2)=\frac{\pi\alpha_{em}^2}{96}
\frac{\left[g_V^{Z^{\prime},l}g_V^{Z,l}+
g_A^{Z^{\prime},l}g_A^{Z,l}\right]}{\sin^4{\theta_W}\cos^4{\theta_W}N_c}
\frac{(Q^2-M_Z^2)(Q^2-M^2_{Z^{\prime}})
+M_Z\Gamma_{Z}M_{Z^{\prime}}\Gamma_{Z^{\prime}}}
{\left[(Q^2-M_{Z^{\prime}}^2)^2 + M_{Z^{\prime}}^2\Gamma_{Z^{\prime}}^2\right]
\left[(Q^2-M_Z^2)^2 + M_Z^2\Gamma_{Z}^2\right]}.\nonumber\\
\ea
where $N_C$ is the number of colours, $\Gamma_{{Z^{\prime}}\rightarrow \bar{l}l}$
is the partial decay width of the gauge boson and the total hadronic widths are defined by
\ba
\Gamma_Z\equiv\Gamma(Z\rightarrow hadrons)=\sum_{i}\Gamma(Z\rightarrow \psi_i\bar{\psi_i})
\nonumber\\
\Gamma_{Z^\prime}\equiv\Gamma(Z^{\prime}\rightarrow
hadrons)=\sum_{i}\Gamma(Z^{\prime}\rightarrow \psi_i\bar{\psi_i}),
\ea
where we refer to hadrons not containing bottom and top quarks (i.e.
$i=u,d,c,s$). We also ignore electroweak corrections of higher order and we have included the top-quark mass and QCD corrections. We have included only tree level
decays into SM fermions, with a total decay rate for the $Z$ and $Z^{\prime}$ which is given by
\ba
&&\Gamma_Z=\sum_{i=u,d,c,s}\Gamma(Z\rightarrow
\psi_i\bar{\psi_i})+\Gamma(Z\rightarrow b\bar{b})
+3\Gamma(Z\rightarrow l\bar{l})+3\Gamma(Z\rightarrow \nu_l\bar{\nu_l})
\nonumber\\
&&\Gamma_{Z^{\prime}}=\sum_{i=u,d,c,s}\Gamma(Z^{\prime}\rightarrow
\psi_i\bar{\psi_i})+\Gamma(Z^{\prime}\rightarrow b\bar{b})
+\Gamma(Z^{\prime}\rightarrow t\bar{t})+3\Gamma(Z^{\prime}\rightarrow l\bar{l})
+3\Gamma(Z^{\prime}\rightarrow \nu_l\bar{\nu_l}).
\nonumber\\
\ea
\begin{figure}[t]
\begin{center}
\includegraphics[scale=0.8]{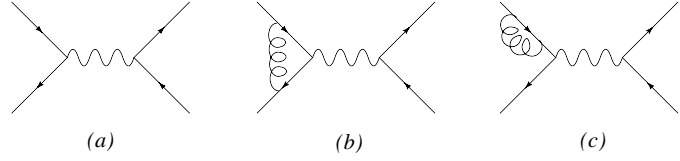}
\caption{\small $q\bar q \rightarrow Z, Z^{\prime}$  at LO and NLO (virtual corrections).}
\label{DYLO}
\end{center}
\end{figure}
\begin{figure}[t]
\begin{center}
\includegraphics[scale=0.8]{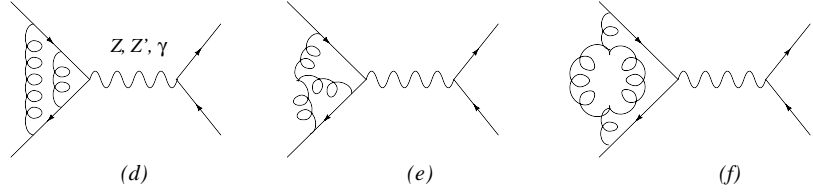}
\caption{\small $q\bar q \rightarrow Z, Z^{\prime}$  at NNLO (virtual corrections).}
\label{DYNLOv}
\end{center}
\end{figure}

Coming to illustrate the contributions included in our analysis, these are shown in some representative
graphs. The complete NNLO expressions of the hard scatterings and the corresponding Feynman diagrams can be found in \cite{Hamberg:1990np}.

\begin{itemize}
\item{\bf SM QCD contributions}
\end{itemize}

\begin{figure}[t]
\begin{center}
\includegraphics[scale=0.8]{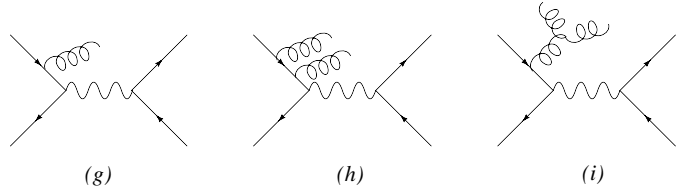}
\caption{\small $q\bar q \rightarrow Z, Z^{\prime}$  with real corrections at NLO $(g)$ and at NNLO $(h)$, $(i)$. }
\label{DYNLOr}
\end{center}
\end{figure}

We show in Fig.~\ref{DYLO} the leading $O(\alpha_w)$ and some typical
next-to-leading order $O(\alpha_w\alpha_s)$ (LO, NLO) contributions to the process
in the annihilation channel (virtual corrections). Examples of higher order
virtual corrections included in the hard scatterings are shown in Fig.~\ref{DYNLOv},
which are of $O(\alpha_s^2 \alpha_w)$, while the corresponding real emissions,
integrated over the final state gluons, are shown in Fig.~\ref{DYNLOr} at NLO (graph g)) and
NNLO (graphs h) and i)).
\begin{figure}[t]
\begin{center}
\includegraphics[scale=0.8]{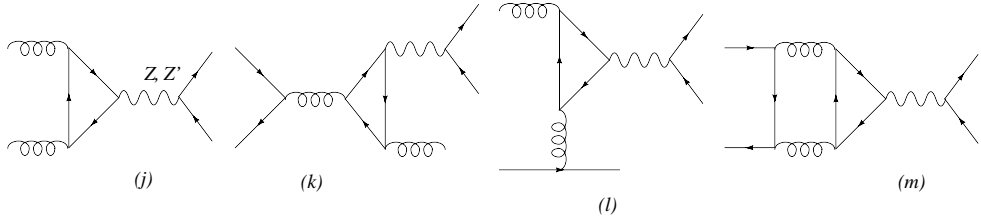}
\caption{\small Anomalous contributions for  $Z^{\prime}$
production in the $gg$, $q \bar q $ and $q g$ sectors at lower orders. }
\label{DYanom}
\end{center}
\end{figure}

%\clearpage

\begin{itemize}
\item{\bf Anomalous  corrections}
\end{itemize}
We shown in Fig.~\ref{DYanom} the leading anomalous corrections to leptoproduction.
At $O(\alpha_s\alpha_w)$ there is a first contribution coming from the interference
between graph j) and the leading order
$q\bar{q}$ annihilation vertex (graph a) of Fig.~\ref{DYLO}. The square of the same
graph appears in the anomalous corrections at $O(\alpha_s^2 \alpha_w)$.
Other contributions that we have included are those due to the exchange
of a physical axion and goldstone modes, which can be removed in the
unitary gauge \cite{Coriano:2008pg}. Of higher order are the contributions
shown in diagram k), l) and n), which contribute via their interference with NLO tree
level graphs. For instance k) interferes with diagram g) of (\ref{DYNLOr}), while m)
interferes with the LO annihilation graph. The analogous contributions in the WZ and GS cases are
obtained by replacing the triangle graph with the GS vertex,
as in Fig.~\ref{GS_AVV}, or, for the WZ case, with
Fig.~\ref{onefig}.
Notice that in Fig.~\ref{onefig}, in the WZ case the anomaly pole is
automatically cancelled by the Ward identity on the lepton pair of the final
state, if the two leptons are taken to be massless at high energy, as is the case.
Then, the only new contributions from the anomaly vertex that survive are those
related to the transverse component of this vertex.  This is an example, as we
have discussed in \cite{Armillis:2008bg}, of a "harmless" anomaly vertex.
A similar situation occurs whenever there is no coupling of the longitudinal
component of the anomaly to the (transverse) external leptonic current.
This property continues to hold also away from the chiral limit, since the
corrections due to the fermion mass in the anomaly have the typical structure
\beq
\Delta_{\mu\nu\rho}(q,k)_{\mbox{\rm\tiny anomaly}}=
\sum_f g_{A,f}^{Z'}e^2 Q_f^2 a_n\frac{(q-k)_{\nu}}{(q-k)^2}\left( \frac{1}{2} - 2 m_f^2 C_0\right)
\epsilon_{\mu\rho\alpha\beta}q^{\alpha}k^{\beta} + \tilde{\Delta}^{trans}\,,
\label{translong}
\eeq
where $\tilde{\Delta}^{trans}$ is the truly transversal component
away from the chiral limit. The most general expression of the coefficient $C_0$ is given
in Eq.(A.8) of ref. \cite{Kniehl:1989qu}. $C_0$ is the scalar 3-point function with a fermion of
mass $m_f$ circulating in the loop.
 In both mechanisms anomaly (strictly massless) effects are comparable
 with the corresponding contributions
coming from the SM for massive fermions.  It should be clear by now that
in the WZ case the anomaly pole is not cancelled, rather an additional
exchange is necessary to re-establish the gauge independence of the S-matrix (the axion).
In DY
this sector does not play a significant role due to the small mass of the lepton pair.
As we have discussed above, the cancellation of the anomaly is due, in this case,
to the Ward identity of the leptonic current and there is no axion exchanged in the $s$-channel.

\subsection{Precision studies on the Z resonance}
\begin{figure}[t]
\begin{center}
\includegraphics[scale=0.8]{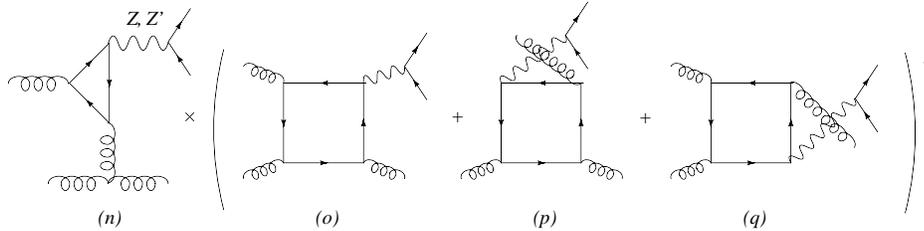}
\caption{\small Anomalous contributions  for the $gg \rightarrow gg$ process mediated by an anomalous $Z^{\prime}$
at higher perturbative orders. }
\label{DYgg}
\end{center}
\end{figure}

The quantification of the corrections due to anomalous abelian gauge structures in
DY requires very high precision, being these of a rather high order. For this reason
we have to identify all the sources of indeterminations in QCD which come from the
factorization/renormalization scale dependence of the cross section, keeping into
account the dependence on $\mu_F$  and $\mu_R$ {\em both} in the DGLAP evolution
{\em and} in the hard scatterings. The set-up of our analysis is similar to that
used for a study of the NNLO DGLAP evolution in previous works
\cite{Cafarella:2007tj,Cafarella:2005zj}, where the study has covered
every source of theoretical error, including the one related to the various
possible resummations of the DGLAP solution, which is about $2-3\%$ in DY and
would be sufficient to swamp away any measurable deviation due to new physics at the LHC.

These previous studies have been focused on the DY distributions on the resonance
peaks, in particular on the peak of the $Z$, where the accuracy at the LHC is of
outmost importance for QCD partonometry. The presence
  of anomalous corrections on the $Z$ peak is due both to the anomalous components of the
$Z$ in the anomalous models and to the interference between the $Z^{\prime}$ and the
$Z$, that we have taken into account. Notice that in DY the treatment of the anomalous
corrections to the $Z$
is drastically simplified if we neglect the (small) mass of the lepton pair, as usual.
In fact, these are due to trilinear (anomaly) vertices which involve the $BBB$,
$BYY $, $B W_3 W_3$ and $B G G$ gauge fields - in the interaction basis - all
of them involving  interactions of the St\"uckelberg field with the corresponding
field-strengths of the gauge fields, such as $b F_B\wedge F_B$,
$b F_Y \wedge F_Y$, $b F_W\wedge F_W$ and $b F_G\wedge F_G$, where $G$ denotes
the gluon field. The only contribution that is relevant for the LHC is then one
obtained by projecting the $b F_G\wedge F_G$ vertex on the physical axion
$\chi$, whose mass is, in principle, a free parameter of the anomalous models.
It is then clear that the axion channel plays a more important role in the
production of the $top$, due to its large mass, than in leptoproduction.
We will now briefly summarize the results for the new contributions in DY,
starting from the non-anomalous ones.

In the $q\bar{q}$ sector we have two contributions involving triangle fermion loops
see Fig.~\ref{DYanom} k,m.
The one depicted in Fig.~\ref{DYanom}m is a two-loop virtual correction with a
$Z$ or a $Z^{\prime}$ boson in the final state, while in
Fig.~\ref{DYanom}k we have a real emission of a gluon in the final state which is integrated out.
The first contribution has been calculated in \cite{Bernreuther:2005rw, Larin:1993tq, Gonsalves:1991qn, Rijken:1995gi},
\ba
&&\Delta^{V}_{q\bar{q}}(x,Q^2,\mu_F^2,m^2) =\delta(1-x)a^{Z'}_q a^{Z'}_Q C_F T_f\frac{1}{2}
\left(\frac{\alpha_s}{\pi}\right)^2\times\nonumber\\
&&\hspace{2cm}\left[\theta(Q^2-4 m^2) G_1(m^2/Q^2)
+\theta(4 m^2-Q^2)G_2(m^2/Q^2)\right)
\ea
where $C_F$ and $ T_f$ are the color factors, $q=u,d,c,s$, $Q=t,b$ and $m$ the mass of the
heavy flavors, while in the massless limit the functions $G_1$ and $G_2$ are given by
\ba
&&G_1(m=0)=3\log\left(\frac{Q^2}{\mu_R^2}\right)-9 +2 \zeta(2)
\nonumber\\
&&G_2(m=0)=0
\ea
and $Q$ represents the invariant mass of the system.
\begin{figure}[t]
\begin{center}
\includegraphics[scale=0.8]{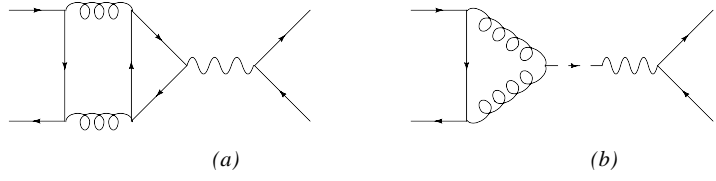}
\caption{\small GS mechanism: anomalous contribution
and counterterm for the $q\bar{q}$ scattering sector.}
\label{qq_AVV_WZ_ff}
\end{center}
\end{figure}
The contribution of Fig.~\ref{DYanom}k in the massless limit is given by
\ba
\Delta^{R}_{q\bar{q}}(x,Q^2,\mu_F^2,m=0)=a^{Z'}_q a^{Z'}_Q C_F T_f\frac{1}{2}
\left(\frac{\alpha_s}{\pi}\right)^2\times
\left\{\frac{(1+x)}{(1-x)_{+}}\left[-2 + 2 x (1-\log(x))\right]\right\},
\ea
while in the $qg$ sector we have the contribution shown in Fig.~\ref{DYanom}l
which is given by
\ba
\Delta_{qg}(x,Q^2,\mu_F^2,m^2)=a^{Z'}_q a^{Z'}_Q T_f^2\frac{1}{2}
\left(\frac{\alpha_s}{\pi}\right)^2\times
\left[\theta(Q^2-4 m^2) H_1(x,Q^2,m^2)+\theta(4 m^2-Q^2)H_2(x,Q^2,m^2)\right]
\ea
with the massless limit of  $H_1(x,Q^2,m^2)$ given by
\ba
H_1(x,Q^2,m=0)=2 x \left[\log\left(\frac{1}{x}\right)\log\left(\frac{1}{x}-1\right)
+Li_2\left(1-\frac{1}{x}\right)\right] + 2 \left(1-\frac{1}{x}\right)
\left[1-2 x \log\left(\frac{1}{x}\right)\right].
\ea
Separating the anomaly-free from the anomalous contributions,
the factorization formula for the invariant mass distribution in DY is given by
\ba
&&\frac{d\sigma}{dQ^2}=\tau \sigma_{\cal Z}(Q^2,M_{\cal Z}^2)\left\{
W_{\cal Z}(\tau,Q^2) + W_{\cal Z}^{anom}(\tau,Q^2)\right\}
\nonumber\\
&&W_{\cal Z}^{anom}(\tau,Q^2)=\sum_{i, j}\int_{\tau}^{1}\frac{d x}{x}\Phi_{i j}(x,\mu_R^2,\mu_F^2)
\Delta_{i j}^{anom}(\frac{\tau}{x},Q^2,\mu_F^2)
\nonumber\\
&&\Delta_{i j}^{anom}(x,Q^2,\mu_F^2)=\Delta^{V}_{q\bar{q}}(x,Q^2,\mu_F^2,m=0)
+\Delta^{R}_{q\bar{q}}(x,Q^2,\mu_F^2,m=0)+\Delta_{qg}(x,Q^2,\mu_F^2,m=0)
\nonumber\\
\ea
that we will be using in our numerical analysis below.

\subsection{Di-lepton production: numerical results}
\begin{figure}[t]
\subfigure[SM vs MLSOM at NLO]{\includegraphics[%
 width=6.5cm,
 angle=-90]{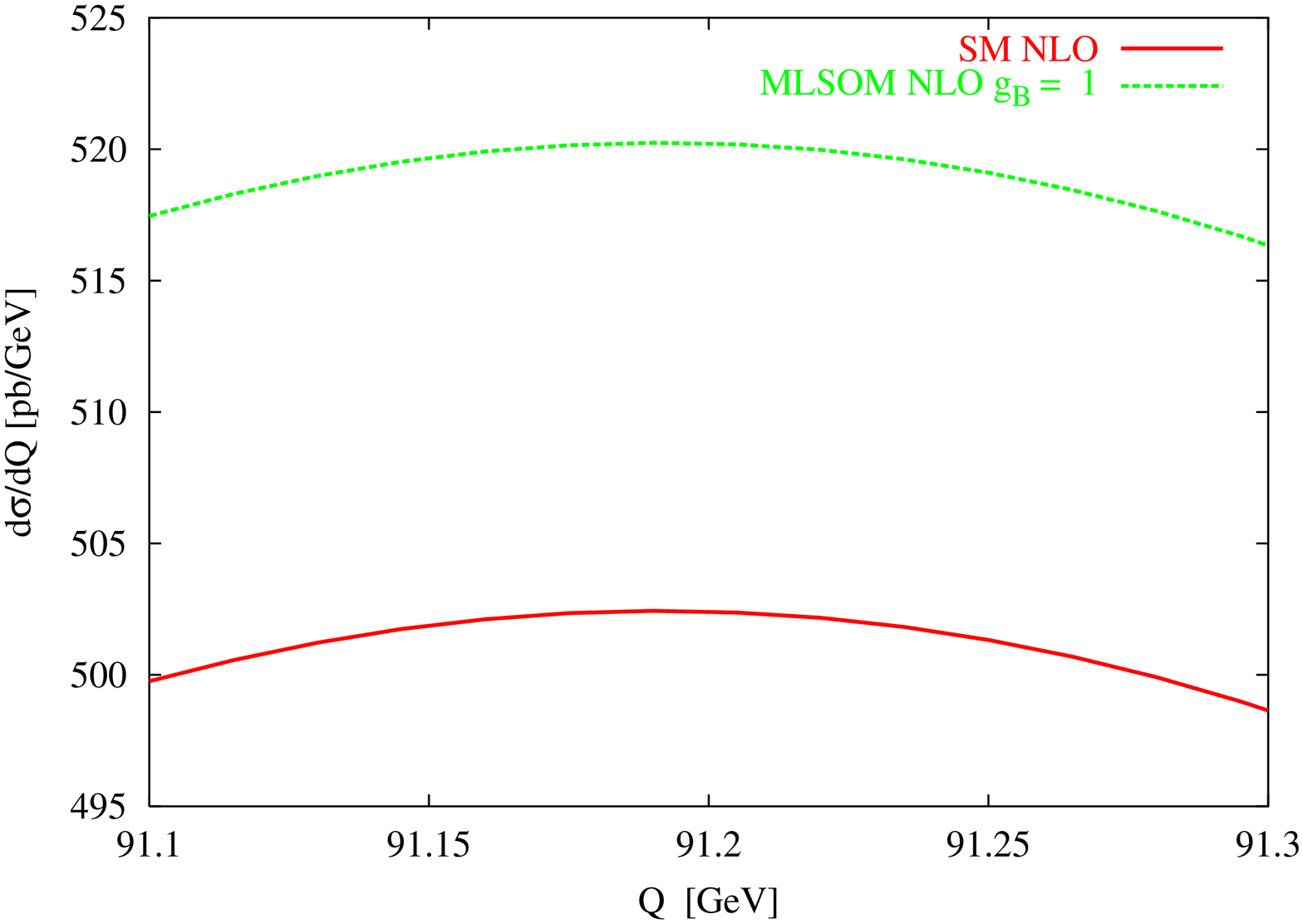}}
\subfigure[SM vs Anomaly free models at NLO]{\includegraphics[%
 width=6.5cm,
 angle=-90]{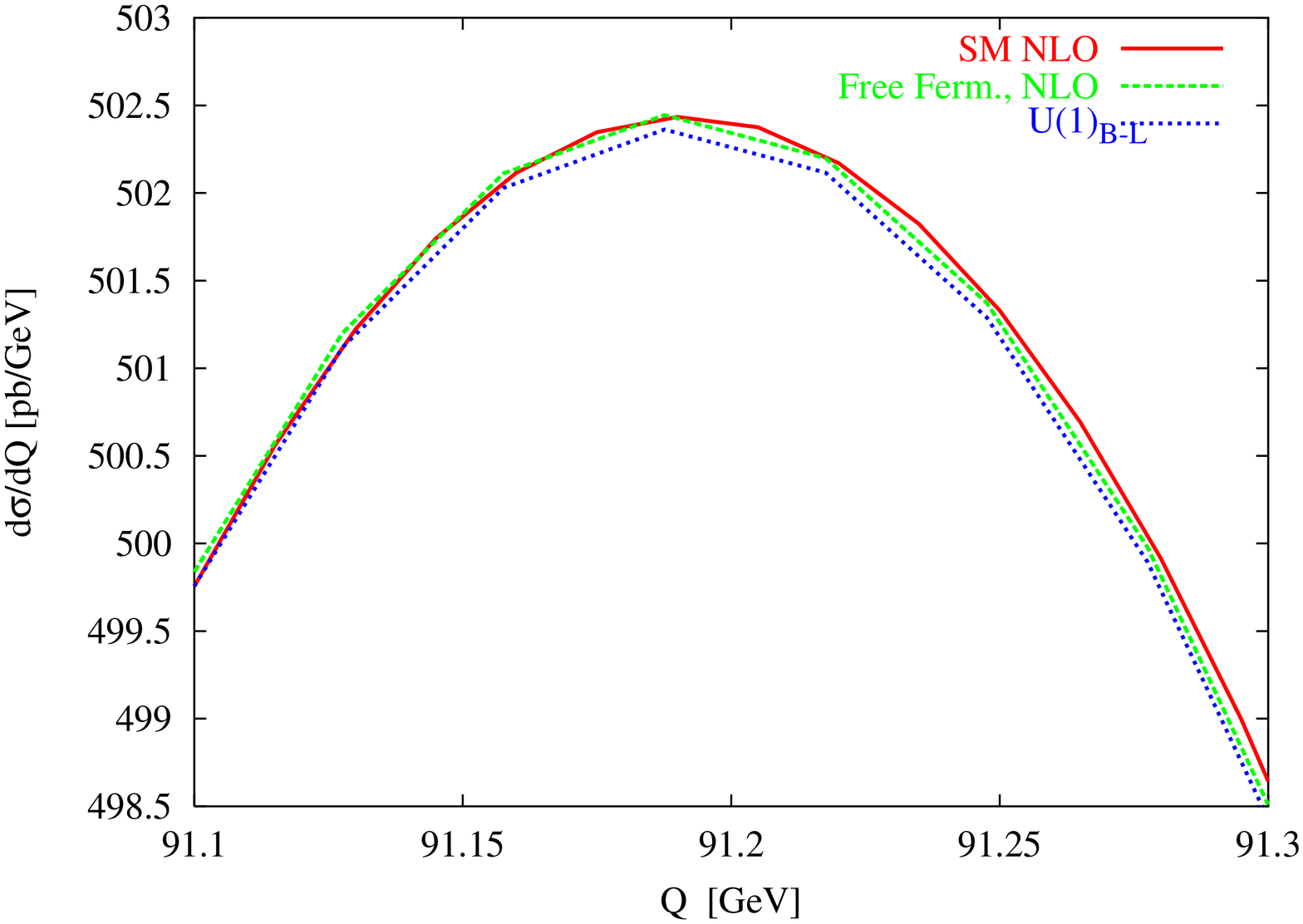}}
\subfigure[SM vs MLSOM at NNLO]{\includegraphics[%
 width=6.5cm,
 angle=-90]{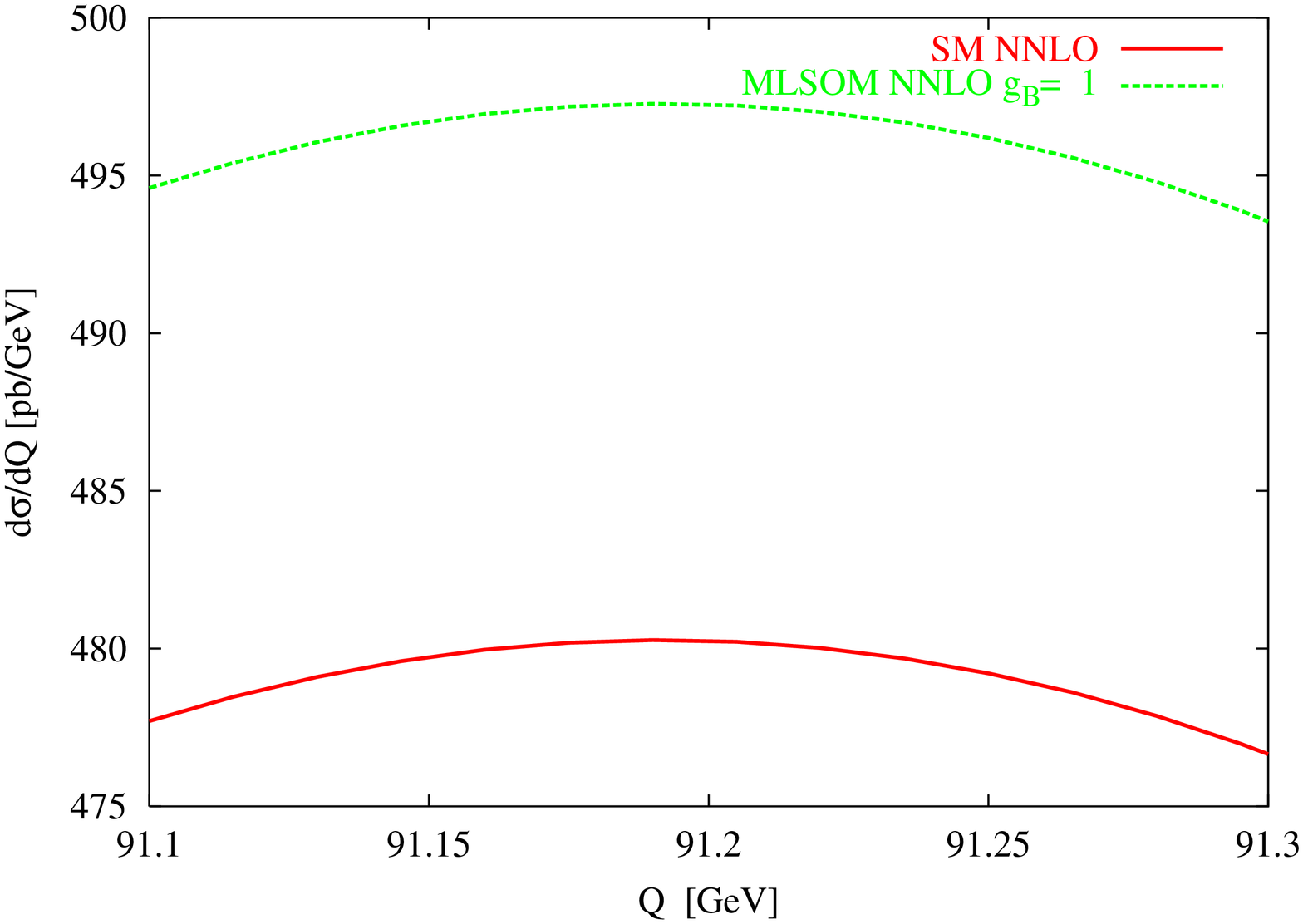}}
\subfigure[SM vs Anomaly free models at NNLO]{\includegraphics[%
 width=6.5cm,
 angle=-90]{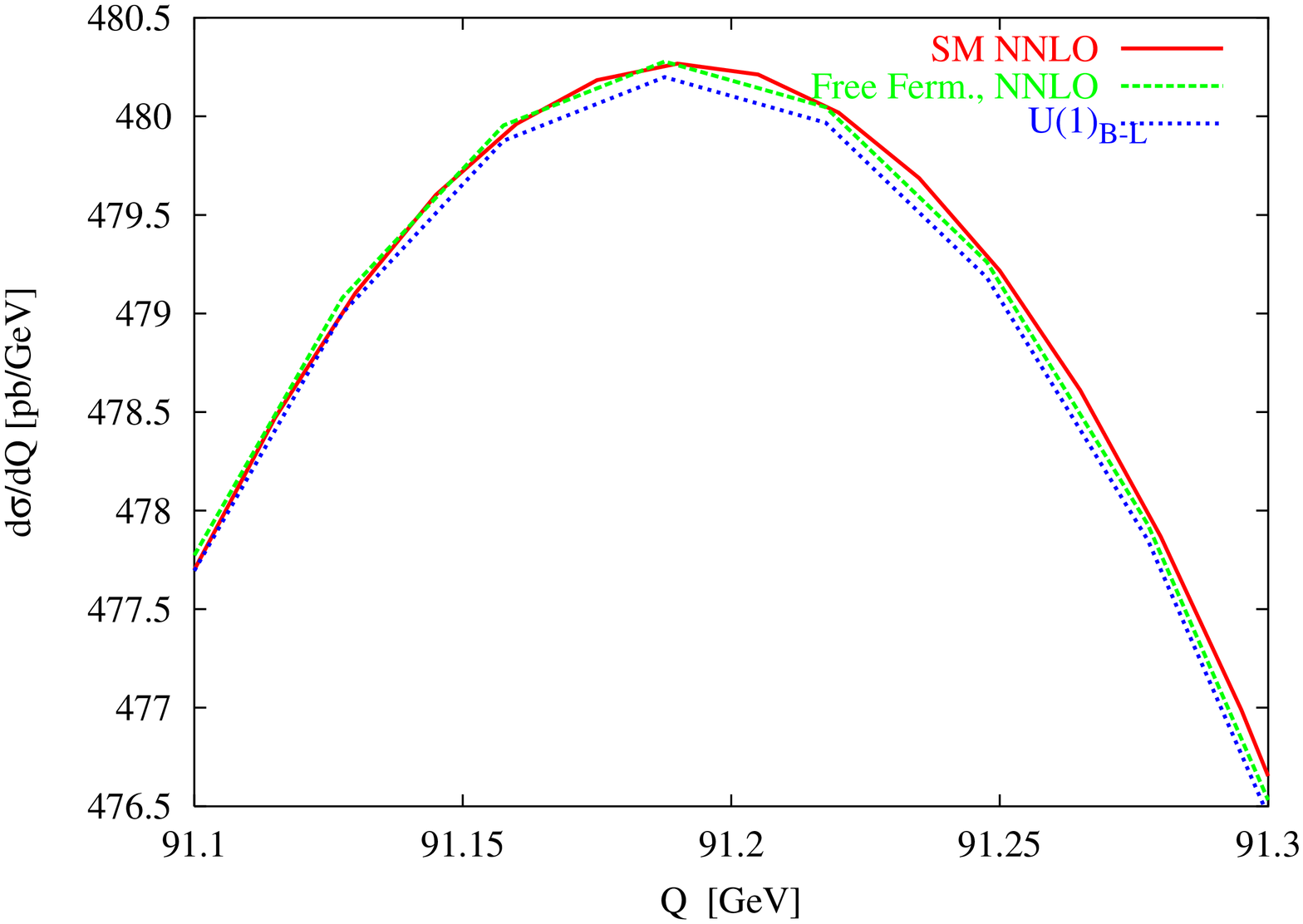}}
\caption{\small Zoom on the $Z$ resonance for anomalous Drell-Yan in
the $\mu_F=\mu_R=Q$ at NLO/NNLO for all the models.}
\label{zpeaknnloall}
\end{figure}

We have used the MRST-2001 set of \textsc{Pdf}'s given in
\cite{Martin:2001es} and \cite{Martin:2002dr}. We start by showing in
Fig.~\ref{zpeaknnloall} various zooms of the differential cross
section on the peak of the $Z$ - for all the models -  both at NLO and at NNLO.
We have kept the factorization and renormalization scales coincident and
equal to $Q$, while the mass of the extra $Z^\prime$ has been chosen around 1 TeV.
The anomaly-free models, from the SM to the three abelian extensions that we
have considered (free fermionic \cite{Coriano:2008wf} and $U(1)_{B-L}$ \cite{Carena:2004xs}
in Fig.~\ref{zpeaknnloall}, while $U(1)_{q+u}$ appears in Tab.~(\ref{tablethird}) )
show that the cross section is more enhanced for the MLSOM, illustrated in
Fig.~\ref{zpeaknnloall}a,c. The plots show a sizeable difference
(at a 3.5  $\%$ level) between
the anomalous and all the remaining anomaly-free models.
A comparison between (a) and (c) indicates, however, that this difference
has to be attributed to the specific
charge assignment of the anomalous model and not to the anomalous partonic sector,
which is present in (c) but not in (a). The anomalous corrections in DY appear at
NNLO and not at NLO, while in both figures the difference between the SM and the
MLSOM remains almost unchanged.

Moving from NLO to NNLO the cross section is reduced. Defining the $K$-factor
\beq
\frac{\sigma_{NNLO} - \sigma_{NLO}}{\sigma_{NLO}}\equiv K_{NLO}
\label{var}
\eeq
in the case of the MLSOM this factor indicates a reduction of about
$4 \%$ on the peak and can be attributed to the NNLO terms in the DGLAP
evolution, rather than to the NNLO corrections to the hard scatterings.
This point can be explored numerically by the (order) variation \cite{Cafarella:2007wd,Cafarella:2007tj}
\ba
\Delta\sigma &\sim& \Delta \hat{\sigma}\otimes \phi + \hat{\sigma}\otimes \Delta \phi \nonumber \\
\Delta\sigma &\equiv& |\sigma_{NNLO} - \sigma_{NLO}|
\ea
which measures the ``error'' change in the hadronic cross section $\sigma$
going from NLO to NNLO ($\Delta \sigma$) in terms of the analogous changes
in the hard scatterings ($\Delta \hat{\sigma}$) and parton luminosities $\Delta{\phi}$).
The dominance of the first or the second term on the rhs of Eq.~(\ref{var}) is an indication
of the dominance of the hard scatterings or of the evolution in moving from lower to higher order.
The same differences emerge also from Tab.~(\ref{table_nlo}) and
(\ref{table_nnlo}). Differences in the resonance region of this size can be considered
marginally relevant for the identification of anomalous components in this observables.
In fact, in \cite{Cafarella:2007tj} a high precision study of this distributions on the
same peak (in the SM case) shows that the total theoretical error is reasonably below
the 4 $\%$ level and can decrease at 1.5-2 $\% $ level when enough statistics will allow
to reduce the experimental errors on the \textsc{Pdf}'s. It is then obvious that the
isolation/identification of a specific model
- whether anomalous or not - appears to be rather difficult from the measurement of
a single observable even with very high statistics, such as the $Z$ resonance.
\begin{table}
\begin{center}
\begin{footnotesize}
\begin{tabular}{|c||c|c|c|c|c|}
\hline
\multicolumn{6}{|c|}{$d\sigma^{nlo}/dQ$ [pb/GeV] for the MLSOM with $M_{1}=1$ TeV, $\tan\beta=40$, Candia evol.}
\tabularnewline
\hline
$Q ~[\textrm{GeV}]$      &
$g_B=0.1$                 &
$g_B=0.36$                 &
$g_B=0.65$                 &
$g_B=1$                   &
$\sigma_{nlo}^{SM}(Q)$\tabularnewline
\hline
\hline
$90.50$&
$3.8551\cdot10^{+2}$&
$3.8711\cdot10^{+2}$&
$3.9106\cdot10^{+2}$&
$3.9902\cdot10^{+2}$&
$3.8543\cdot10^{+2}$
\tabularnewline
\hline
$90.54$&
$3.9712\cdot10^{+2}$&
$3.9877\cdot10^{+2}$&
$4.0284\cdot10^{+2}$&
$4.1105\cdot10^{+2}$&
$3.9704\cdot10^{+2}$
\tabularnewline
\hline
$90.59$&
$4.0861\cdot10^{+2}$&
$4.1030\cdot10^{+2}$&
$4.1449\cdot10^{+2}$&
$4.2294\cdot10^{+2}$&
$4.0852\cdot10^{+2}$
\tabularnewline
\hline
$90.63$&
$4.1988\cdot10^{+2}$&
$4.2162\cdot10^{+2}$&
$4.2592\cdot10^{+2}$&
$4.3461\cdot10^{+2}$&
$4.1979\cdot10^{+2}$
\tabularnewline
\hline
$90.68$&
$4.3084\cdot10^{+2}$&
$4.3263\cdot10^{+2}$&
$4.3705\cdot10^{+2}$&
$4.4596\cdot10^{+2}$&
$4.3075\cdot10^{+2}$
\tabularnewline
\hline
$90.99$&
$4.9041\cdot10^{+2}$&
$4.9245\cdot10^{+2}$&
$4.9749\cdot10^{+2}$&
$5.0766\cdot10^{+2}$&
$4.9031\cdot10^{+2}$
\tabularnewline
\hline
$91.187$&
$5.0254\cdot10^{+2}$&
$5.0463\cdot10^{+2}$&
$5.0981\cdot10^{+2}$&
$5.2024\cdot10^{+2}$&
$5.0243\cdot10^{+2}$
\tabularnewline
\hline
$91.25$&
$5.0143\cdot10^{+2}$&
$5.0352\cdot10^{+2}$&
$5.0869\cdot10^{+2}$&
$5.1911\cdot10^{+2}$&
$5.0133\cdot10^{+2}$
\tabularnewline
\hline
$91.56$&
$4.6103\cdot10^{+2}$&
$4.6296\cdot10^{+2}$&
$4.6772\cdot10^{+2}$&
$4.7732\cdot10^{+2}$&
$4.6094\cdot10^{+2}$
\tabularnewline
\hline
$91.77$&
$4.1178\cdot10^{+2}$&
$4.1350\cdot10^{+2}$&
$4.1776\cdot10^{+2}$&
$4.2635\cdot10^{+2}$&
$4.1170\cdot10^{+2}$
\tabularnewline
\hline
$92.0$&
$3.5297\cdot10^{+2}$&
$3.5444\cdot10^{+2}$&
$3.5810\cdot10^{+2}$&
$3.6547\cdot10^{+2}$&
$3.5289\cdot10^{+2}$
\tabularnewline
\hline
\end{tabular}
\end{footnotesize}
\end{center}
\caption{\small Invariant mass distributions at NLO for the MLSOM and the SM around the peak of the $Z$.
The mass of the anomalous extra $Z^{\prime}$ is taken to be 1 TeV with $\mu_F=\mu_R= Q$.}
\label{table_nlo}
\end{table}

\begin{figure}
\subfigure[      ]{\includegraphics[%
 width=6.5cm,
 angle=-90]{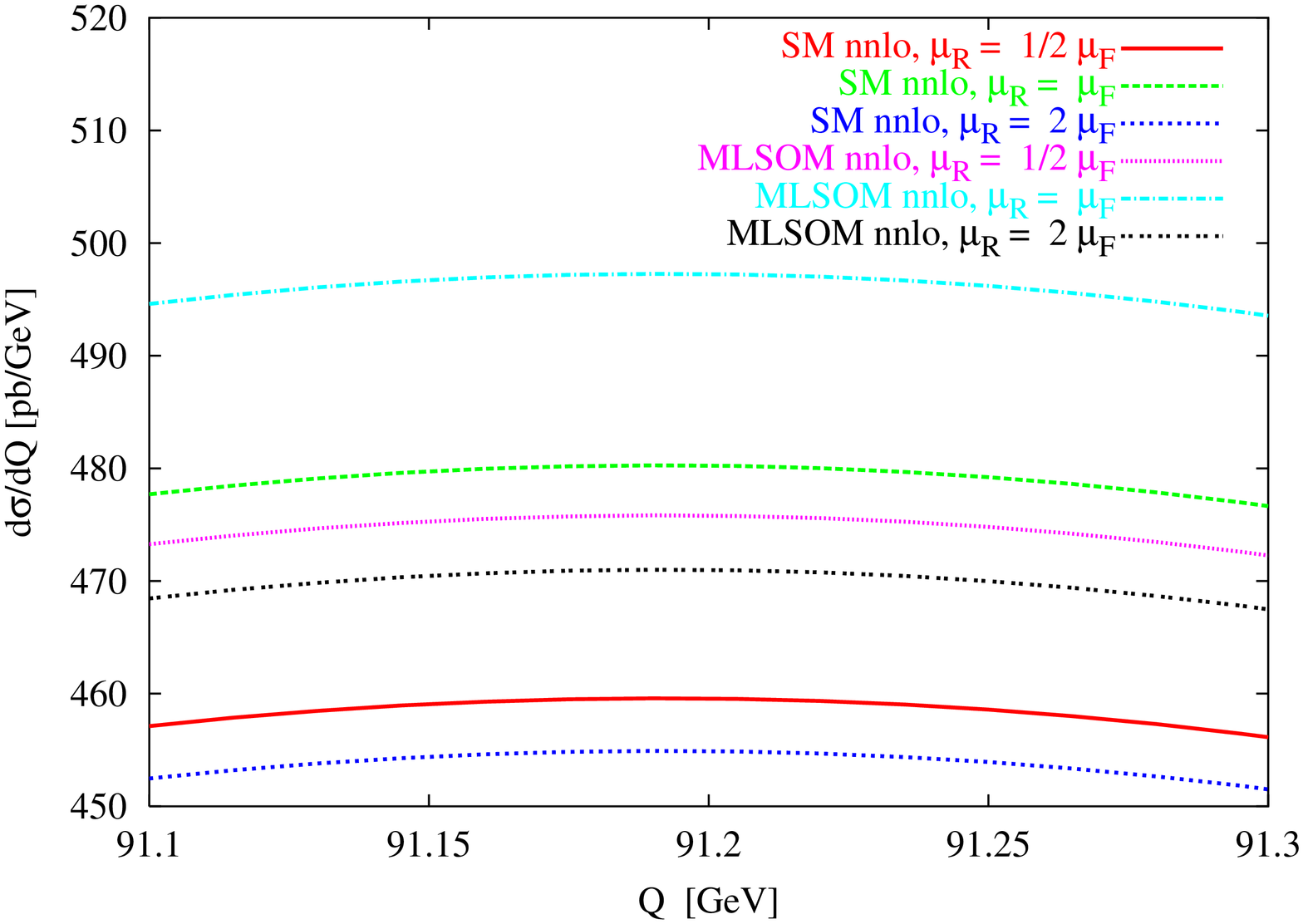}}
\subfigure[        ]{\includegraphics[%
 width=6.5cm,
 angle=-90]{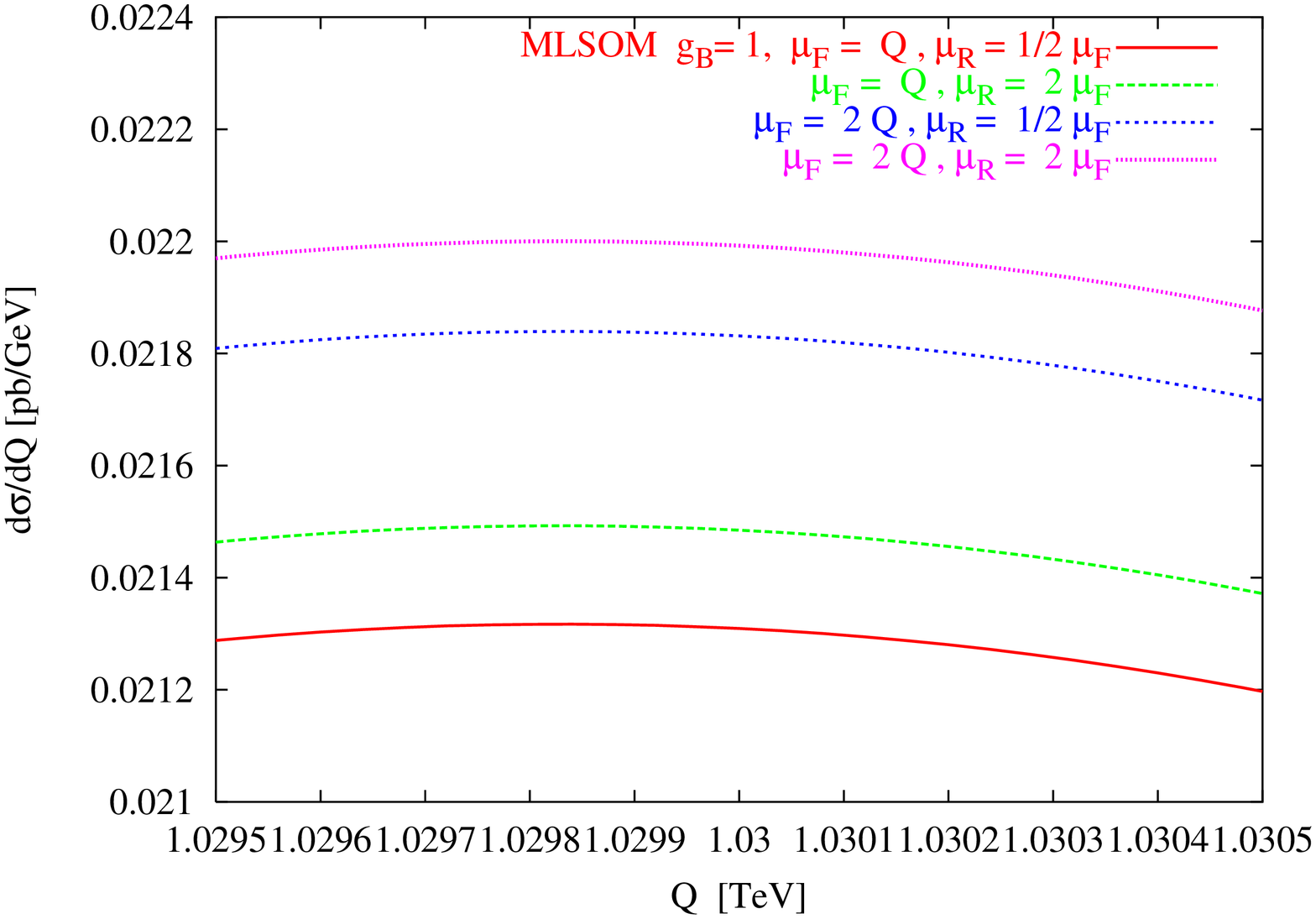}}
\caption{\small Zoom on the $Z$ resonance for anomalous Drell-Yan for varying
factorization and renormalization scales at NNLO for the SM and the MLSOM.
Results are shown for $Q\sim 91$ GeV (a) and 1 TeV  (b) both for $\sqrt{S}= 14$ TeV.}
\label{factren}
\end{figure}

\begin{table}
\begin{center}
\begin{footnotesize}
\begin{tabular}{|c||c|c|c|c|c|}
\hline
\multicolumn{6}{|c|}{$d\sigma^{nnlo}/dQ$ [pb/GeV] for the MLSOM with $M_{1}=1$ TeV, $\tan\beta=40$, \textsc{Candia} evol.}
\tabularnewline
\hline
$Q ~[\textrm{GeV}]$      &
$g_B=0.1$                 &
$g_B=0.36$                 &
$g_B=0.65$                 &
$g_B=1$                   &
$\sigma_{nnlo}^{SM}(Q)$\tabularnewline
\hline
\hline
$90.50$&
$3.6845\cdot10^{+2}$&
$3.6997\cdot10^{+2}$&
$3.7374\cdot10^{+2}$&
$3.8132\cdot10^{+2}$&
$3.6835\cdot10^{+2}$
\tabularnewline
\hline
$90.54$&
$3.7956\cdot10^{+2}$&
$3.8112\cdot10^{+2}$&
$3.8500\cdot10^{+2}$&
$3.9282\cdot10^{+2}$&
$3.7945\cdot10^{+2}$
\tabularnewline
\hline
$90.59$&
$3.9054\cdot10^{+2}$&
$3.9215\cdot10^{+2}$&
$3.9615\cdot10^{+2}$&
$4.0419\cdot10^{+2}$&
$3.9043\cdot10^{+2}$
\tabularnewline
\hline
$90.63$&
$4.0132\cdot10^{+2}$&
$4.0298\cdot10^{+2}$&
$4.0708\cdot10^{+2}$&
$4.1535\cdot10^{+2}$&
$4.0121\cdot10^{+2}$
\tabularnewline
\hline
$90.68$&
$4.1180\cdot10^{+2}$&
$4.1351\cdot10^{+2}$&
$4.1772\cdot10^{+2}$&
$4.2621\cdot10^{+2}$&
$4.1169\cdot10^{+2}$
\tabularnewline
\hline
$90.99$&
$4.6879\cdot10^{+2}$&
$4.7073\cdot10^{+2}$&
$4.7554\cdot10^{+2}$&
$4.8523\cdot10^{+2}$&
$4.6866\cdot10^{+2}$
\tabularnewline
\hline
$91.187$&
$4.8040\cdot10^{+2}$&
$4.8239\cdot10^{+2}$&
$4.8733\cdot10^{+2}$&
$4.9727\cdot10^{+2}$&
$4.8027\cdot10^{+2}$
\tabularnewline
\hline
$91.25$&
$4.7935\cdot10^{+2}$&
$4.8134\cdot10^{+2}$&
$4.8627\cdot10^{+2}$&
$4.9619\cdot10^{+2}$&
$4.7922\cdot10^{+2}$
\tabularnewline
\hline
$91.56$&
$4.4076\cdot10^{+2}$&
$4.4259\cdot10^{+2}$&
$4.4713\cdot10^{+2}$&
$4.5628\cdot10^{+2}$&
$4.4064\cdot10^{+2}$
\tabularnewline
\hline
$91.77$&
$3.9371\cdot10^{+2}$&
$3.9535\cdot10^{+2}$&
$3.9941\cdot10^{+2}$&
$4.0759\cdot10^{+2}$&
$3.9360\cdot10^{+2}$
\tabularnewline
\hline
$92.0$&
$3.3750\cdot10^{+2}$&
$3.3891\cdot10^{+2}$&
$3.4239\cdot10^{+2}$&
$3.4942\cdot10^{+2}$&
$3.3741\cdot10^{+2}$
\tabularnewline
\hline
\end{tabular}
\end{footnotesize}
\end{center}
\caption{\small Invariant mass distributions at NNLO for the MLSOM and the SM around the peak of the $Z$.
The mass of the anomalous extra $Z^{\prime}$ is taken to be 1 TeV with $\mu_F=\mu_R = Q$.}
\label{table_nnlo}
\end{table}
The evolution of the \textsc{Pdf}'s has been performed with \textsc{Candia}
\cite{Cafarella:2008du} which allows independent variations of  $\mu_F$ and
$\mu_R$ in the initial state. This analysis is shown in
Fig.~\ref{factren}, where we vary $\mu_F$ up to $2 Q$, while we have taken
$1/2 \mu_F \leq \mu_R \leq 2 \mu_F$. We observe that by increasing
both scales there is an enhancement
in the result and this is due to the logarithms $\ln{\mu_R^2/\mu_F^2}$ and
$\ln{Q^2/\mu_F^2}$, contained in the hard scatterings.
The scale variations induce changes of about $4\%$ in the SM
case at NNLO and about $3.5 \%$ in the MLSOM on the peak of the
$Z$. Notice that the variations are not symmetric as we vary the
scales and the percentual changes refer to the maximum variability.
This typical scale dependence is universal for all the studies presented
so far on the peak of the $Z$ and is a limitation of the parton model prediction.
After a large data taking, optimal choices for the \textsc{Pdf}'s and for $\mu_R$
and $\mu_F$ will allow a considerable reduction of this indetermination.
In subfigure (\ref{factren}, b) we repeat the same analysis, for the same c.m. energy,
this time for $Q\sim 1$ TeV, on the $Z^{\prime}$ resonance in the MLSOM, for a sizeable
coupling of the anomalous gauge boson,  $g_B=1$. Compared to the value on
the $Z$ peak, the reduction of the cross section is by a factor of
$2\times 10^4$. Also in this interval the variation of the differential
cross section with the two scales is around $3 \%$.
\begin{table}
\begin{center}
\begin{footnotesize}
\begin{tabular}{|c||c|c|c|c|}
\hline
\multicolumn{5}{|c|}{$\sigma_{tot}^{nnlo}$ [fb], $\sqrt{S}=14$ TeV, $M_{1}=1$ TeV, $\tan\beta=40$}
\tabularnewline
\hline
$g_z$&
MLSOM&
$U(1)_{B-L}$&
$U(1)_{q+u}$&
$Free Ferm.$
\tabularnewline
\hline
\hline
$0.1$ & $5.982$& $3.575$& $2.701$& $1.274$  \\
       & $0.173$& $0.133$& $0.177$& $0.122$  \\
       & $0.277$& $0.445$& $0.252$& $0.017$
\tabularnewline
\hline
$0.36$ & $106.674$& $105.567$& $53.410$& $42.872$  \\
       & $2.248$& $1.733$& $2.308$& $1.583$  \\
       & $4.937$& $13.138$& $4.991$& $0.586$
\tabularnewline
\hline
$0.65$ & $240.484$& $143.455$& $108.344$& $51.155$  \\
       & $7.396$& $5.700$& $7.592$& $5.205$  \\
       & $11.127$& $17.853$& $10.124$& $0.699$
\tabularnewline
\hline
$1$ & $532.719$& $317.328$& $239.401$& $113.453$  \\
       & $17.810$& $13.720$& $18.274$& $12.530$  \\
       & $24.639$& $39.491$& $22.370$& $1.550$
\tabularnewline
\hline
\end{tabular}
\end{footnotesize}
\end{center}
\caption{\small Total cross sections,  widths and
$\sigma_{tot}\times BR(Z\rightarrow l\bar{l})$, where
$BR(Z\rightarrow l\bar{l})=\Gamma_{Z^{\prime}\rightarrow l\bar{l}}/\Gamma_{Z^{\prime}}$,  for the MLSOM and three anomaly-free extensions of the SM; they are all shown as functions of the coupling constant.}
\label{tablethird}
\end{table}
We have added a table (Tab.\ref{tablethird}) in which we show results
for the total cross sections for the various models at the $Z$ peak. In the first
line of each column we show the results for
the total cross section in $[fb]$, in the 2nd line the total width
$\Gamma_{Z^{\prime}}$, expressed in $GeV$ and
in the 3rd line the observable $\sigma_{tot}\times BR(Z\rightarrow l\bar{l})$, where
$BR(Z\rightarrow l\bar{l})=\Gamma_{Z^{\prime}\rightarrow l\bar{l}}/\Gamma_{Z^{\prime}}$.
These quantities refer to the value of the coupling constant $g_z$ listed in the first column.
\footnote{Notice that we have chosen $g_z=g_B$ for the MLSOM.}

\begin{figure}[t]
\subfigure[  ]{\includegraphics[%
 width=6.5cm,
 angle=-90]{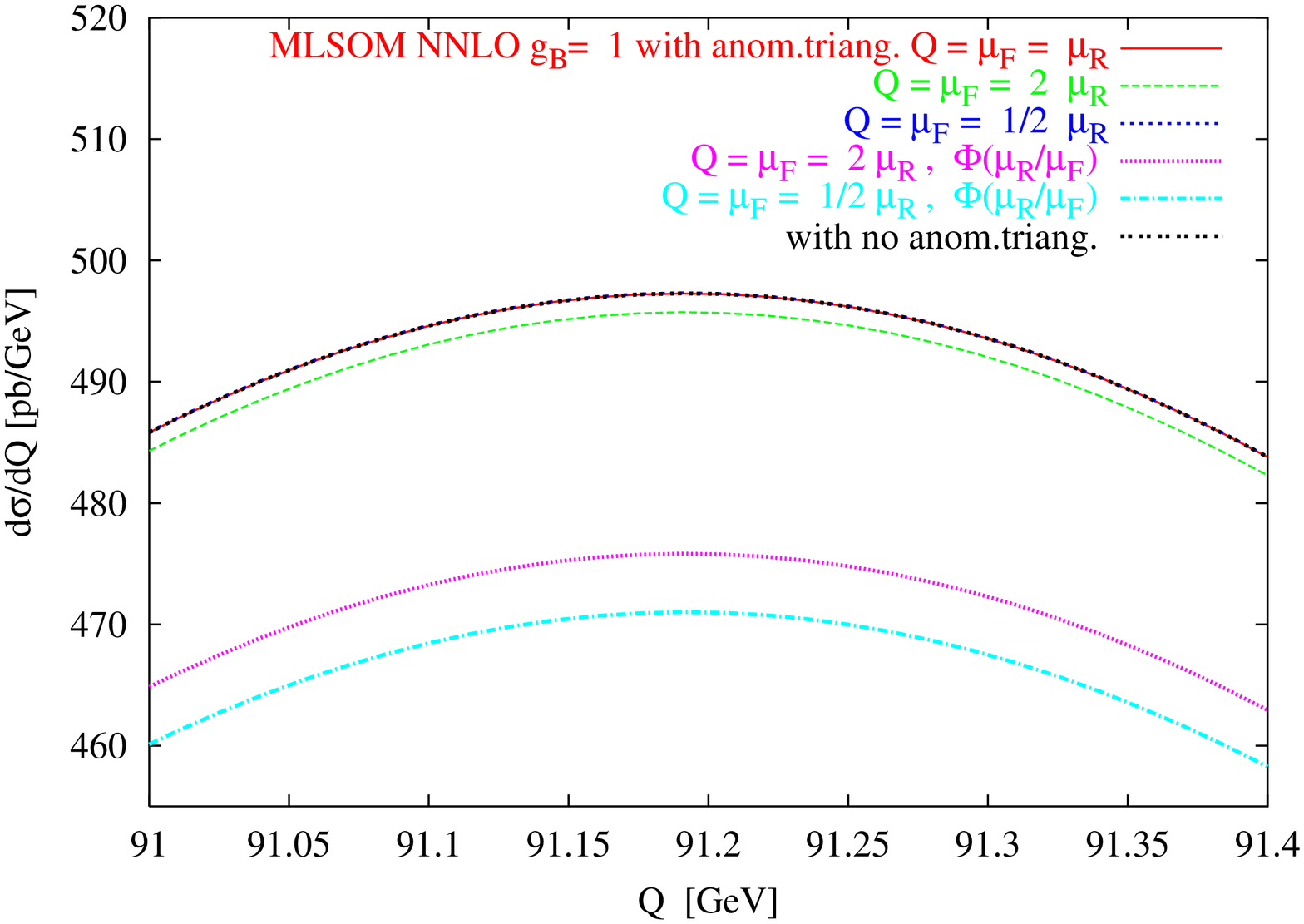}}
\subfigure[  ]{\includegraphics[%
 width=6.5cm,
 angle=-90]{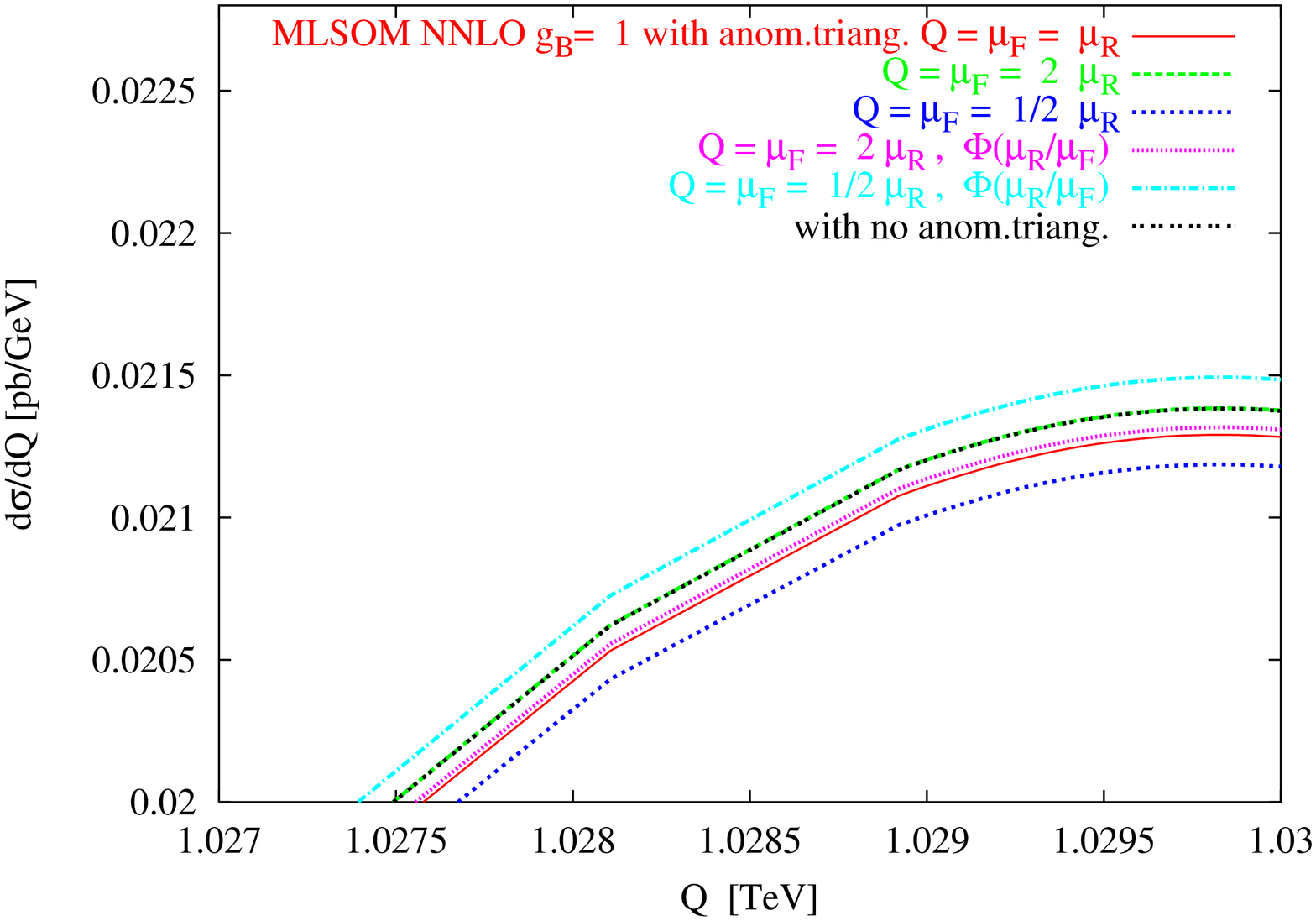}}
\caption{\small  Plot of the DY invariant mass distributions on the
peak of the $Z$ (a) and of the $Z^{\prime}$ (b). Shown are the total
contributions of the MLSOM and those in which the anomalous terms have been
removed. The variation of the result on $\mu_F$ and $\mu_R$ is included
both in the hard scatterings and in the luminosities ($\Phi(\mu_F/\mu_R)$).}
\label{mlsomef0}
\end{figure}

We show in Fig.~\ref{mlsomef0}a,b two plots of the results
for the MLSOM of the DY cross section on the peak of the Z and
of the extra $Z^{\prime}$, where we vary the scales both in the
hard scatterings and in the parton luminosities. We have added and
subtracted the anomalous sector in order to estimate their size respect
to the remaining contributions. As we have already pointed out, the
scale variability at NNLO is larger than the
changes induced on the result by the anomalous graphs.
The anomalous effects are more visible at large
$Q$ (subfig. (b)), on the resonance of the extra $Z^{\prime}$,
and are due to the behaviour of the anomalous components at large-$x$ due to a growing $Q$.

\begin{figure}[t]
\subfigure[  ]{\includegraphics[%
 width=6.5cm,
 angle=-90]{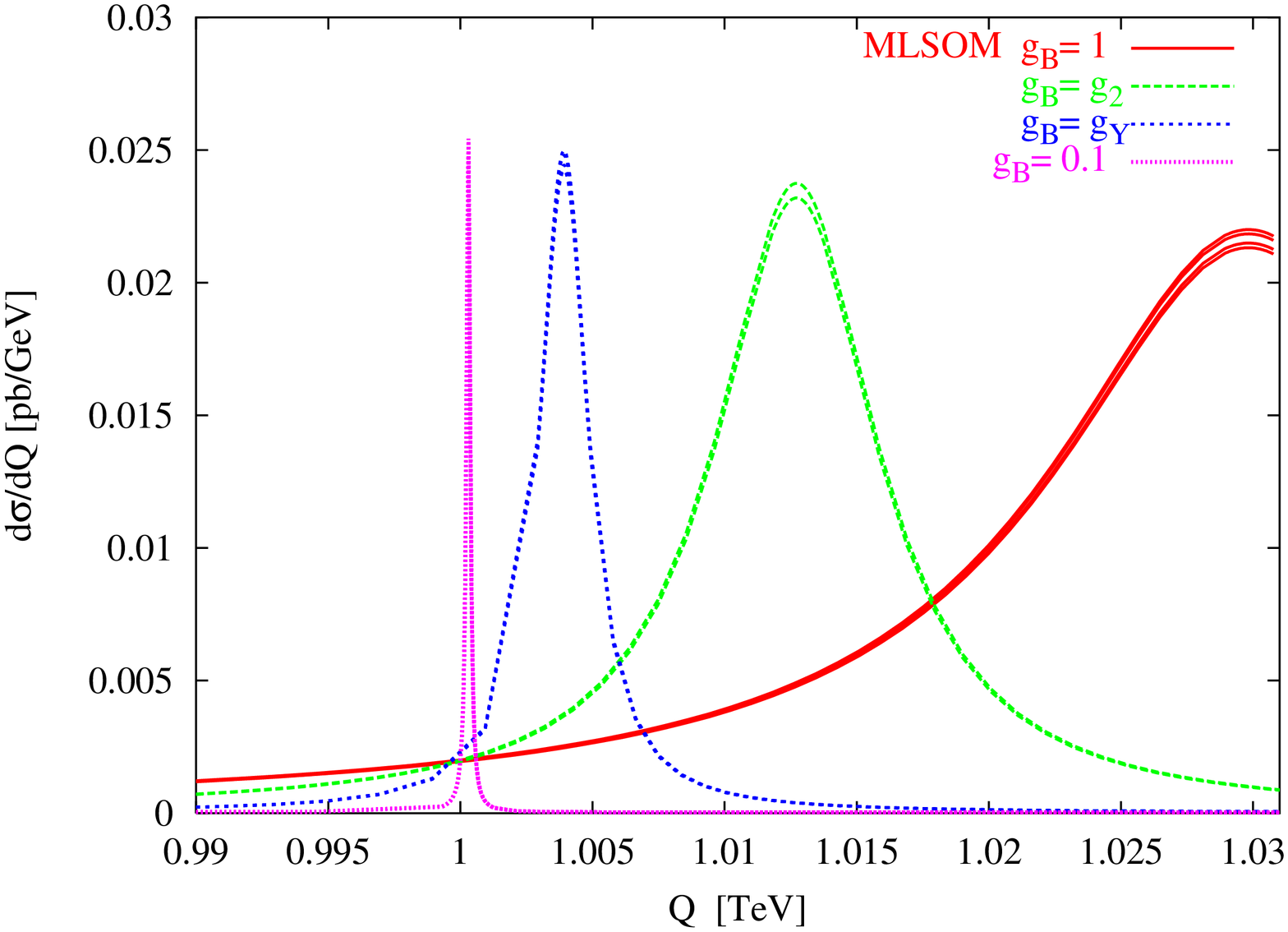}}
\subfigure[  ]{\includegraphics[%
 width=6.5cm,
 angle=-90]{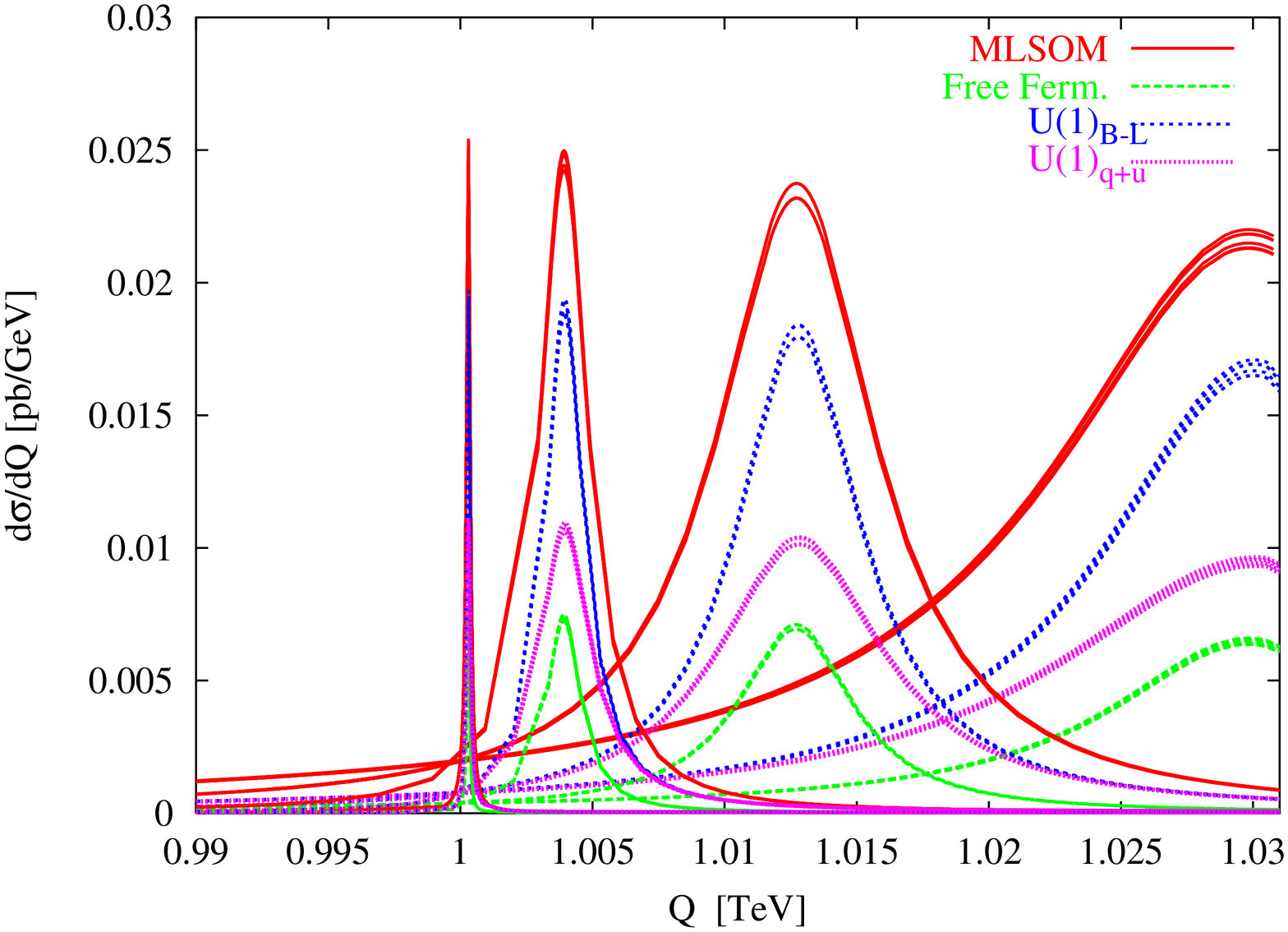}}
\caption{\small  (a) Anomalous $Z'$ resonances obtained by varying $g_B$. (b) Comparisons among anomalous Drell-Yan in the MLSOM versus several anomaly-free models.}
\label{mlsomef}
\end{figure}

In Fig.~\ref{mlsomef} we show in (a) a plot of the MLSOM for
different values of $g_B$ and for different values of $\mu_R$ and $\mu_F$.
The first peak (purple line) corresponds to $g_B=0.1$ the 2nd (blue line)
to $g_B=g_Y$ and so on. As $g_B$ grows the width of each peak
gets larger but the peak-value of the cross section decreases.
Different choices of $g_B$ correspond to slightly different values of
the mass of the extra $Z^{\prime}$ because of the relation between the
St\"uckelberg mass $M_1$ and $M_{Z^{\prime}}$ given in eq. (\ref{corr}).
For a fixed value of the coupling, the effects due to the variations
of the scales become visible only
for $g_B=1$ and in this case they are around 2-3\%.
In the case $g_B=1$  (red line), the uppermost lines correspond
to the choice $\mu_F=2 Q$, $\mu_R=1/2 \mu_F$ and $\mu_R= 2 \mu_F$,
while the lowermost lines correspond to the choice
$\mu_F=Q$, $\mu_R=1/2 \mu_F$ and $\mu_R= 2 \mu_F$.
Again, we notice that if we increase $\mu_F$ and $\mu_R$ the cross section
grows.

In Fig.~\ref{mlsomef}b we show the result of a comparison between
the MLSOM and the anomaly-free extensions. We have also included the $\mu_R/\mu_F$ scale dependence,
which appears as a band, and the variations with respect to $g_B$.
As shown in this figure, the red lines correspond to the MLSOM, the
blue lines to the $U(1)_{B-L}$ model, the green lines to the free fermionic model
and the purple lines to $U(1)_{q+u}$. Right as before, the first peak
corresponds to $g_B=0.1$, the 2nd to $g_B=g_Y$ etc.
The peak-value of the anomalous model is the largest of all, with a cross section which is
around $0.022$ [pb/GeV], the free fermionic appears to be the smallest with
a value around $0.006$ [pb/GeV].

\section{Direct Photons with GS and WZ interactions}
The analysis of $pp\to \gamma \gamma $ proceeds similarly to the DY
case, with a numerical investigation of the background and of the anomalous signal at parton level.

We start classifying the strong/weak interference
effects that control the various sectors of the process and then identify
the leading contributions due to the presence of anomaly diagrams.
\begin{figure}[ht]
\begin{center}
\includegraphics[scale=0.8]{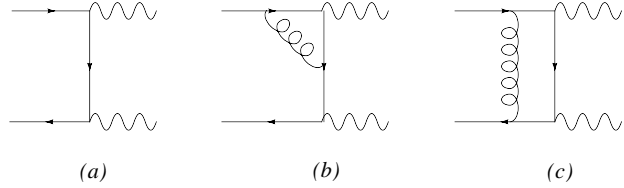}
\caption{\small $q\bar q$ sector for the process $q \bar q \rightarrow \g \g$ including virtual corrections at LO (a) and NLO (b,c).}
\label{qqbarvirtual}
\end{center}
\end{figure}
\begin{figure}[ht]
\begin{center}
\includegraphics[scale=0.8]{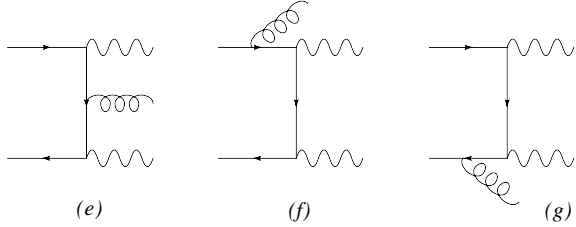}
\caption{\small Real emissions for $q \bar q \rightarrow \g \g$ at NLO.}
\label{qqbarreal}
\end{center}
\end{figure}
\begin{figure}[ht]
\begin{center}
\includegraphics[scale=0.8]{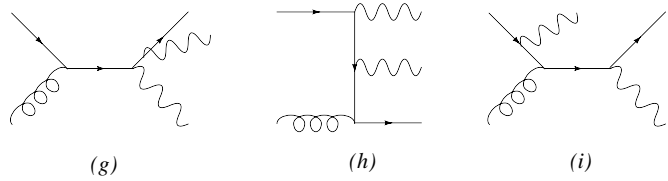}
\caption{\small $q g $ sector for the process $q g \rightarrow \g \g$. }
\label{qgall}
\end{center}
\end{figure}
\begin{figure}[ht]
\begin{center}
\includegraphics[scale=1.0]{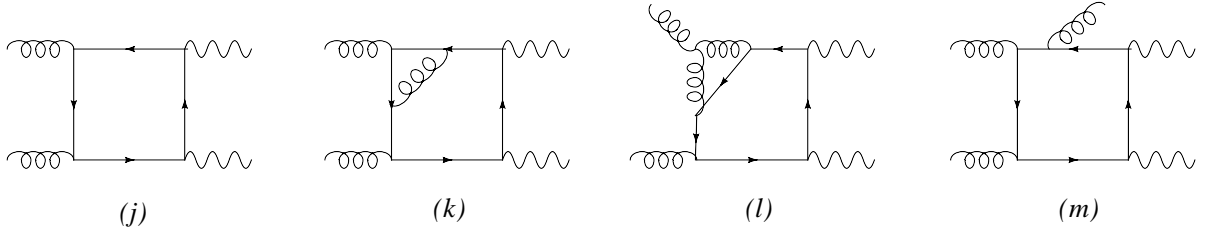}
\caption{\small $gg $ sector for the process $gg \rightarrow \g \g$ with virtual and real radiative corrections. }
\label{ggall}
\end{center}
\end{figure}
\begin{figure}[h]
\begin{center}
\includegraphics[scale=1.0]{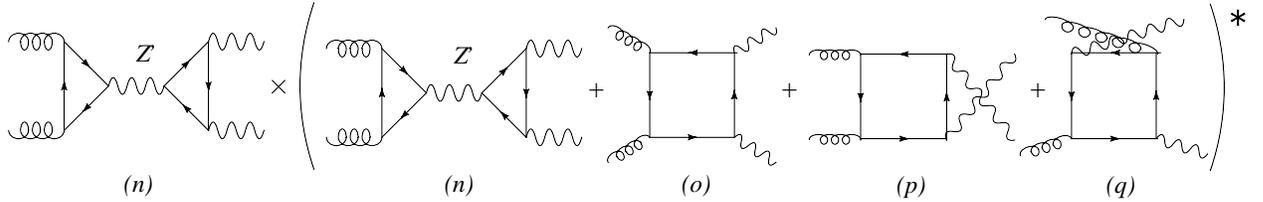}
\caption{\small Anomalous contributions for $ gg \rightarrow \g \g$ involving the BIM amplitude and its interference with the box graphs. }
\label{nopq}
\end{center}
\end{figure}
\begin{figure}[h]
\begin{center}
\includegraphics[scale=1.0]{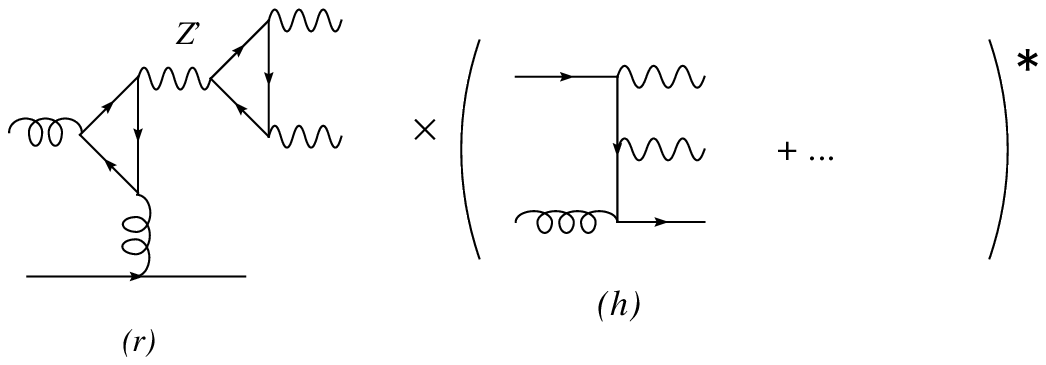}
\caption{\small Total amplitude for $ qg \rightarrow \g \g$. }
\label{qgtwodelta}
\end{center}
\end{figure}
\begin{figure}[h]
\begin{center}
\includegraphics[scale=1.0]{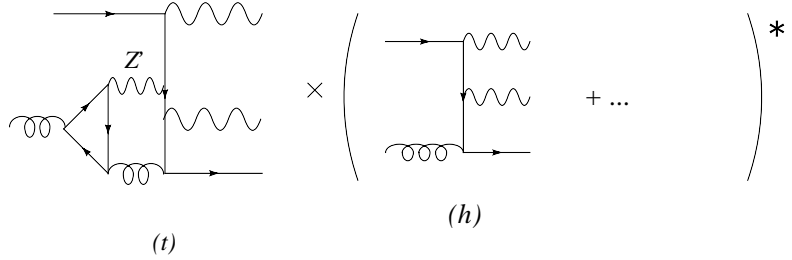}
\caption{\small Another configuration for the total amplitude of the  $ qg \rightarrow \g \g$ process. }
\label{delta2}
\end{center}
\end{figure}
\begin{figure}[t]
\begin{center}
\includegraphics[scale=0.8]{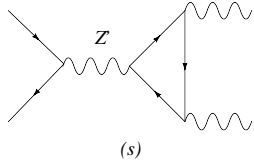}
\caption{\small Single diagram with an exchanged $Z^{\prime}$ boson in the s-channel. }
\label{s}
\end{center}
\end{figure}

\begin{figure}[t]
\begin{center}
\includegraphics[scale=1.1]{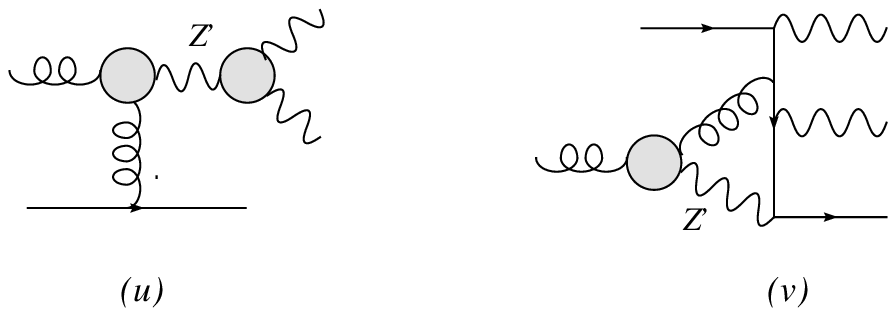}
\caption{\small Generic representation of the $qg\rightarrow
\g\g$ process in the presence of a GS vertex of the $AVV$ type.}
\label{qgGS}
\end{center}
\end{figure}

Direct photons are one of the possible channels to detect anomalous gauge
interactions, although, as we are going to see, also in this case the
anomalous signal remains rather small.
Direct photons are produced by partonic interactions rather than as a
result of the electromagnetic decay of hadronic states. At leading order
(LO) they carry the transverse momentum of the hard scatterers,
offering a direct probe of the underlying quark-gluon dynamics.
The two main channels in $pp$ collisions are the annihilation $q \bar{q}$
and Compton ($q\, g$), the second one being roughly $80 \%$ of  the entire
signal at large $p_T$  (from $p_T=4$ GeV on). The annihilation channel is
subleading, due to the small antiquark densities in the proton. The cross
section is also strongly suppressed (by a factor of approximately $10^{-3}$)
compared to the jet cross section. For this reason the electromagnetic decay
of the produced hadrons is a significant source of background, coming mostly
from the $\pi^0\to \gamma \gamma$ decay. In this case the angular opening of
the two photons is in general rather small, due to the small pion mass, and
proportional to their energy asymmetry
$((E_1-E_2)/(E_1 + E_2))$. A limitation in the granularity of the detector,
therefore,  may cause the two photons to be unresolved experimentally, giving
a spurious signal for direct photons. Even in the presence
of enough granularity in the detector, very asymmetric decays can also cause
the failure of the experimental
apparatus to resolve the low energy photon. A second source of background is
due to $\eta\to \gamma\gamma$ decay, which is about $20 \%$ of the pion contribution.
These two contributions account for almost all the background to direct photons.
The overall signal to background, though small, is supposed to raise with an
increasing $p_T$. This is due to the steepening of the
$\pi$ and $\eta$ spectra -at higher $Q^2$- respect to the $p_T$ spectra of
the parent jet, so there are smaller fractions of these particles that can
fragment into photons. The reduction of the background can be performed either
using reconstruction of the contributions due to the pions by measurements of
the invariant mass of the pair in the final state, or by imposing isolation cuts.
In the isolation procedure one can eliminate events with more than 2 particle in
the final state, considering that emissions from fragmentation are usually accompanied
by a large multiparticle background.
The selection of appropriate isolation cuts are one of the way to render the anomalous
signal more significant, considering that the tagged photon signal, although being of
higher order in $\alpha_s$ (NNLO in QCD), is characterized by a two-photon-only final
state. As we are going to see this signal is non-resonant, even in the presence of
an $s$-channel exchange, due to the anomaly.

We show in Fig.~\ref{qqbarvirtual} a partial list of the various background
contributions to the DP channel in
$pp $ collisions.
We show the leading order (LO) contribution in diagram (a)
with some of the typical virtual corrections included in (b)
and (c). These involve the $q\bar{q}$ sector giving a cross section of the form
\beq
\sigma_{q\bar{q}}= \alpha_{em}^2 ( c_1 + c_2 \alpha_s).
\eeq
These corrections are the NLO ones in this channel.
The infrared safety of the process is guaranteed at the same perturbative order
by the real emissions in Fig.~\ref{qqbarreal} with an integrated gluon in the final
state, which are also of $O(\alpha_{em}^2 \alpha_s)$.

A second sector is the $qg$ one, which is shown in Fig.~\ref{qgall}, also of the same
order $(O(\alpha_s \alpha_{em}))$. These corrections are diagrammatically
the NLO ones. In general, the NLO prediction for this process
are improved by adding a part of the NNLO
(or $O(\alpha_{em}^2\alpha_s^2)$) contributions, such as
the box contribution (j) of the $gg$ sector which is of
higher order ($O(\alpha_{em}^2 \alpha_s^2)$) in $\alpha_s$,
the reason being that these contributions have been shown to be
sizeable and comparable with the genuine NLO ones. All these corrections have
been computed long ago \cite{Coriano:1996us} and implemented independently
in a complete Monte Carlo in \cite{Binoth:1999qq,Binoth:2001vm} with a more
general inclusion of the fragmentation.
More recently, other NNLO contributions have been added
to the process, such as those involving the gg sector through $O(\alpha_{em}^2 \alpha_s^3)$,
\beq
\sigma_{gg}= \alpha_{em}^2( d_1 \alpha_s^2 + d_2 \alpha_s^3),
\eeq
shown in graphs $(k),(l),(m)$.
The other sectors have not yet been computed with the same
accuracy, for instance in the $qq$ and $q\bar{q}$ channels
they involve 2 to 4 emission amplitudes which need to be
integrated over 2 gluons. For instance, graph (m) is a real
emission in $\sigma_{gg}$ which is needed to cancel the
infrared/collinear singularities of the virtual ones at the same order.

\begin{figure}[t]
\begin{center}
\includegraphics[scale=0.9]{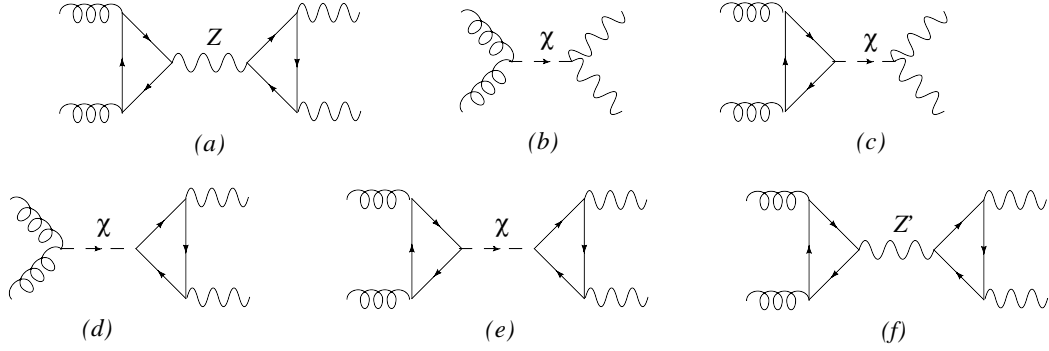}
\caption{\small Complete list of amplitudes included in the type of graphs shown in Fig.~\ref{nopq}. They also have the exchange of a physical axion and contributions proportional to the mass of the internal fermion.}
\label{qgonedelta}
\end{center}
\end{figure}

The anomalous contributions are shown in Figs.~\ref{nopq}, \ref{qgtwodelta}, \ref{delta2}, \ref{s}
and \ref{qgGS}.\footnote{As stated in the previous sections, the diagram in
Fig.~\ref{s} vanish because of a Ward identity.}
One of the most important partonic
process, in this case, is the BIM amplitude shown in diagram (n)
which violates unitarity at very high energy. In the WZ case, where the axion $b$
appears in trilinear interactions not as a virtual state (such as in $(b F\wedge F)$)
this amplitude has two main properties;
1) it is well defined and finite in the chiral limit even
for on-shell physical gluons/photons and
2) it is non-resonant and can grow beyond the unitarity limit.
Working in the chiral limit, its expression is given by the Dolgov-Zakharov
limit of the anomaly amplitude (Eqs. (\ref{massiveT}) and (\ref{a6}), with $m_f=0$),
which appears both in the production mechanism of the extra $Z^{\prime}$,
$(gg\to Z^{\prime}) $, where the $Z^\prime$ in the $s$-channel is virtual, and in
its decay into two photons.
We recall that the presence of an anomaly in the initial and in the final state
cancels the resonance of a given channel \cite{Coriano:2008pg}.
For simplicity we consider amplitude Fig.~\ref{nopq}n, assuming to have photons both in the
initial and in the final state. The amplitude is given by
\ba
A_{BIM}&=&   \frac{a_n}{k^2} k^{\lambda} \varepsilon[\mu,\nu,k_1,k_2]  \, \frac{- i}{k^2 - M^2}
\left( g^{\lambda \lambda^\prime}
- \frac{k^\lambda k^{\lambda^\prime} }{ M^2 }   \right)  \frac{a_n}{k^2}
( - k^{\lambda^\prime}) \varepsilon[\mu^\prime,\nu^\prime,k_1^\prime,k_2^\prime]   \nonumber\\
&=&   \frac{a_n}{k^2}  \varepsilon[\mu,\nu,k_1,k_2]  \, \frac{- i}{ k^2 - M^2 }   \frac{k^{ \lambda^\prime}( M^2 - k^2 ) }{ M^2 }  \,  \frac{a_n}{k^2}
( - k^{\lambda^\prime}) \varepsilon[\mu^\prime,\nu^\prime,k_1^\prime,k_2^\prime]   \nonumber\\
&=&   \frac{a_n}{k^2}  \varepsilon[ \mu,\nu,k_1,k_2]  \, \left(  \frac{- i k^2 }{  M^2 }  \right) \,  \frac{a_n}{k^2}
 \varepsilon[\mu^\prime,\nu^\prime,k_1^\prime,k_2^\prime]   \nonumber\\
&=&  - \frac{a_n}{M}  \varepsilon[ \mu,\nu,k_1,k_2]  \,\,  \frac{ i }{ k^2 }  \,\,  \frac{a_n}{M}
 \varepsilon[\mu^\prime,\nu^\prime,k_1^\prime,k_2^\prime].
\label{resBIM}
\ea
where $M$ denotes, generically, the mass of the anomalous gauge boson
in the $s$-channel. If we multiply this amplitude by the external polarizators of
the photons, square it and perform the usual averages, one finds that it grows
quadratically with energy.  The additional contributions in the $s$-channel that
accompany this amplitude are shown in Fig.~\ref{qgonedelta}.
The exchange of a massive axion (Fig.~\ref{qgonedelta}b),
due to a mismatch between the coupling and the parameteric dependence
between Fig.~\ref{qgonedelta}a and b, does not erase the growth
(see the discussion in the appendix).
This mismatch is at the origin of the unitarity bound for this
theory analyzed in \cite{Coriano:2008pg}.
The identification of this
scale in the context of QCD is quite subtle, since the lack of unitarity in a partonic
process implies a violation of unitarity also at hadron level, but at
a different scale compared to the partonic one, which needs to be
determined numerically directly from the total hadronic cross
section $\sigma_{pp}$. Overall, the convolution of a BIM amplitude
with the parton distributions will cause a suppression of the rising partonic contributions,
due to the small gluon density at large Bjorken $x$. Therefore,
the graphs do not generate a large anomalous signal in this channel.
However, the problem of unitarizing the theory by the inclusion of
higher dimensional operators beyond the minimal dimension-5 operator $b F\wedge F$ remains.

The anomalous terms, beside the $(\textrm{n}){(\textrm{n})}^* $ contribution with the
exchange of an extra $Z/Z^\prime$ which carry an anomalous component,
which is $O(\alpha_{em}^2 \alpha_s \alpha_w^2)$, include the
interference between the $s/t/u$ box diagrams of $gg\to\gamma\gamma$
with the same BIM amplitude (n). In the $gg$ sector
the anomalous terms give, generically, an expression of the form
\beq
\sigma_{gg}^{an}= \alpha_{em}^2(a_1\alpha_s^2\alpha_w + a_2 \alpha_s^2 \alpha_w^2),
\eeq
with the first contribution coming from $(\textrm{n}){(\textrm{o})}^* $
and from the interference with the ($gg\to\gamma \gamma$) box diagram,
while the second from $(\textrm{n}){(\textrm{n})}^* $.
Other contributions which appears at $O(\alpha_{em}^2\alpha_s \alpha_w)$
are those shown in Fig.~\ref{qgtwodelta} which involve 2 anomaly diagrams (r) and their
interference with the NLO real emission diagram of type (m). These contributions are phase-space
suppressed. If we impose isolation cuts on the amplitude, we can limit our analysis, for the anomalous signal, only to 2-to-2 processes.

\subsection{Helicity amplitudes: massless box diagrams and anomalous interferences}

Moving to the computation of the anomalous contributions to
$g(p_1,\pm)+g(p_2,\pm)\rightarrow \gamma(k_1,\pm)+\gamma(k_2,\pm)$,
coming from the 2-to-2 sector we identify the
following non-vanishing helicity amplitudes for the diagrams shown in
Fig.~\ref{nopq}, with the usual conventions
\ba
&&s=(p_1+p_2)^2
\nonumber\\
&&t=(p_1-k_1)^2=-s/2(1-\cos{\theta})
\nonumber\\
&&u=(p_1-k_2)^2=-s/2(1+\cos{\theta}),
\ea
where $\theta$ is the angle between $\vec{p}_1$ and $\vec{k}_1$, and we obtain
\ba
M^{{\cal Z}}_{++++}&=&\frac{s}{(2\pi)^4 M_{\cal Z}^2},
\nonumber\\
M^{{\cal Z}}_{----}&=&\frac{s}{(2\pi)^4 M_{\cal Z}^2},
\nonumber\\
M^{{\cal Z}}_{++--}&=& i\,\varepsilon[k_1,k_2,p_1,p_2]
\frac{{\left( t^2 - u^2 \right) }^3 - s^2\,\left( t^4 - u^4 \right)}{8 M_{\cal Z}^2\,{\pi }^4\,s^3\,t^2\,u^2}
\nonumber\\
&-&\frac{ {\left( t^2 - u^2 \right) }^4 -2\,s^2\,{\left( t^2 - u^2 \right) }^2\,
\left( t^2 + u^2 \right) + s^4\,\left( t^4 + u^4 \right)}{32\,M_{\cal Z}^2\,{\pi }^4\,s^3\,t^2\,u^2},
\nonumber\\
M^{{\cal Z}}_{--++}&=& -i\,\varepsilon[k_1,k_2,p_1,p_2]
\frac{{\left( t^2 - u^2 \right) }^3 - s^2\,\left( t^4 - u^4 \right)}{8 M_{\cal Z}^2\,{\pi }^4\,s^3\,t^2\,u^2}
\nonumber\\
&-&\frac{ {\left( t^2 - u^2 \right) }^4 -2\,s^2\,{\left( t^2 - u^2 \right) }^2\,
\left( t^2 + u^2 \right) + s^4\,\left( t^4 + u^4 \right)}{32\,M_{\cal Z}^2\,{\pi }^4\,s^3\,t^2\,u^2},
\nonumber\\
M^{\chi}_{++++}&=&\frac{16 s^2}{(s-M_{\chi}^2)},
\nonumber\\
M^{\chi}_{----}&=&\frac{16 s^2}{(s-M_{\chi}^2)},
\nonumber\\
M^{\chi}_{++--}&=&-32\, i \varepsilon[k_1,k_2,p_1,p_2]
\frac{s^2 \left( t^4 - u^4 \right)-\left( t^2 - u^2 \right)^3 }
{s^2 \left( s -M_{\chi}^2 \right) \,t^2\,u^2}
\nonumber\\
&-&8\,\frac{\left( t^2 - u^2 \right)^4 - 2\,s^2\,\left( t^2 - u^2 \right)^2\,
\left( t^2 + u^2 \right) + s^4\,\left( t^4 + u^4 \right)}{s^2\,
\left( s-M_{\chi}^2 \right) \,t^2\,u^2},
\nonumber\\
M^{\chi}_{--++}&=& 32\, i \varepsilon[k_1,k_2,p_1,p_2]
\frac{s^2 \left( t^4 - u^4 \right)-\left( t^2 - u^2 \right)^3 }
{s^2 \left( s -M_{\chi}^2 \right) \,t^2\,u^2}
\nonumber\\
&-&8\,\frac{\left( t^2 - u^2 \right)^4 - 2\,s^2\,\left( t^2 - u^2 \right)^2\,
\left( t^2 + u^2 \right) + s^4\,\left( t^4 + u^4 \right)}{s^2\,
\left( s-M_{\chi}^2 \right) \,t^2\,u^2},
\ea
where again ${\cal Z}$ indicates generically either a $Z$ or a $Z^{\prime}$
and the $M^{\chi}$ refers to the
contributions with the exchange of a axi-higgs.
In the above formulas we have omitted all the coupling constants to obtain
more compact results. The helicity amplitudes for the massless box contribution
have been computed in \cite{Bern:2001df} and are given by
\ba
&&M^{box}_{--++}=1,
\nonumber\\
&&M^{box}_{-+++}=1,
\nonumber\\
&&M^{box}_{++++}=-\frac{1}{2} {t^2+u^2\over s^2}
\Bigl[ \ln^2\Bigl({t\over u}\Bigr) + \pi^2 \Bigr]
- {t-u\over s} \ln\Bigl({t\over u}\Bigr) - 1,
\nonumber \\
&&M_{+--+}^{box} =- {1\over2} {t^2+s^2\over u^2}
\ln^2\Bigl(-{t\over s}\Bigr)
- {t-s\over u} \ln\Bigl(-{t\over s}\Bigr) - 1
\nonumber \\
&& \null \hskip 2 cm
- i \pi \biggl[ {t^2+s^2\over u^2} \ln\Bigl(-{t\over s}\Bigr)
+ {t-s\over u} \biggr],
\nonumber \\
&&M_{+-+-}^{box}(s,t,u)=M_{+--+}^{box}(s,u,t),
\ea
giving a differential cross section
\beq
{d\sigma^{box} \over d\cos\theta} = {\alpha_{em}^2 \alpha_{s}^2 N_c^2\over 64 \pi \, s}
\left[\sum_f Q_f^2\right]^2   \Bigl\{
     | M_{--++}^{box} |^2 + 4 \, | M_{-+++}^{box} |^2
   + | M_{++++}^{box} |^2 + | M_{+--+}^{box} |^2 + | M_{+-+-}^{box} |^2
   \Bigr\}.
\label{Box_cross}
\eeq
The interference terms are listed in Fig.~\ref{nopq} and the interference differential cross
section is given by
\ba
{d\sigma^{int} \over d\cos\theta} = \sum_{{\cal Z}=Z,Z^{\prime}} {d  \sigma^{ {\cal Z}, box} \over d\cos\theta}  +
{d\sigma^{\chi, box} \over d\cos\theta},
\ea
where
\ba
 {d  \sigma^{ {\cal Z}, box} \over d\cos\theta} &=& \frac{1}{256 \pi s}   \sum_{q}
  \frac{1}{2}    c_1^q     \sum_{q^\prime} \frac{1}{2}  c_2^{ q^\prime }
 \sum_f Q_f^2 \alpha_{em}\alpha_{s} N_c
{\mbox {Re}}\left[2 M_{++++}^{\cal Z} M_{++++}^{* box} + (M_{++--}^{\cal Z} + M_{--++}^{\cal Z})M_{++--}^{* box}
\right]\nonumber\\
&=& - \frac{1}{8192 \pi^5 M^2_{\cal Z} } \sum_{q}
  \frac{1}{2}    c_1^q     \sum_{q^\prime} \frac{1}{2}  c_2^{ q^\prime }   \sum_f Q_f^2 \alpha_{em}\alpha_{s} N_c
 \left[   \left(       \cos^2 \theta + 1  \right)
   \log^2  \left(  \frac{ 1 -   \cos\theta  }{  1+   \cos\theta    }  \right)    \right.    \nonumber\\
 &&    +   4  \cos\theta  \log \left( \frac{ 1-   \cos\theta }{ 1 +  \cos\theta   } \right)
 + \left(  \cos^2\theta + 1 \right)  \pi^2 + 8   \Big],
\ea
\ba
 {d  \sigma^{ \chi, box} \over d\cos\theta} &=& \frac{ g^{\chi}_{GG} g^{\chi}_{\g\g} }{256 \pi s}
\sum_f Q_f^2 \alpha_{em}\alpha_{s} N_c   \,{\mbox Re}
\left[2 M_{++++}^{\chi} M_{++++}^{* box} + (M_{++--}^{\chi} + M_{--++}^{\chi})M_{++--}^{* box}
\right]
\nonumber\\
&=&  - \frac{s}{32 \pi ( s- M_\chi^2) } g^{\chi}_{GG} g^{\chi}_{\g \g}
\sum_f Q_f^2 \alpha_{em}\alpha_{s} N_c
    \left[  \left(    \cos^2 \theta + 1 \right)
   \log^2  \left(  \frac{ 1 -  \cos\theta   }{ 1 +  \cos\theta  } \right)     \right.   \nonumber\\
&&  +4  \cos\theta \log \left( \frac{ 1 -    \cos\theta  }{ 1+ \cos\theta  }  \right)
    + \left( 1+  \cos^2\theta  \right)
   \pi^2 + 8  \Big] ,
\ea

Here $M_{++--}^{box}=M_{--++}^{box}$ and we have introduced the parameters
of the model, explicitly given in \cite{Coriano:2008pg}. These correspond
to the chiral asymmetries $(D_{BYY})$ of the anomalous $U(1)_B$ in the
$BYY$ mixed anomalous diagram whose anomaly is $a_n$; rotation matrix
elements from the interaction eigenstates to the physical basis after
electroweak symmetry breaking ($O^A$); rotation matrix elements in the CP-odd sector
($O^{\chi}$), the strong coupling constant $g_3$, the anomalous coupling
constant  $g_B$ and the axial-vector coupling of the neutral gauge bosons
($\mathcal{Z}=Z,Z^{\prime}$) to a quark flavour $q$ ($g_{A,q}^{\cal Z}$).
These parameters combine to define the coupling of the physical axion to
the various gauge fields ($g^{\chi}_{gg}, g^{\chi}_{\g\g}$)

\begin{eqnarray}
&&g_{A,q}^{\cal Z} = \frac{1}{2}(Q_{\cal Z}^{R,q}-Q_{\cal Z}^{L,q}),
\nonumber\\
&&g^{\chi}_{gg}=\frac{D}{M_1}O^{\chi}_{31},
\nonumber\\
&&g^{\chi}_{\g\g}=\left[\frac{F}{M_1}(O^{A}_{W\g})^2
+\frac{C_{YY}}{M_1}(O^{A}_{Y\g})^2\right]O^{\chi}_{31}
\end{eqnarray}
and
\ba
&&D=ig_B g_3^2 a_n D_B^{(L)},\hspace{1cm} D_B^{(L)}=-\frac{1}{8}\sum_f Q_{B,f}^{L}
\nonumber\\
&&F=ig_B g_2^2 \frac{a_n}{2} D_B^{(L)},
\nonumber\\
&&C_{YY}=ig_B g_Y^2 \frac{a_n}{2} D_{BYY},\hspace{1cm}
D_{BYY}=-\frac{1}{8}\sum_f \left[Q_{B,f}^{L}(Q_{Y,f}^{L})^2-Q_{B,f}^{R}(Q_{Y,f}^{R})^2\right].
\nonumber\\
\ea
More details can be found in \cite{Coriano:2008pg, Coriano:2007fw, Coriano:2007xg, Armillis:2007tb}.

\subsection{ Numerical analysis for direct photons}

\begin{figure}[t]
\begin{center}
\includegraphics[width=7cm,angle=-90]{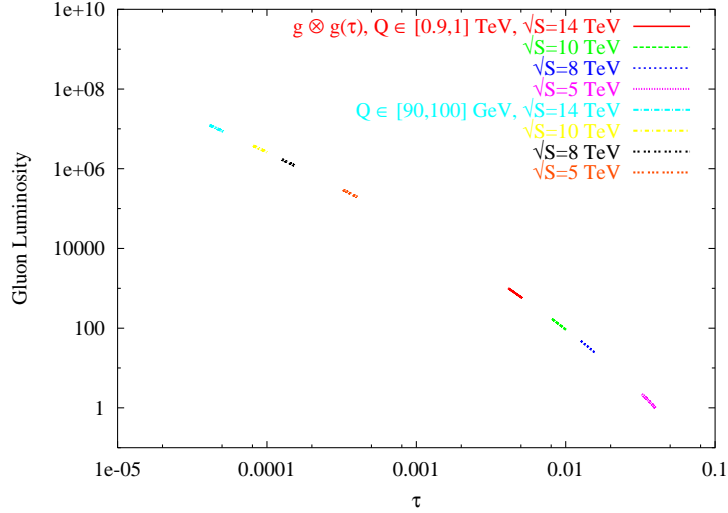}
\caption{\small The NNLO gluon luminosity as a function of $\tau=Q^2/S$  for various value of the invariant mass of the photon pair ($Q$) and energy of the hadronic beam ($S$) at the LHC evolved with \textsc{Candia}.}
\label{Gluons}
\end{center}
\end{figure}

In our numerical implementation of double prompt photon production
we compare the size of the anomalous
corrections respect to the SM background evaluated by a Monte Carlo
\cite{Binoth:1999qq, Bern:2002jx}.
Since the anomalous signal is small compared to that of the SM, we have extracted
both for the SM case and the anomalous case the $gg$ sector and compared them at
hadron level by convoluting the partonic contributions with
the \textsc{Pdf}'s (see Fig.~\ref{Gluons}).
In this comparison, the SM sector is given by the graphs shown in
Fig.~\ref{ggall} plus the interference graphs shown in  Fig.~\ref{nopq}. In the
SM case this second set of graphs contributes proportionally to the mass of the
heavy quarks in the anomaly loop. At high energy the hard scatterings coming from
this interference are essentially due to the mass of the top quark running inside
a BIM amplitude and are, therefore, related to heavy quark effects.
In the anomalous case the same set of graphs is considered, but now the
anomaly contributions are explicitly included. The hadronic differential
cross section due to the anomalous interactions for massless quarks
is given by
\ba
&&\frac{d\sigma}{d Q}=\int_{0}^{2\pi}d\phi \int_{-1}^{1} d\cos{\theta}\,\,
\frac{\tau}{4 Q}\int_{\tau}^{1}\frac{d x}{x}\Phi_{gg}(\frac{\tau}{x})\Delta(x,\theta),
\nonumber\\
&&\Phi_{gg}(y)=\int_{y}^{1}\frac{d z}{z}g(y/z)g(z),
\nonumber\\
&&\Delta(x)=\delta(1-x)\left[\frac{d\sigma_{Z}}{d\cos{\theta}}
+\frac{d\sigma_{Z'}}{d\cos{\theta}}+\frac{d\sigma_{\chi}}{d\cos{\theta}}
+\frac{d\sigma_{int}}{d\cos{\theta}}\right]
\nonumber\\
&&\frac{d\sigma_{int}}{d\cos{\theta}}=\frac{d\sigma^{Z,box}}{d\cos{\theta}}+
\frac{d\sigma^{Z',box}}{d\cos{\theta}}+\frac{d\sigma^{\chi,box}}{d\cos{\theta}}.
\ea
The contributions which are part of this sector due to exchange
of a $Z$ or a $Z^{\prime}$ and a $\chi$ (see (a), (b) and (f) of the BIM set
in Fig.~\ref{qgonedelta}) are those labelled above, while $\sigma_{int}$
refers to the interferences shown in Fig.~\ref{nopq}, with the inclusion
of a $Z^{\prime}$ and a physical axion
(such as Fig.~\ref{qgonedelta}b).

Defining
\ba
&&\sigma_{gg\to \gamma \gamma} \equiv  \int_{0}^{2\pi}d\phi \int d\cos{\theta} \,\, \Delta(x,\theta)
\ea
the hadronic cross section takes the form of a product of the gluon luminosity and the partonic $gg\to\gamma \gamma$ cross section
\ba
\frac{d\sigma}{dQ}=\frac{Q}{4 S}\sigma_{gg\to \gamma\gamma} \Phi(\tau).
\label{factor2}
\ea

\subsection{The $gg$ sector}

\begin{figure}[h]
\begin{center}
\includegraphics[width=7cm, angle=-90]{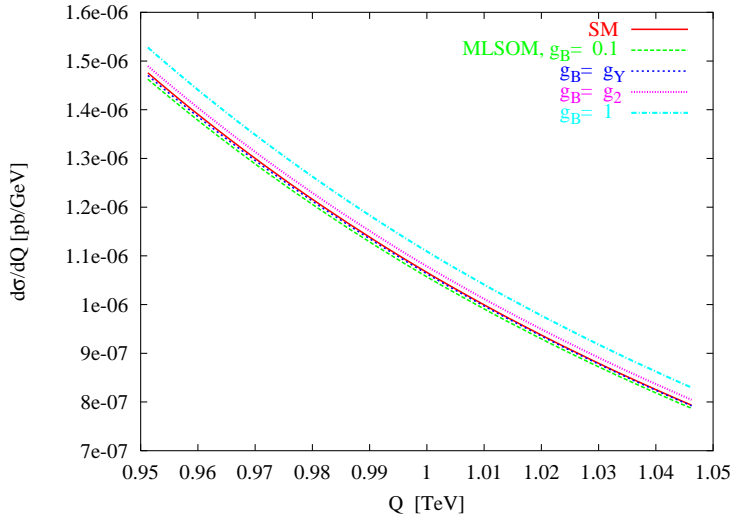}
\caption{\small Comparison plots for the gluon sector in the SM and in the anomalous model for a resonance of
1 TeV. The box-like contributions are not included,
while they appear in the interference with the BIM amplitudes.}
\label{new2}
\end{center}
\end{figure}

Coming to the analysis of the gluon fusion sector, the result of this study
is shown in Fig.~\ref{new2}
where we plot the gluon contribution to the hadronic cross section
for both the SM and the MLSOM, having chosen $M_1=1$ TeV.
We have used the MRST99 set of parton distributions
to generate the NNLO gluon luminosity
with $\alpha_s(M_Z)=0.1175$, $Q=\frac{1}{2}\mu_R$ and $\sqrt{S}=14$ TeV.
We have chosen $\tan\beta=40$ and different values of $g_B$. The size of the
cross section is around $10^{-6}$ [pb/GeV] - right on
the mass of the resonance - for both models, with a difference that grows as
we rise the coupling constant for the anomalous $U(1)$ ($g_B$).
We have chosen four possible values for $g_B$: a small parametric value
($g_B=0.1$); equal to the coupling of the hypercharge $g_Y$ at the same scale
($g_B=g_Y$) or to
the $SU(2)_w$ coupling $g_2$  ($g_B=g_2$) or, finally, parameterically
sizeable, with $g_B=1$. In the interference graphs used for this comparison
between the anomalous signal of the MLSOM and the SM  (in this second case
the BIM amplitudes contribute via the heavy quark mass in the loops) we have
included, beside the BIM amplitude, the entire set of contributions shown in
Fig.~\ref{qgonedelta}, with the exchange of a $Z$, a
$Z^{\prime}$ and the axi-Higgs $\chi $.
We have chosen a light St\"uckelberg axion with $m_\chi=30$ GeV.
For $Q$ around $m_\chi$ the
anomalous signal grows quite substantially, as we are going to show next.
The overall flatness of the result in
this region - the width of the interval is just 5 GeV in Fig.~\ref{new2}
- shows that the corrections are non-resonant and rather small.
They are also overlapping for both models and the extraction of additional
information concerning the anomalous sector appears to be very difficult.
Before we get into a more detailed analysis of the various contributions
to this sector, we mention that we have performed isolation
cuts on the cross section of the SM background in DP with a choice
of $R=0.4$  for the radius of the cone of isolation of the photons.
This is defined in terms of an azimuthal
angle $\phi$ and pseudorapidity $\eta=\ln{\tan{\theta/2}}$, with a
maximal value of transverse energy $E_{Tmax}=15$ GeV in the cone,
as implemented in \cite{Bern:2002jx}.
\begin{figure}
\subfigure[  ]{\includegraphics[%
 width=6.5cm,
 angle=-90]{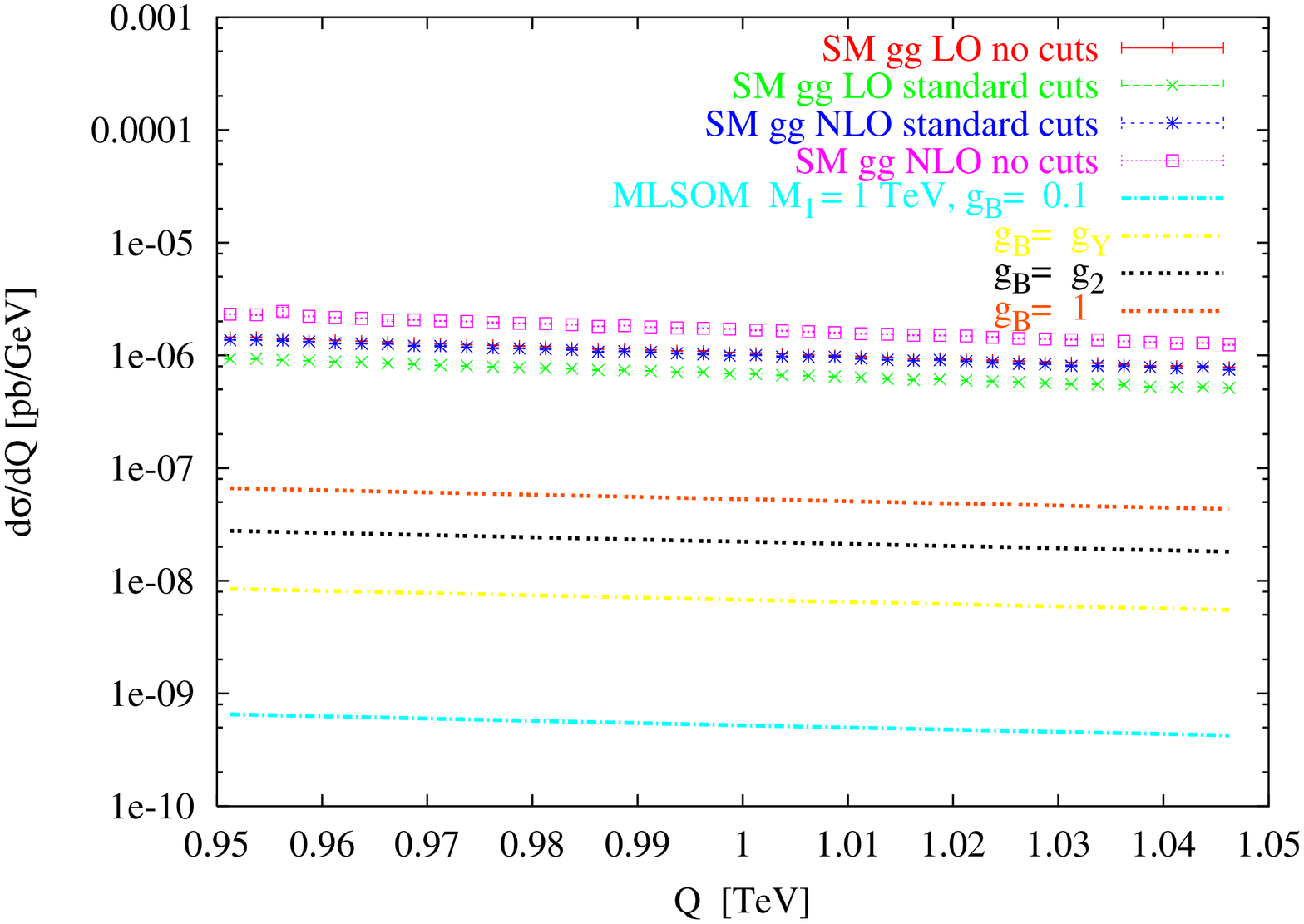}}
\subfigure[  ]{\includegraphics[%
 width=6.5cm,
 angle=-90]{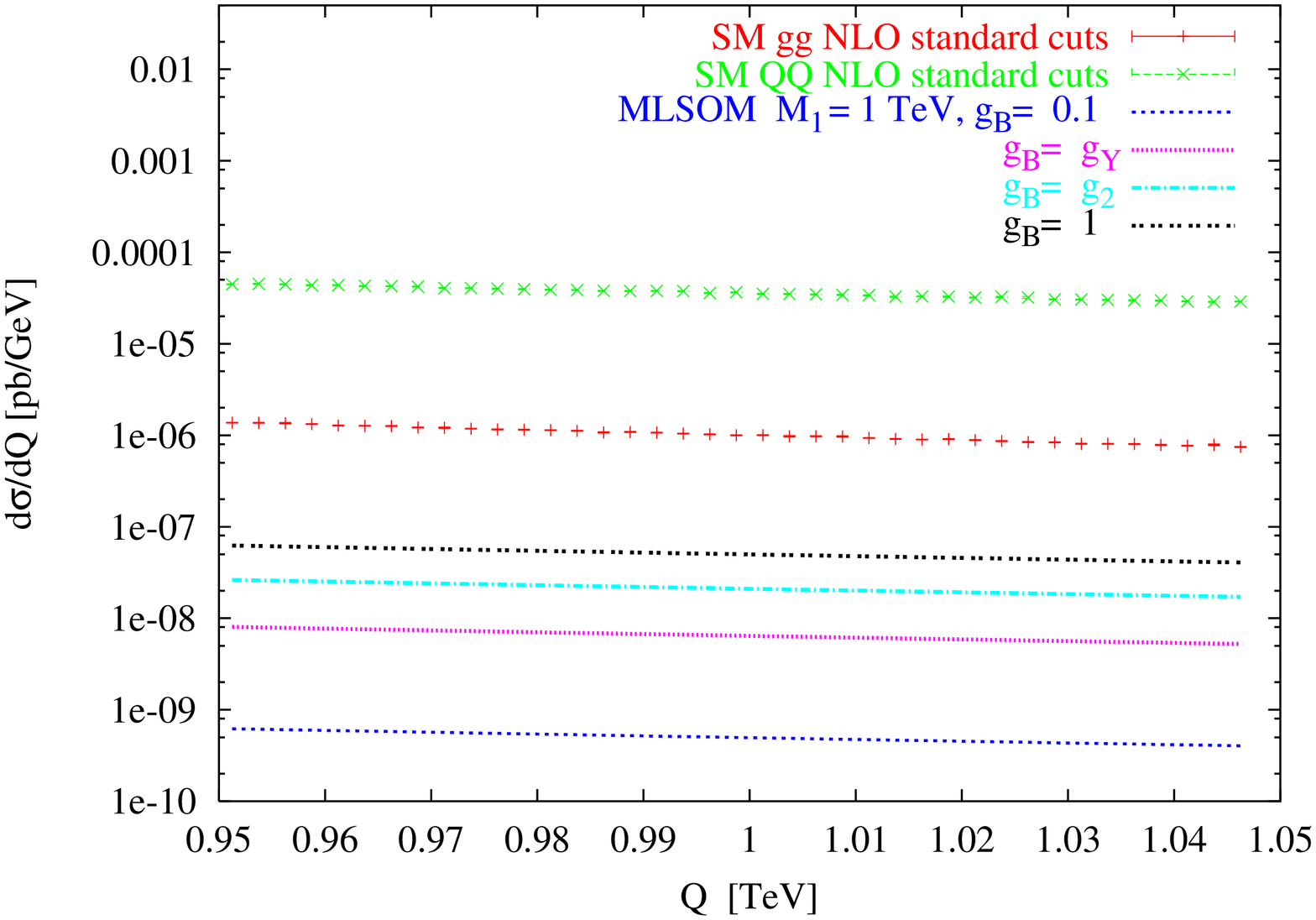}}
\caption{\small (a): SM contributions for the gluon-gluon channel obtained with the Monte
Carlo \textsc{Gamma2MC}. These are indicated by dotted
lines and include all the interferences and the box graphs.
Shown are also the anomalous contributions of the MLSOM (no box).
(b): as in (a) but we have included the Monte Carlo results for the SM qq channel at NLO.}
\label{ab}
\end{figure}
We show a more detailed investigation of the results for the various contributions
in the $gg$ sector in Fig.~\ref{ab}a, b. The dotted lines are the results obtained by the Monte Carlo and include
both 2-to-2 and 2-to-3 contributions (pure QCD) with and without cuts, computed at LO and at NLO. The size of these contributions is around $2\times 10^{-6}$ pb/GeV in the SM case. We show in the same subfigure the
anomalous corrections in the MLSOM, which vary between $10^{-9}$ and $10^{-7}$ pb/GeV. Therefore, for
$g_B\sim1$, the anomalous sector of the MLSOM (the square of the box terms here are not included for the MSLOM) is suppressed by a factor of 10 respect to the signal from the same sector coming from the SM.
In subfig.~(b) we show the same contributions but we include in the SM also the quark channel (shown separately from the gluon channel), which is around $10^{-4}$ pb/GeV. Therefore, the quark sector overshadows the anomalous corrections by a factor of approximately $10^{3}$, which are difficult to extract at this value of the invariant mass.

\begin{figure}
\subfigure[  ]{\includegraphics[%
width=6.5cm,
 angle=-90]{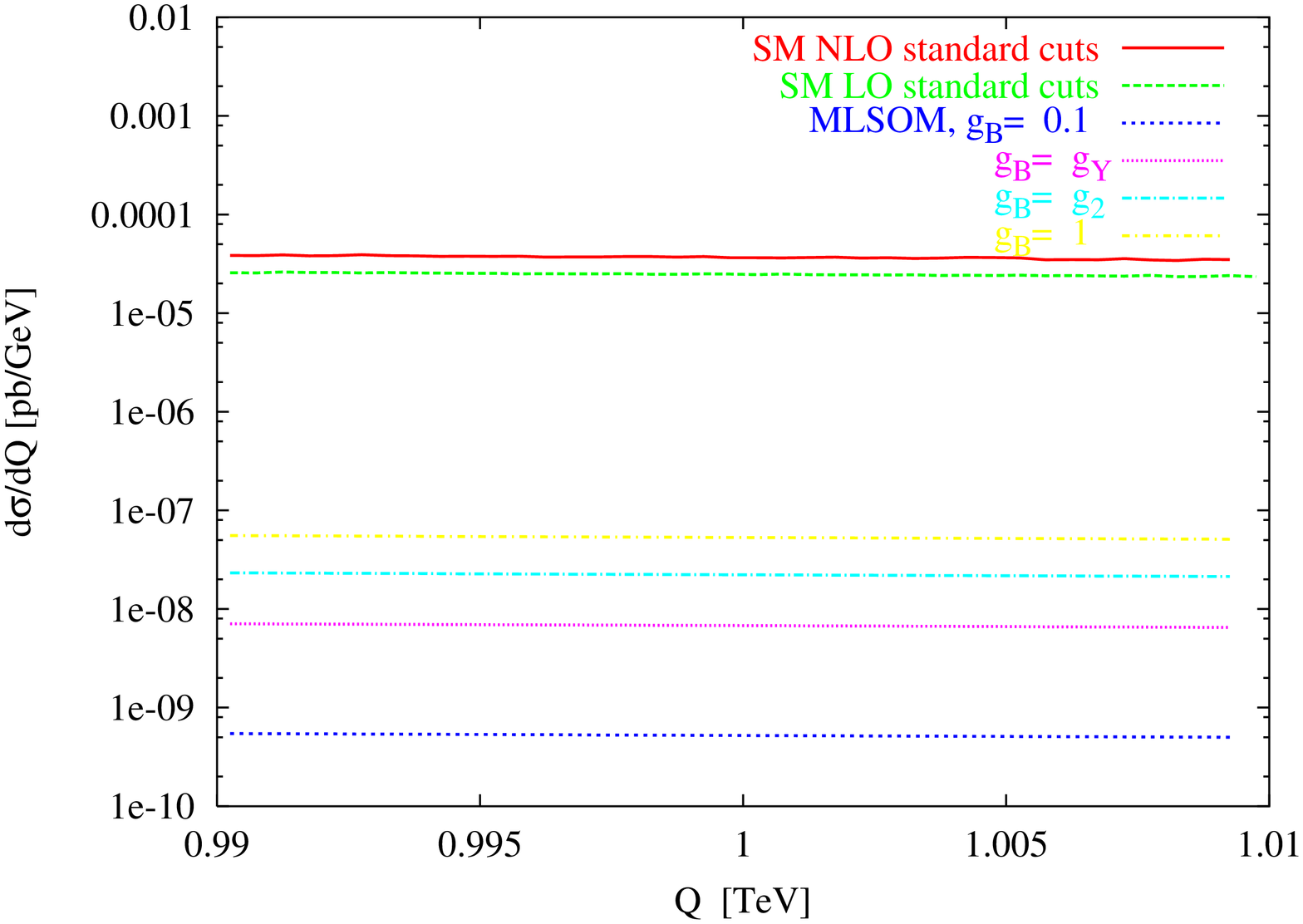}}
\subfigure[  ]{\includegraphics[%
 width=6.5cm,
 angle=-90]{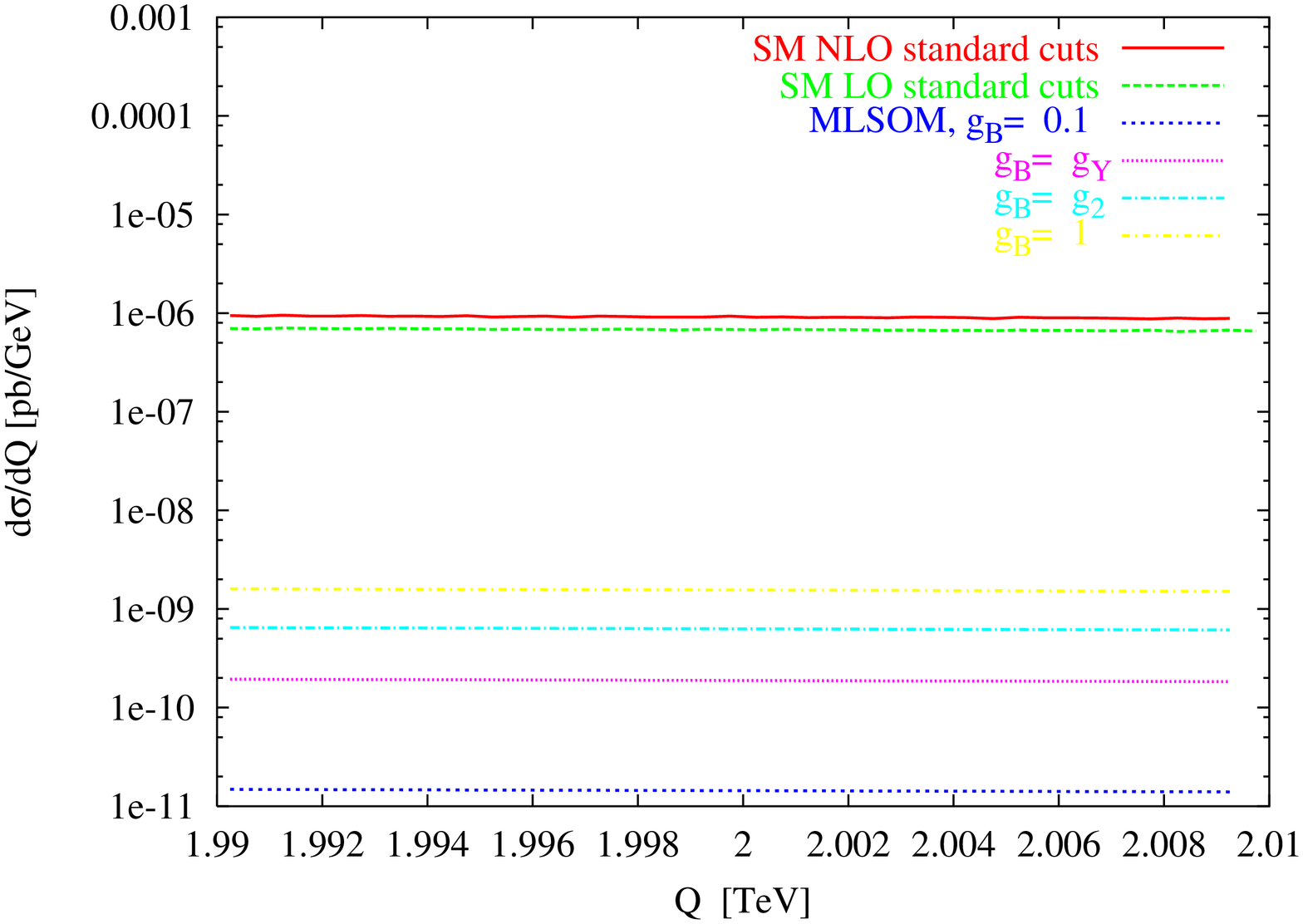}}
\caption{\small (a) Plot of the invariant mass distribution for $Q\sim 1 TeV$ and for a  St\"uckelberg mass of $1 TeV$
showing a comparison between the anomalous contributions (anomalous massless and massive BIM +
interference with the box) and the SM contributions. All the sectors at 1-loop and 2-loops have been obtained with the Monte Carlo \textsc{Gamma2MC}.
(b) As in the previous figure but the invariant mass distribution has been chosen around $2 TeV$.}
\label{new678}
\end{figure}

A comparison between the differential cross section  obtained by
the Monte Carlo and the anomalous contributions is shown in Fig.~\ref{new678}a,
from which one can see that the anomalous components are down by a factor of $10^3-10^7$
respect to the background, depending on the value of the anomalous coupling
$g_B$.

\begin{figure}
\subfigure[  ]{\includegraphics[%
width=6.5cm,
 angle=-90]{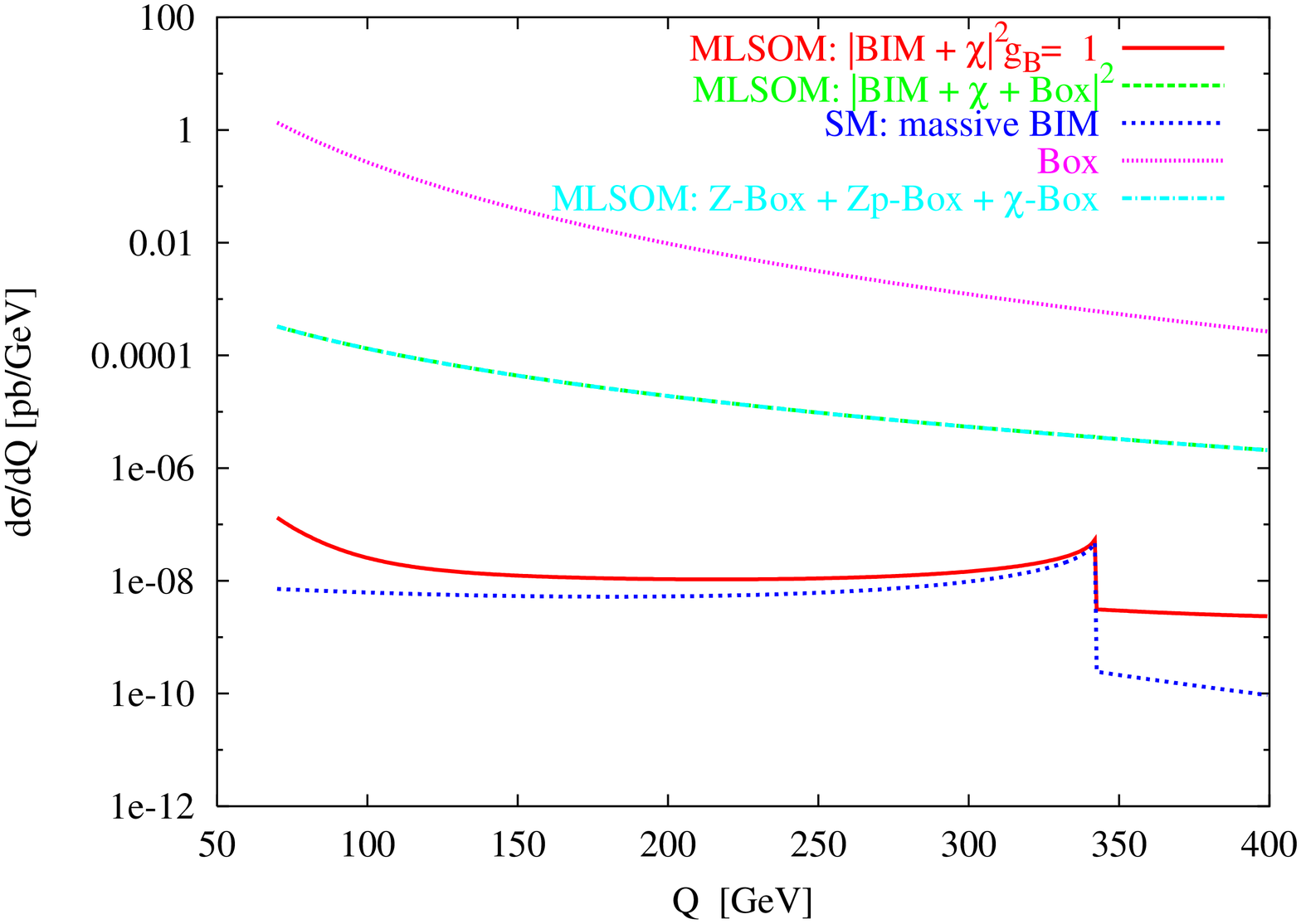}}
\subfigure[  ]{\includegraphics[%
 width=6.5cm,
 angle=-90]{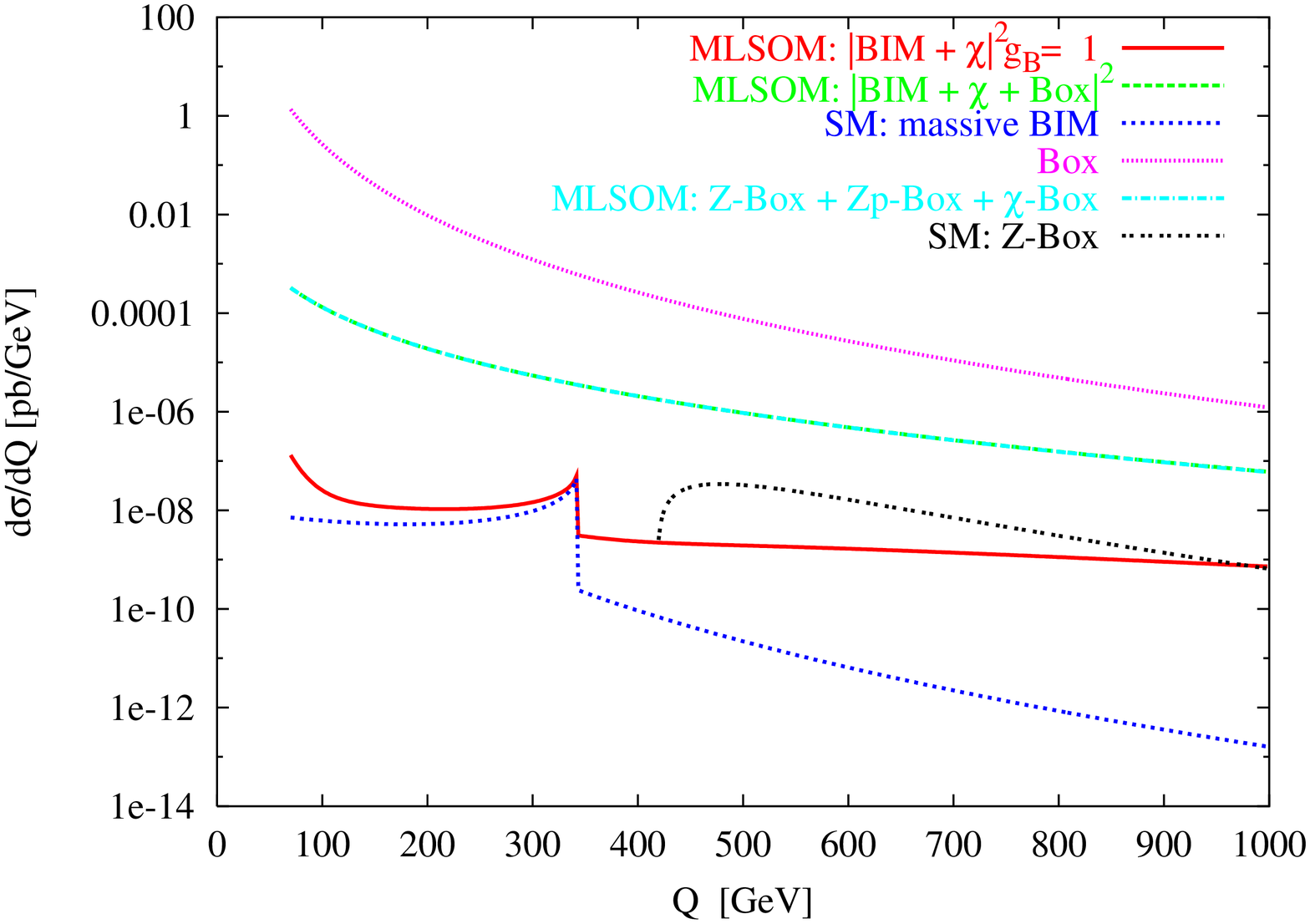}}
\caption{\small Plots of the various anomalous and box-like
components in the SM and MLSOM sector  at (a) lower $Q$ and (b) higher $Q$. }
\label{ex12}
\end{figure}

A similar comparison between anomalous signal and background
for a 1 TeV extra gauge boson but at larger invariant mass
(2 TeV) of the di-photon is shown in Fig.~\ref{new678}b. In both cases the
ratio between the size of the anomalous signal and the background
is $10^{-3}$, showing the large suppression as in the other regions.
\begin{itemize}
\item{\bf Anatomy of the gluon sector}
\end{itemize}
We show in Fig.~\ref{ex12}a and b) two plots that illustrate the size
of the various contributions to the
$gg$ sector in the SM and in the anomalous model. We have separated
these contributions into several
components in the MLSOM and SM cases.

In the anomalous model we have ``pure BIM-like'' amplitudes:

1) the square of the ``BIM set''  shown in Fig.~\ref{qgonedelta},
which contains the $s$-channel exchanges of the $Z$, of the
$Z^{\prime}$ and of the $\chi$. In Figs.~\ref{ex12}a and b these contributions are indicated as
``BIM + $\chi$''.

Analogously, in the SM case we have that

1$'$) the BIM amplitudes contribute away from the chiral limit due to
the exchange of the $Z$ gauge boson and with top/bottom  quarks
running inside each of the two loops. In Figs.~\ref{ex12}a and b
the contributions in the SM are denoted by ``SM: massive BIM'', and are
just obtained by squaring the single BIM amplitude of Fig.~\ref{qgonedelta}a.
In the SM this component is sizeable around the
threshold $s=4m_t^2$, with $m_t$ being the mass of the top quark,
which explains the cusp in the figure around this energy value.

The interferences sets of the MLSOM  (``BIM-Box'') in which the
BIM amplitudes interfere with the box graphs:

2) these are denoted as ``Z-Box + Zp-Box + $\chi$-Box''.
The three box diagrams are those shown in Fig.~\ref{nopq},
corresponding to graphs (o), (p) and (q) in this figure.

In the SM we have similar contributions. These are generated by

2$'$) interfering the same box graphs mentioned above with the graph  in
Fig.~\ref{qgonedelta}a. In the SM case, as we have already mentioned, these
contributions are due to heavy quarks, having neglected the mass of the leptons
and of the light quarks. This interference is denoted as ``Z-Box''.

Finally,

3)
In both cases we have the squared ``Box'' contributions, which are
just made from the o), p) and q) diagrams of Fig.~\ref{nopq}.

From a look at Figs.~\ref{ex12}a and b it is quite evident
that the contributions obtained by squaring the three box diagrams are
by far the most important at $\sqrt{S}=14$ TeV. We have tried to
cover both the region $Q < 400$ GeV and the region $400 < Q < 1000$ GeV.
The second largest contributions, in these two plots, are those due to the
interference between the BIM amplitudes and the box. These contributions
are enhanced because of the presence
of the box amplitude. Notice that in the SM case this interference is
small and negative for $Q < 400$ GeV
(see Fig.~\ref{ex12}a) and is not reported. In the second region
($400 < Q< 1000 $ GeV) this contributions gets sizeable
around the two-particle cut of the scalar triangle diagram due
to the top quark in the loop and shows up as a steepening in the
line labeled as ``Z-box'' in Fig.~\ref{ex12}b.

The ``pure anomalous'' contributions, due to the squared BIM amplitudes
are down by a factor of about $10^{5}$ respect to the dominant box contributions.
A similar trend shows up also in the SM case. Around the 2-particle cut the
BIM amplitudes both in the SM and in the MLSOM have a similar behaviour,
but differ substantially away from this point in the larger-$Q$ and smaller-$Q$ region.
These are the two regions
where the effects of the anomaly are more apparent.
For instance, for $Q< 200$ GeV the steep rise of
the anomalous contributions of the
``BIM + $\chi$'' line is due to the fact that $Q$ is getting closer to the
resonance of the axion $\chi$. In the SM the same region is characterized
by a suppression by the ratio ${m_f}^2/Q^2$. At larger $Q$ values,
the growth of the anomalous contributions are due to the bad behaviour of the
anomaly, which grows quadratically with energy, as we have discussed above.
This trend is more visible in Fig.~\ref{ex12}b (red line) in the region
$Q> 800$ GeV.
The possibility of increasing the di-photon (gluon-fusion)
signal is related to the possibility of imposing suitable
kinematical constraints in the analysis of the final state
- or phase-space cuts - in the experimental analysis.
To achieve this goal there are several features of the process
that should be kept into account. The first is the sharp decrease
of the cross section at large invariant mass $Q$ of the di-photon final state;
the second is its increase with $\sqrt{S}$, the collision energy of the two colliding protons.
In the first case we have an increase of the relevant parameter
$\tau=Q^2/S$ characterizing the gluon density; in the second
case this is reduced considerably, giving a fast enhancement of the gluon luminosity.
\begin{figure}[t]
\begin{center}
\includegraphics[width=7cm,angle=-90]{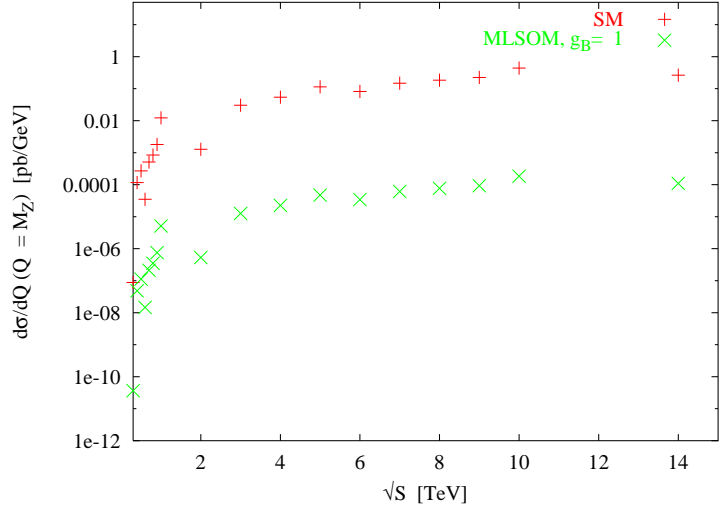}
\caption{\small Double photon invariant mass cross section at $Q=M_Z$ plotted as a function of the energy.
Here the SM contributions include the massive BIM + Box + interference, while the
MLSOM contributions include the massless and massive BIM + interference.}
\label{sqrts}
\end{center}
\end{figure}

\begin{figure}[t]
\begin{center}
\includegraphics[width=7cm,angle=-90]{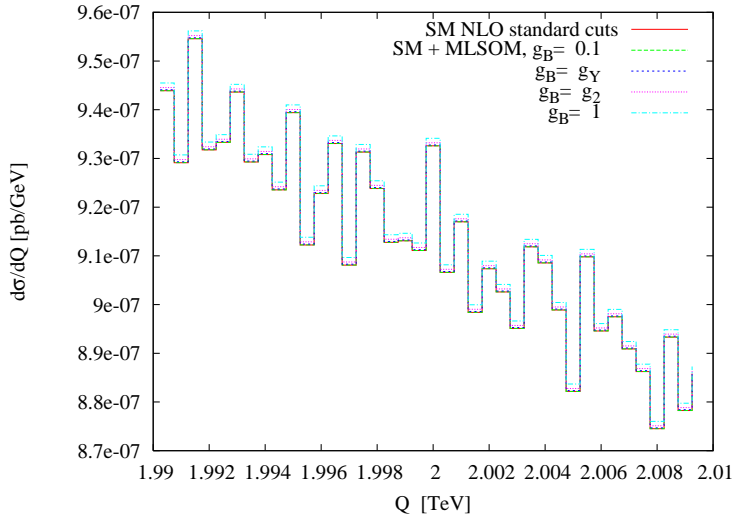}
\caption{\small Binned cross sections for the SM and the MLSOM. The small anomalous corrections are the thick lines at the top of each bin.}
\label{merge}
\end{center}
\end{figure}
 Since bounds on the coupling of the anomalous models may
come both from a combined analysis of both DY and of the
di-photon cross section, we have chosen $Q$ around the
$Z$ peak (small $Q$ option) and the same luminosity evaluated
on the peak of the extra $Z^{\prime}$ (large $Q$ option), both as
a function of the collisional energy $S$.  This is shown in Fig.~\ref{sqrts}.
The anomalous signal grows with $\sqrt{S}$ but the non-unitary growth is not apparent at LHC
energies after convolution with the parton densities; on the other end,
the growths in the SM and in the anomalous sector of the MLSOM appear to be,
at LHC energies, quite similar.

If an extra $Z^{\prime}$
is found in DY, then the analysis of di-photon both on the peak
of the $Z$ and on the peak of the
$Z^{\prime}$ could be used to set reasonable bounds on the coupling
of the underlying model, trying to uncover the possible presence of
an anomalous signal. A $\tau\sim 10^{-4}$ can be reached at the
$Z$ peak for a collision energy of 10 TeV, which causes an enhancement
of the anomalous cross section by a factor of approximately $10^4$
respect to the same anomalous contribution measured for $Q\sim 1$ TeV.

In a final figure we show the shapes of the distributions for the SM and the MLSOM using the Monte Carlo
with a binning of the cross section in both cases. This is provided in Fig.~\ref{merge},
where the colored lines on top of each bin indicate the small anomalous signal in the cross section.

\section{Summary and Conclusions }

Both DY and DP have some special features, being characterized by a clean final state. In DY the identification of a new resonance in the neutral current sector would bring to the immediate conclusion that an extra $Z^{\prime}$ is present in the spectrum, but would give not specific indication concerning its true nature. Current experimental bounds constrain the mixing of a possible extra neutral component with the Z gauge boson, with a mass which should be larger than 800 GeV, rendering the future search of extra neutral interactions, at least for DY, quite delicate, being the allowed mass range at the tail of the invariant mass distribution of this process. For this reason, the identification of a restricted mass range for the extra gauge boson would be crucial for its discovery, but unfortunately, there is no extension of the SM that comes with a definite prediction for it.

In intersecting brane models one encounters a similar indetermination and for this reason in our construction the St\"uckelberg mass has been assumed to be a free parameter.
If an extra s-channel resonance is found, then the investigation of its specific properties would require much more effort and several years of data collection at the LHC, especially for larger mass values (above 1.5 TeV or so) of the extra $Z^\prime$.  A sizeable width of the resonance could then allow to study the V-A structure of the coupling, though this study is quite complex. In most of the models studied so far the corresponding
widths are expected to be quite small ( $\ll 30-40$ GeV)
even for sizeable couplings ($g_B\sim O(1)$) \cite{Coriano:2008wf}, probably below or barely close to the limits of the effective resolution of the detector.

Under these conditions, deciding over the true nature of the extra $Z^\prime$, whether anomalous or not, would then
be far more challenging and would require a parallel study of several independent channels. For this reason we have analyzed two processes which are both affected by anomalous contributions
and could be used for correlated studies of the same interaction.

 We have seen that changes in the factorization/renormalization scales both in the hard scatterings and in the evolution of the PDF's can easily overshadow the anomalous corrections, making a NLO/NNLO analysis
 truly necessary.
 We have concentrated our investigation on an extra
$Z^{\prime}$ of 1 TeV in mass and searched for anomalous
effects in the invariant mass distributions  on the $Z$ peak,
at 1 TeV and for large $Q$ values (up to 2 TeV's). From our analysis the large
suppression of the anomalous signal compared to the QCD background
is rather evident, although there are ways to improve on our results.
In fact, our analysis has been based on the characterization of inclusive observables, which are not sensitive
to the geometric structure of the final state. In principle, the shape of the final state event could be
resolved at a finer level of detail, by analyzing, for instance, the rapidity correlations between the
di-photon and a jet, or other similar less inclusive cross sections, as in other cases \cite{Chang:1997sn}.   Obviously, these types of studies are theoretically challenging and require an excellent knowledge of the QCD background at NNLO for these observables. This precision could be achievable only by a combination of analytical and numerical  methods at such perturbative order, implemented in a flexible Monte Carlo program. This and other  related issues are left to future studies.

\centerline{\bf Acknowledgements}
We thank Nikos Irges for discussions. We thank Carl Schmidt for help
on the use of the \textsc{Gamma2MC} code for the
NNLO study of the QCD background in di-photon and J.P.
Guillet for exchanges concerning \textsc{Diphox}.
The work of C.C. was supported in part
by the European Union through the Marie Curie Research and
Training Network ``Universenet'' (MRTN-CT-2006-035863)
and by The Interreg II Crete-Cyprus Program.

\section{Appendix}

The decay rates into leptons for the $Z$ and the $Z^{\prime}$ are
universal and are given by
\ba
&&\Gamma({\cal Z}\rightarrow l\bar{l})=\frac{g^2}{192\pi c_w^2}
M_{{\cal Z}}\left[(g_{V}^{{\cal Z},l})^2+(g_{A}^{{\cal Z},l})^2\right]=
\frac{\alpha_{em}}{48 s_w^2 c_w^2}M_{{\cal Z}}\left[(g_{V}^{{\cal Z},l})^2+
(g_{A}^{{\cal Z},l})^2\right]\,,
\nonumber\\
&&\Gamma({\cal Z}\rightarrow \psi_i\bar{\psi_i})=\frac{N_c\alpha_{em}}{48
s_w^2 c_w^2}
M_{{\cal Z}}\left[(g_{V}^{{\cal Z},\psi_i})^2+(g_{A}^{{\cal
Z},\psi_i})^2\right]\times\nonumber\\
&&\hspace{3cm}\left[1+ \frac{\alpha_s(M_{{\cal Z}})}{\pi}
+1.409\frac{\alpha_s^2(M_{{\cal
Z}})}{\pi^2}-12.77\frac{\alpha_s^3(M_{\cal Z})}{\pi^3}\right],\,
\ea
where $i=u,d,c,s$ and ${\cal Z}=Z,Z^{\prime}$.

For the $Z^{\prime}$ and $Z$ decays into heavy quarks we obtain
\ba
&&\Gamma({\cal Z}\rightarrow b\bar{b})=\frac{N_c\alpha_{em}}{48 s_w^2 c_w^2}
M_{{\cal Z}}\left[(g_{V}^{{\cal Z},b})^2+(g_{A}^{{\cal
Z},b})^2\right]\times\nonumber\\
&&\hspace{3cm}\left[1+ \frac{\alpha_s(M_{{\cal
Z}})}{\pi}+1.409\frac{\alpha_s^2(M_{{\cal Z}})}{\pi^2}
-12.77\frac{\alpha_s^3(M_{\cal Z})}{\pi^3}\right]\,,
\nonumber\\
&&\Gamma({\cal Z}\rightarrow t\bar{t})=\frac{N_c\alpha_{em}}{48 s_w^2 c_w^2}
M_{{\cal Z}}\sqrt{1 - 4 \frac{m_t^2}{M_{{\cal Z}}^{2}} }\times\nonumber\\
&&\hspace{3cm}\left[(g_{V}^{{\cal Z},t})^2\left(1 + 2
\frac{m_t^2}{M_{{\cal Z}}^{2}}\right)
+(g_{A}^{{\cal Z},t})^2\left(1 - 4 \frac{m_t^2}{M_{{\cal
Z}}^{2}}\right)\right]\times\nonumber\\
&&\hspace{3cm}\left[1+ \frac{\alpha_s(M_{{\cal
Z}})}{\pi}+1.409\frac{\alpha_s^2(M_{{\cal Z}})}{\pi^2}
-12.77\frac{\alpha_s^3(M_{\cal Z})}{\pi^3}\right]\,.\nonumber\\
\ea

\section{Appendix. A comment on the unitarity breaking in WZ lagrangeans}

\begin{figure}[t]
\begin{center}
\includegraphics[scale=0.8]{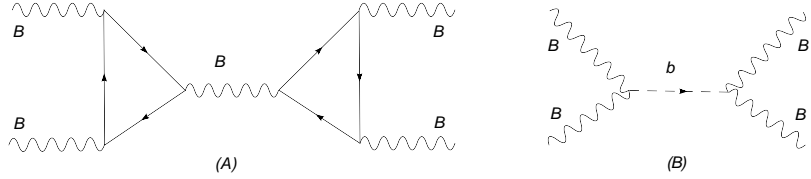}
\caption{\small BIM amplitude for a toy model with a $BBB$ anomaly and $\chi$ exchange diagram.}
\label{unitarity}
\end{center}
\end{figure}

\begin{figure}[t]
\begin{center}
\includegraphics[scale=0.8]{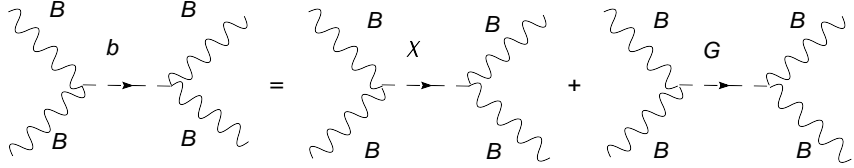}
\caption{\small Decomposition of the St\"uckelberg axion in a goldstone boson and a physical axion}
\label{rotationb}
\end{center}
\end{figure}

In the framework of an Abelian $AB$-model, see \cite{Coriano:2007fw} for more details,
the presence of an untamed growth of the amplitude for a 2-to-2
process can be simplified as follows. In this
simple model $A$ is vector-like and $B$ is axial-vector like.
We also have a single chiral fermion, with an uncancelled $BBB$
anomaly that requires an axion-like interaction $b F_B\wedge F_B$ of the St\"uckelberg
$(b)$ with the $B$ gauge field for the restoration of gauge invariance.
The two contributions to $BB\to BB$ are the BIM amplitude see Fig.~\ref{unitarity},\ref{rotationb}
for the process mediated by a B boson and a similar one mediated by a physical
axion $\chi$. If we neglect the Yukawa couplings
(we take the fermion to be massless) the only contributions
involved are those already shown in Fig.~\ref{qgonedelta}, diagrams (a) and (b),
but with different couplings.
\ba
A_{BIM}^{\mu \nu \mu^\prime \nu^\prime}= (g_B)^3 \Delta^{\lambda \mu \nu}(-k_1, -k_2) \frac{- i}{k^2-M_B^2} \left( g^{\lambda \lambda^\prime} - \frac{k^\lambda k^{\lambda^\prime}}{M_B^2} \right) (g_B)^3  \Delta^{\lambda^\prime \mu^\prime \nu^\prime}(k_1, k_2)
\ea
where $M_B=\sqrt{M_1^2 + (2 g_B v)^2}$ is the mass of the gauge boson in the s-channel after symmetry breaking.
The exchange of the physical axion gives
\ba
B^{\mu \nu \mu^\prime \nu^\prime}= 4 \times \left( \frac{4}{M} \alpha_1 C_{BB} \right)^2   \varepsilon[\mu, \nu, k_1, k_2]
\frac{i}{k^2-m_\chi^2}  \varepsilon[\mu^\prime, \nu^\prime, k^\prime_1, k^\prime_2]
\label{axi_exchange}
\ea
where the overall factor of 4 in front is a symmetry factor,
the coefficient $\alpha_1=\frac{2 g_B v}{M_B}$ comes from the rotation of
the $b$ axion over the axi-higgs $\chi$, and the coefficient
$C_{BB}$ has been determined from the condition of gauge invariance
of the anomalous effective action before symmetry breaking
\ba
C_{BB}= \frac{i g_B^3}{3!} \frac{1}{4} a_n \frac{M}{M_1}.
\ea
 The anomaly diagrams are longitudinal, taking the DZ form
\ba
A_{BIM}^{\mu \nu \mu^\prime \nu^\prime}(a_n)= (g_B)^3 \frac{a_n}{k^2} (-k^\lambda) \varepsilon[\mu, \nu, k_1,k_2] \frac{- i}{k^2-M_B^2} \left( g^{\lambda \lambda^\prime} - \frac{k^\lambda k^{\lambda^\prime}}{M_B^2} \right) (g_B)^3 \frac{a_n}{k^2} (k^{\lambda^\prime}) \varepsilon[\mu^\prime, \nu^\prime, k^\prime_1,k^\prime_2].
\ea
It is easy to recognize in this anomalous amplitude the same
structure of the amplitude for the axi-Higgs exchange in Eq.~(\ref{axi_exchange}).
If we add up these two amplitudes and impose the cancellation
of the two amplitudes (which is a fine tuning) we obtain the condition
\ba
\frac{1}{M_B^2} + \frac{(2 g_B v)^2}{M_B^2 M_1^2} = 0
\ea
in terms of the mass of the $B$ and the St\"uckelberg mass $M_1$,
which does not have a real solution. This is a simple example that conveys the
issue of the presence of unitarity limit for (local) WZ interactions.

\bibliographystyle{h-elsevier3}
%\bibliography{PhenomenalSubmissionJHEP}

\end{document}